\documentclass[12pt,reqno]{amsart}
\allowdisplaybreaks[4]
\usepackage{graphicx} 
\usepackage{amssymb}
\usepackage{amsmath}
\usepackage{cite}
\usepackage{subfigure}
\usepackage{graphicx}
\usepackage{epstopdf}
\newcommand{\pa}{\partial}

\begin{document}

\title[Darboux]{The rational solutions of the mixed nonlinear Schr\"odinger  equation}
\author{Jingsong He $^{1*}$,  Shuwei Xu $^{2}$, Yi Cheng $^{2}$
 }

\thanks{$^*$ Corresponding author: hejingsong@nbu.edu.cn,jshe@ustc.edu.cn}

 \maketitle
\dedicatory {\mbox{\hspace{0.3cm}}$^{1}$Department of Mathematics, Ningbo
University, Ningbo, Zhejiang 315211, P.\ R.\ China\\
\mbox{\hspace{0.75cm}}$^{2}$School of Mathematical
Sciences, USTC, Hefei, Anhui 230026, P.\ R.\ China\\}
\begin{abstract}
The mixed nonlinear Schr\"odinger (MNLS) equation is a model for the
propagation of the Alfv\'en wave in plasmas and the ultrashort light
pulse in optical fibers with two nonlinear effects of
self-steepening and self phase-modulation(SPM), which is also the
first non-trivial flow of the integrable Wadati-Konno-Ichikawa(WKI)
system. The determinant representation $T_n$ of a n-fold Darboux
transformation(DT) for the MNLS equation is presented. The
smoothness of the solution $q^{[2k]}$ generated by $T_{2k}$ is
proved for the two cases ( non-degeneration and  double-degeneration
) through the iteration and determinant representation. Starting
from a periodic seed(plane wave), rational solutions  with two
parameters $a$ and $b$ of the MNLS equation are constructed by the
DT and the Taylor expansion. Two parameters denote the contributions
of two nonlinear effects in solutions. We show an unusual result:
for a given value of $a$, the increasing value of $b$ can damage
gradually the localization of the rational solution, by analytical
forms and figures. A novel two-peak rational solution with  variable
height and a non-vanishing boundary is also obtained.
 \end{abstract}
\mbox{}\hspace{2cm}Darboux transformation.

{\bf PACS(2010) numbers}: 05.45.Yv, 42.65.Tg, 52.35.Bj,94.05.Fg,



\section{Introduction}
Rogue wave (RW) has been introduced and become an interesting objective
in the investigation of oceanography \cite{Kharif1,Kharif2}(and references therein),
starting with modeling a short-lived large amplitude wave in ocean.
Recently, rogue waves  have also been observed in photonic crystal
fibers \cite{Optical1,Optical2}, in space plasmas\cite{derman,
Moslem,shukla}, in Bose-Einstein condensates \cite{konotop1},
in water tanks\cite{akhmediev1,akhmediev2,akhmediev3},
and so on.

One of widely  accepted prototypes of  rogue wave in one dimensional
space and time is considered as Peregrine soliton\cite{Peregrine,Dysthe1,
shrira} of  the nonlinear Schr\"odinger  equation (NLS),
which usually takes the form of a single dominant peak accompanied by
one deep cave at each side in a plane with a nonzero boundary.
In other words, the characteristic property of the RW is localization in
both space and time directions in a nonzero plane.
The existence of this  solution is due to modulation instability
of the NLS equation \cite{Peregrine,zakharov1,zakharov2
,akhmediev4,akhmediev5}. In consequence, different patterns of the RW  will occur
when two or more breathers with different relative phase shifts
collide with each other\cite{Dubard2,Gaillard,triplets,ohtayang,Circular,guo,hegenerating2012,Gaillard2}.
One of the possible generating mechanisms for
rogue waves is through the creation of breathers possessing a
particular frequency, which is realized theoretically by choosing a
special eigenvalue in breathers\cite{hegenerating2012}.
Recently,by applying Darboux transformation(DT) \cite{matveev,hedt,kenji1,steduel},
 the rogue waves \cite{xuhe,xxuhe,gzxwhe2014} of derivative nonlinear Schr\"odinger
 equation (DNLS) are also given in the form of "Peregrine soliton".

In the field of optics, the nonlinear terms in the NLS and DNLS  denote the effects of
phase-modulation (SPM) and self-steepening, respectively. So it is natural and worthwhile
to look for an integrable equation with these two terms from the points of view of
mathematics and physics. There is indeed such an integrable equation-a mixed NLS (MNLS)
equation \cite{KMio,Zabolotskii}\begin{equation}\label{MDNLS}
q_{t}-iq_{xx}+a(q^{\ast}q^2)_{x}+ib q^{\ast}q^2=0,
\end{equation}
in physics. Here $q$ represents a complex field envelope and
asterisk denotes complex conjugation, $a$ and $b$ are two
non-negative constants, and subscript $x$ (or $t$) denotes partial
derivative with respect to $x$ (or $t$). The MNLS equation is used to
model the propagations of the Alfv\'en waves in plasmas\cite{KMio} and
the ultrashort light pulse in optical fibersc\cite{Zabolotskii}.
Moreover, the MNLS equation can also be given by following
coupled system
\begin{equation}\label{sy1}
r_{t}+ir_{xx}-a(r^2q)_{x}+ibr^2q=0,
\end{equation}
\begin{equation}\label{sy2}
q_{t}-iq_{xx}-a(rq^2)_{x}-ibrq^2=0,
\end{equation}
under a condition $r=-q^\ast$. This coupled system is nothing but the
first non-trivial flow of the Wadati-Konno-Ichikawa(WKI) system\cite{MikiWadati1}, and
the corresponding Lax pair is given by the WKI spectral problem and a time flow \cite{MikiWadati1}
\begin{equation}\label{sys11}
 \pa_{x}\psi=(-aJ\lambda^2+Q_1\lambda+Q_0)\psi=U\psi,
\end{equation}
\begin{equation}\label{sys22}
\pa_{t}\psi=(-2a^2J\lambda^4+V_{3}\lambda^3+V_{2}\lambda^2+V_{1}\lambda+V_0)\psi=V\psi,
\end{equation}
with
\begin{equation}
     J= \left( \begin{array}{cc}
      i &0 \\
      0 &-i\\
   \end{array} \right),\nonumber\\
  \quad Q_1=\left( \begin{array}{cc}
     \sqrt{2b} &aq \\
     ar & -\sqrt{2b}\\
  \end{array} \right),\nonumber\\
   \quad Q_0=\left( \begin{array}{cc}
     0 &i\sqrt{\dfrac{b}{2}}q \\
     i\sqrt{\dfrac{b}{2}}r & 0\\
  \end{array} \right),\nonumber\\
\end{equation}
\begin{equation}
\psi=\left(\begin{array}{c}
      \phi \\
      \varphi\\
     \end{array} \right),\nonumber\\
   \quad V_{3}=\left( \begin{array}{cc}
     4a\sqrt{2b} &2a^2q \\
     2a^2r & -4a\sqrt{2b}\\
  \end{array} \right),\nonumber\\  \quad V_{2}=\left( \begin{array}{cc}
     4ib-ia^2rq &3ia\sqrt{2b}q \\
      3ia\sqrt{2b}r& -(4ib-ia^2rq)\\
  \end{array} \right),\nonumber\\
\end{equation}
\begin{equation}
  V_{1}=\left(\mbox{\hspace{-0.3cm}}\begin{array}{cc}
a\sqrt{2b}qr &\mbox{\hspace{-0.3cm}}-2bq+ia q_x+a^2rq^2 \\
 -2br-ia r_x+a^2 q r^2&\mbox{\hspace{-0.3cm}}-a\sqrt{2b}qr\\
\end{array}\mbox{\hspace{-0.2cm}}\right),\nonumber\\
V_{0}=\left(\mbox{\hspace{-0.3cm}}\begin{array}{cc}
\dfrac{1}{2}ibqr &\mbox{\hspace{-0.3cm}}\sqrt{\dfrac{1}{2}b}(-q_x+iarq^2) \\
 \sqrt{\dfrac{1}{2}b}(r_x+iar^2q)&\mbox{\hspace{-0.3cm}}-\dfrac{1}{2}ibqr\\
\end{array}\mbox{\hspace{-0.3cm}}\right).\nonumber\\
\end{equation}
Here $\lambda\in \mathbb{C}$, is called the eigenvalue(or spectral parameter), and
$\psi$ is called the eigenfunction associated with  $\lambda$ of the
WKI system. Equations(\ref{sy1}) and (\ref{sy2}) are equivalent to
the integrability condition  $U_{t}-V_{x}+[U,V]=0$ of (\ref{sys11})
and (\ref{sys22}). In addition, the MNLS
 equation can be generated from other systems (or equations )
 in the literatures\cite{AKundu,Seenuvasakumaran,QingDing,Kundu}.
The MNLS equation is exactly solved by the inverse scattering method
under the non-vanishing boundary condition \cite{TKawata}.  Later,
Periodic solutions of the MNLS equation are be analyzed in terms of
the Riemann's $\theta$ functions\cite{Chowdhury}. Some special
solutions, such as breather solutions, of this equation are also
discussed in Ref\cite{mihalache,Doktorov2}. At the same time, the
solutions of the MNLS equation have  been constructed via  Backlund
or Darboux
transformation\cite{Rangwala,OCWrigh,TiechengXia,HaiQiangZhang} and
the Hirota method\cite{Shanliang,MinLi}. What is more, using the
matrix Riemann-Hilbert factorization approach, asymptotic analysis
of the MNLS equation is discussed in Ref.\cite{Kitaevy,Vartaniana}.
Considering many wave propagation phenomena described by integrable
equations in some ideal conditions, the effects of small
perturbations on the MNLS equation are study by the direct soliton
perturbation theory\cite{perturbation} and the perturbation theory
based on the inverse scattering
transform\cite{Shchesnovich,VMLashkin}. Recently, the semiclassical
analysis of the MNLS has been studied in Ref.\cite{miller1,miller2,
miller3}.

In light of the above results, two questions arise  naturally.
First, is there a rational solution of the MNLS equation which can
be generated from a periodic seed by DT and Taylor expansion?
Second, how the localization of this rational solution is affected
by the two nonlinear effects through the $a$ and $b$? For the first
question, the first order and the second order rational solutions
are given explicitly by the determinant representation of the DT and
Taylor expansion with respect to the degenerate eigenvalues. To
answer the second question, according to a common understanding of
the role for the nonlinear effects in wave propagation, one
reasonable conjecture is that the localization of this solution will
be enhanced because of the appearance of the two nonlinear effects.
However we shall show an unusual result: for a given value of $a$,
the increasing value of $b$ can damage gradually the localization of
the rational solution.

The organization of this paper is as follows. In section 2, we provide a relatively simple
approach to DT for the WKI system, and then the expressions of the $q^{[n]}$, $r^{[n]}$
and $\psi_j^{[n]}$ of  the WKI system are generated by n-fold Darboux transformation.
The reduction of DT for the WKI system to the MNLS equation is also discussed by
choosing paired eigenvalues and eigenfunctions. In section 3, the smoothness of the
solutions $q^{[2k]}$ generated by $T_{2k}$ is proved for the two cases (non-degeneration and
double-degeneration). In section 4, we present the rational solutions
of the MNLS and discuss its localized properties for a given value of $a$.
Finally, we summary our results in section 5.

\section{Darboux transformation}

\indent Inspired by the results of the DT for the NLS \cite{matveev,hedt} and the DNLS
\cite{xuhe,xxuhe,kenji1,steduel}, the main task of this section is to present a detailed derivation of
the Darboux transforation of the MNLS and the determinant
representation of the n-fold transformation.  It is easy to see that the
spectral problem (\ref{sys11}) and (\ref{sys22}) are transformed
to
  \begin{equation}\label{bh1}
{\psi^{[1]}}_{x}=U^{[1]}~\psi^{[1]},\ \ U^{[1]}=(T_{x}+T~U)T^{-1}.
\end{equation}
\begin{equation}\label{bh2}
{\psi^{[1]}}_{t}=V^{[1]}~\psi^{[1]}, \ \ V^{[1]}=(T_{t}+T~V)T^{-1}.
\end{equation}\\
under a gauge transformation \begin{equation}\label{bh3}
\psi^{[1]}=T~\psi.
\end{equation}
By cross differentiating (\ref{bh1}) and (\ref{bh2}), we obtain
\begin{equation}\label{bh4}
{U^{[1]}}_{t}-{V^{[1]}}_{x}+[{U^{[1]}},{V^{[1]}}]=T(U_{t}-V_{x}+[U,V])T^{-1}.
\end{equation}
This implies that, in order to make eqs.(\ref{sy1}) and eq.(\ref{sy2}) invariant under the
transformation (\ref{bh3}), it is crucial to search a matrix $T$ so that  $U^{[1]}$, $V^{[1]}$have
the same forms as $U$, $V$. At the same time the old potential(or
seed solution)($q$, $r$) in spectral matrixes $U$, $V$ are mapped
into new potentials
(or new solution)($q^{[1]}$, $r^{[1]}$) in transformed spectral
matrixes $U^{[1]}$, $V^{[1]}$. Next, it is necessary to parameterize the matrix $T$ by the eigenfunctions associated with the seed solution.\\
\\
{2.1 \bf One-fold Darboux transformation of the WKI system}
\\
\\
\indent Without losing any generality, let  Darboux
matrix $T$ be  the  form  of
\begin{equation}\label{TT}
 T_{1}=T_{1}(\lambda;\lambda_1)=\left( \begin{array}{cc}
a_{1}&0 \\
0 &d_{1}\\
\end{array} \right)\lambda+\left( \begin{array}{cc}
i\dfrac{\sqrt{2b}}{2a}a_{1}&b_{0}\\
c_{0} &i\dfrac{\sqrt{2b}}{2a}d_{1}\\
\end{array} \right).
\end{equation}
Here $a_{1}, d_{1}, b_0$ and $c_0$ are undetermined function of ($x$, $t$), which
will be parameterized by the eigenfunction associated with $\lambda_1$ and seed ($q,r$)
in the WKI spectral problem. Refer to Appendix I for detail derivation.

First of all, we introduce  $n$
eigenfunctions $\psi_j$ as
\begin{eqnarray}
&&\psi_{j}=\left(
\begin{array}{c}\label{jie2}
 \phi_{j}   \\
 \varphi_{j}  \\
\end{array} \right),\ \ j=1,2,....n,\phi_{j}=\phi_{j}(x,t,\lambda_{j}), \
\varphi_{j}=\varphi_{j}(x,t,\lambda_{j}). \label{jie1}
\end{eqnarray}

\noindent {\bf Theorem 1.}{\sl  The  elements of one-fold DT  are  parameterized by the
eigenfunction $\psi_1$ associated with
$\lambda_1$ as
 \begin{eqnarray}\label{DT1aibi}
&&d_{1}=\dfrac{1}{a_{1}}, \ a_{1}=-\frac{\varphi_{1}}{\phi_{1}}\exp(-i(\dfrac{b}{a}x+\dfrac{b^2}{a^2}t)),\nonumber\\
&&b_{0}=(\lambda_1+i\dfrac{\sqrt{2b}}{2a})\exp(-i(\dfrac{b}{a}x+\dfrac{b^2}{a^2}t)), \ c_{0}=(\lambda_1+i\dfrac{\sqrt{2b}}{2a})\exp(i(\dfrac{b}{a}x+\dfrac{b^2}{a^2}t), \nonumber\\
\end{eqnarray}
then
\begin{eqnarray} T_1(\lambda;\lambda_1)=\left(
\begin{array}{cc}
-\frac{\varphi_{1}}{\phi_{1}}\exp(-i(\dfrac{b}{a}x+\dfrac{b^2}{a^2}t))(\lambda+i\dfrac{\sqrt{2b}}{2a})& (\lambda_1+i\dfrac{\sqrt{2b}}{2a})\exp(-i(\dfrac{b}{a}x+\dfrac{b^2}{a^2}t))\\
(\lambda_1+i\dfrac{\sqrt{2b}}{2a})\exp(i(\dfrac{b}{a}x+\dfrac{b^2}{a^2}t)
&-\frac{\phi_{1}}{\varphi_{1}}\exp(i(\dfrac{b}{a}x+\dfrac{b^2}{a^2}t))(\lambda+i\dfrac{\sqrt{2b}}{2a})
\end{array}  \right).  \label{DT1matrix}
\end{eqnarray}
$T_1$ implies following new solutions
\begin{eqnarray}\label{sTT}
&&q^{[1]}=(\frac{\varphi_{1}}{\phi_{1}})^2\exp(-2i(\dfrac{b}{a}x+\dfrac{b^2}{a^2}t))q-2i\frac{\varphi_{1}}{\phi_{1}}(\lambda_1+i\dfrac{\sqrt{2b}}{2a})\exp(-2i(\dfrac{b}{a}x+\dfrac{b^2}{a^2}t)),\nonumber\\
&&r^{[1]}=(\frac{\phi_{1}}{\varphi_{1}})^2\exp(2i(\dfrac{b}{a}x+\dfrac{b^2}{a^2}t))r+
2i\frac{\phi_{1}}{\varphi_{1}}(\lambda_1+i\dfrac{\sqrt{2b}}{2a})\exp(2i(\dfrac{b}{a}x+\dfrac{b^2}{a^2}t)),\nonumber\\
\end{eqnarray}
and corresponding new eigenfunction
\begin{equation}
\psi^{[1]}_j= \left(
\begin{array}{c}
\dfrac{1}{\phi_1}\left|\begin{array}{cc}
-(\lambda_j+i\dfrac{\sqrt{2b}}{2a})\phi_j & \varphi_j\\
-(\lambda_1+i\dfrac{\sqrt{2b}}{2a})\phi_1 & \varphi_1
 \end{array}\right| \exp(-i(\dfrac{b}{a}x+\dfrac{b^2}{a^2}t))\\ \\
\dfrac{1}{\varphi_1} \left|\begin{array}{cc}
-(\lambda_j+i\dfrac{\sqrt{2b}}{2a})\varphi_j & \phi_j\\
-(\lambda_1+i\dfrac{\sqrt{2b}}{2a})\varphi_1 & \phi_1
 \end{array}\right|\exp(i(\dfrac{b}{a}x+\dfrac{b^2}{a^2}t))
\end{array}
\right).
\end{equation}
} {\bf Proof.} See appendix II.\\
It is straightforward to verify that $T_1$ annihilate its generating function, i.e., $\psi^{[1]}_1=0$. \\
\\
{\bf 2.2 n-fold Darboux transformation for WKI system}
\\
\\
\indent
The main result in this subsection is the determinant representation of the
n-fold DT for WKI system. To this purpose, set
\cite{kenji1}
$$
\begin{array}{cc}
\textbf{D}=&\left\{\left.\left(\begin{array}{cc}
a& 0\\
0&d
  \end{array} \right)\right| a,d \text{ are complex functions of}\ x\ \text{and}\ t  \right\},\\
\textbf{A}=&\left\{\left.\left(\begin{array}{cc}
0& b\\
c&0
  \end{array} \right)\right| b,c \text{ are complex functions of}\ x\ \text{and}\ t  \right\}.
\end{array}
$$

 According to the form of $T_1$ in eq.(\ref{TT}), the n-fold DT should be the form of
\begin{equation}\label{tnss}T_{n}=\widetilde{T_{n}}(\widetilde{\lambda};\widetilde{\lambda_1},\widetilde{\lambda_2}, \cdots,\widetilde{\lambda_n})=\sum_{l=0}^{n}P_{l}\widetilde{\lambda}^{l}, \;
\end{equation}
with $ \widetilde{\lambda}=\lambda+i\dfrac{\sqrt{2b}}{2a},
 \widetilde{\lambda_j}=\lambda_j+i\dfrac{\sqrt{2b}}{2a}$, $\lambda_i\not=
\lambda_j$ if $i\not=j$ and
\begin{eqnarray}
\label{tnsss}
P_{n}=\left( \mbox{\hspace{-0.2cm}}
\begin{array}{cc}
a_{n}\mbox{\hspace{-0.3cm}}&0 \\
0 \mbox{\hspace{-0.3cm}}&d_{n}\\
\end{array}  \mbox{\hspace{-0.2cm}}\right)\in \textbf{D},\ P_{n-1}=\left( \mbox{\hspace{-0.2cm}}
\begin{array}{cc}
0 \mbox{\hspace{-0.3cm}}&b_{n-1} \\
c_{n-1} \mbox{\hspace{-0.3cm}}&0\\
\end{array}  \mbox{\hspace{-0.2cm}}\right)\in \textbf{A},\ P_{l}\in \textbf{D}&\textrm{\small \mbox{\hspace{-0.3cm}}(if $l-n$ is even)}
,\ P_{l}\in \textbf{A}&\textrm{\small \mbox{\hspace{-0.3cm}}(if $l-n$ is odd)}. \nonumber
\end{eqnarray}
In order to get diagonal matrix and anti-matrix coefficients in $\widetilde{T_{n}}$, we introduce $\widetilde{\lambda}$ and $\widetilde{\lambda_j}$ by a shift. Here $P_{0}$ is a constant matrix, $P_i(1\leq i\leq n)$ is the function of $x$ and $t$.
In particular, $P_0 \in \textbf{D}$ if $n$ is even and $P_0 \in \textbf{A}$ if $n$ is odd, which
leads to a separate discussion on the determinant representation of $T_n$ in the following
by means of its kernel.

Specifically, from algebraic equations,
\begin{equation}\label{ttnss}
\psi_{k}^{[n]}=\widetilde{T_{n}}(\widetilde{\lambda};\widetilde{\lambda_1},\widetilde{\lambda_2}, \cdots,\widetilde{\lambda_n})|_{\lambda=\lambda_k}\psi_{k}=\sum_{l=0}^{n}P_{l}\widetilde{\lambda_{k}}^{l}\psi_{k}=0,
k=1,2,\cdots,n,
\end{equation}
coefficients $P_i$ are solved by Cramer's rule. Thus we get determinant representation of the $T_n$.

\noindent {\bf Theorem2.} {\sl (1) For $n=2k (k=1,2,3,\cdots)$, the n-fold
DT of the WKI system can be expressed by
\begin{equation}
\label{fss1}T_{n}=\widetilde{T_{n}}(\widetilde{\lambda};\widetilde{\lambda_1},\widetilde{\lambda_2}, \cdots,\widetilde{\lambda_n}) =\left(
\begin{array}{cc}
\dfrac{\widetilde{(T_{n})_{11}}}{W_{n}}& \dfrac{\widetilde{(T_{n})_{12}}}{W_{n}}\\ \\
\dfrac{\widetilde{(T_{n})_{21}}}{\widetilde{W_{n}}}& \dfrac{\widetilde{(T_{n})_{22}}}{\widetilde{W_{n}}}\\
\end{array} \right),
\end{equation}
with
\begin{equation}\label{fsst1}
W_{n}=\begin{vmatrix}
\widetilde{\lambda_{1}}^{n-1}\phi_{1}&\widetilde{\lambda_{1}}^{n-2}\varphi_{1}&\ldots&\widetilde{\lambda_{1}}\phi_{1}&\varphi_{1}\\
\widetilde{\lambda_{2}}^{n-1}\phi_{2}&\widetilde{\lambda_{2}}^{n-2}\varphi_{2}&\ldots&\widetilde{\lambda_{2}}\phi_{2}&\varphi_{2}\\
\vdots&\vdots&\vdots&\vdots&\vdots\\
\widetilde{\lambda_{n-1}}^{n-1}\phi_{n-1}&\widetilde{\lambda_{n-1}}^{n-2}\varphi_{n-1}&\ldots&\widetilde{\lambda_{n-1}}\phi_{n-1}&\varphi_{n-1}\\
\widetilde{\lambda_{n}}^{n-1}\phi_{n}&\widetilde{\lambda_{n}}^{n-2}\varphi_{n}&\ldots&\widetilde{\lambda_{n}}\phi_{n}&\varphi_{n}\nonumber\\
\end{vmatrix},
\end{equation}
\begin{equation}\label{fsst2}
\widetilde{(T_{n})_{11}}=\begin{vmatrix}
\widetilde{\lambda}^{n}&0&\ldots&\widetilde{\lambda}^{2}&0&1\\
\widetilde{\lambda_{1}}^{n}\phi_{1}&\widetilde{\lambda_{1}}^{n-1}\varphi_{1}&\ldots&\widetilde{\lambda_{1}}^{2}\phi_{1}&\widetilde{\lambda_{1}}\varphi_{1}&\phi_{1}\\
\widetilde{\lambda_{2}}^{n}\phi_{2}&\widetilde{\lambda_{2}}^{n-1}\varphi_{2}&\ldots&\widetilde{\lambda_{2}}^{2}\phi_{2}&\widetilde{\lambda_{2}}\varphi_{2}&\phi_{2}\\
\vdots&\vdots&\vdots&\vdots&\vdots&\vdots\\
\widetilde{\lambda_{n-1}}^{n}\phi_{n-1}&\widetilde{\lambda_{n-1}}^{n-1}\varphi_{n-1}&\ldots&\widetilde{\lambda_{n-1}}^{2}\phi_{n-1}&\widetilde{\lambda_{n-1}}\varphi_{n-1}&\phi_{n-1}\\
\widetilde{\lambda_{n}^{n}}\phi_{n}&\widetilde{\lambda_{n}}^{n-1}\varphi_{n}&\ldots&\widetilde{\lambda_{n}}^{2}\phi_{n}&\widetilde{\lambda_{n}}\varphi_{n}&\phi_{n}\nonumber\\
\end{vmatrix},
\end{equation}
\begin{equation}\label{fsst3}
\widetilde{(T_{n})_{12}}=\begin{vmatrix}
0&\widetilde{\lambda}^{n-1}&\ldots&0&\widetilde{\lambda}&0\\
\widetilde{\lambda_{1}}^{n}\phi_{1}&\widetilde{\lambda_{1}}^{n-1}\varphi_{1}&\ldots&\widetilde{\lambda_{1}}^{2}\phi_{1}&\widetilde{\lambda_{1}}\varphi_{1}&\phi_{1}\\
\widetilde{\lambda_{2}}^{n}\phi_{2}&\widetilde{\lambda_{2}}^{n-1}\varphi_{2}&\ldots&\widetilde{\lambda_{2}}^{2}\phi_{2}&\widetilde{\lambda_{2}}\varphi_{2}&\phi_{2}\\
\vdots&\vdots&\vdots&\vdots&\vdots&\vdots\\
\widetilde{\lambda_{n-1}}^{n}\phi_{n-1}&\widetilde{\lambda_{n-1}}^{n-1}\varphi_{n-1}&\ldots&\widetilde{\lambda_{n-1}}^{2}\phi_{n-1}&\widetilde{\lambda_{n-1}}\varphi_{n-1}&\phi_{n-1}\\
\widetilde{\lambda_{n}}^{n}\phi_{n}&\widetilde{\lambda_{n}}^{n-1}\varphi_{n}&\ldots&\widetilde{\lambda_{n}}^{2}\phi_{n}&\widetilde{\lambda_{n}}\varphi_{n}&\phi_{n}\nonumber\\
\end{vmatrix},
\end{equation}
\begin{equation}\label{fsst4}
\widetilde{W_{n}}=\begin{vmatrix}
\widetilde{\lambda_{1}}^{n-1}\varphi_{1}&\widetilde{\lambda_{1}}^{n-2}\phi_{1}&\ldots&\widetilde{\lambda_{1}}\varphi_{1}&\phi_{1}\\
\widetilde{\lambda_{2}}^{n-1}\varphi_{2}&\widetilde{\lambda_{2}}^{n-2}\phi_{2}&\ldots&\widetilde{\lambda_{2}}\varphi_{2}&\phi_{2}\\
\vdots&\vdots&\vdots&\vdots&\vdots\\
\widetilde{\lambda_{n-1}}^{n-1}\varphi_{n-1}&\widetilde{\lambda_{n-1}}^{n-2}\phi_{n-1}&\ldots&\widetilde{\lambda_{n-1}}\varphi_{n-1}&\phi_{n-1}\\
\widetilde{\lambda_{n}}^{n-1}\varphi_{n}&\widetilde{\lambda_{n}}^{n-2}\phi_{n}&\ldots&\widetilde{\lambda_{n}}\varphi_{n}&\phi_{n}\nonumber\\
\end{vmatrix},
\end{equation}
\begin{equation}\label{fsst5}
\widetilde{(T_{n})_{21}}=\begin{vmatrix}
0&\widetilde{\lambda^{n-1}}&\ldots&0&\widetilde{\lambda}&0\\
\widetilde{\lambda_{1}}^{n}\varphi_{1}&\widetilde{\lambda_{1}}^{n-1}\phi_{1}&\ldots&\widetilde{\lambda_{1}}^{2}\varphi_{1}&\widetilde{\lambda_{1}}\phi_{1}&\varphi_{1}\\
\widetilde{\lambda_{2}}^{n}\varphi_{2}&\widetilde{\lambda_{2}}^{n-1}\phi_{2}&\ldots&\widetilde{\lambda_{2}}^{2}\varphi_{2}&\widetilde{\lambda_{2}}\phi_{2}&\varphi_{2}\\
\vdots&\vdots&\vdots&\vdots&\vdots&\vdots\\
\widetilde{\lambda_{n-1}}^{n}\varphi_{n-1}&\widetilde{\lambda_{n-1}}^{n-1}\phi_{n-1}&\ldots&\widetilde{\lambda_{n-1}}^{2}\varphi_{n-1}&\widetilde{\lambda_{n-1}}\phi_{n-1}&\varphi_{n-1}\\
\widetilde{\lambda_{n}}^{n}\varphi_{n}&\widetilde{\lambda_{n}}^{n-1}\phi_{n}&\ldots&\widetilde{\lambda_{n}}^{2}\varphi_{n}&\widetilde{\lambda_{n}}\phi_{n}&\varphi_{n}\nonumber\\
\end{vmatrix},
\end{equation}
\begin{equation}\label{fsst6}
\widetilde{(T_{n})_{22}}=\begin{vmatrix}
\widetilde{\lambda}^{n}&0&\ldots&\widetilde{\lambda}^{2}&0&1\\
\widetilde{\lambda_{1}}^{n}\varphi_{1}&\widetilde{\lambda_{1}}^{n-1}\phi_{1}&\ldots&\widetilde{\lambda_{1}}^{2}\varphi_{1}&\widetilde{\lambda_{1}}\phi_{1}&\varphi_{1}\\
\widetilde{\lambda_{2}}^{n}\varphi_{2}&\widetilde{\lambda_{2}}^{n-1}\phi_{2}&\ldots&\widetilde{\lambda_{2}}^{2}\varphi_{2}&\widetilde{\lambda_{2}}\phi_{2}&\varphi_{2}\\
\vdots&\vdots&\vdots&\vdots&\vdots&\vdots\\
\widetilde{\lambda_{n-1}}^{n}\varphi_{n-1}&\widetilde{\lambda_{n-1}}^{n-1}\phi_{n-1}&\ldots&\widetilde{\lambda_{n-1}}^{2}\varphi_{n-1}&\widetilde{\lambda_{n-1}}\phi_{n-1}&\varphi_{n-1}\\
\widetilde{\lambda_{n}}^{n}\varphi_{n}&\widetilde{\lambda_{n}}^{n-1}\phi_{n}&\ldots&\widetilde{\lambda_{n}}^{2}\varphi_{n}&\widetilde{\lambda_{n}}\phi_{n}&\varphi_{n}\nonumber\\
\end{vmatrix}.
\end{equation}
(2) For $n=2k+1(k=1,2,3,\cdots)$,
\begin{equation}\label{fss2}
T_{n}=\widetilde{T_{n}}(\widetilde{\lambda};\widetilde{\lambda_1},\widetilde{\lambda_2}, \cdots,\widetilde{\lambda_n})=\left(
\begin{array}{cc}
\dfrac{\widehat{(T_{n})_{11}}}{Q_{n}}& \dfrac{\widehat{(T_{n})_{12}}}{Q_{n}}\\ \\
\dfrac{\widehat{(T_{n})_{21}}}{\widehat{Q_{n}}}& \dfrac{\widehat{(T_{n})_{22}}}{\widehat{Q_{n}}}\\
\end{array} \right),
\end{equation}
with
\begin{equation}\label{fsstt1}
Q_{n}=\begin{vmatrix}
\widetilde{\lambda_{1}}^{n-1}\phi_{1}&\widetilde{\lambda_{1}}^{n-2}\varphi_{1}&\ldots&\widetilde{\lambda_{1}}^{2}\phi_{1}&\widetilde{\lambda_{1}}\varphi_{1}&\phi_{1}\\
\widetilde{\lambda_{2}}^{n-1}\phi_{2}&\widetilde{\lambda_{2}}^{n-2}\varphi_{2}&\ldots&\widetilde{\lambda_{2}}^{2}\phi_{2}&\widetilde{\lambda_{2}}\varphi_{2}&\phi_{2}\\
\vdots&\vdots&\vdots&\vdots&\vdots&\vdots\\
\widetilde{\lambda_{n}}^{n-1}\phi_{n}&\widetilde{\lambda_{n}}^{n-2}\varphi_{n}&\ldots&\widetilde{\lambda_{n}}^{2}\phi_{n}&\widetilde{\lambda_{n}}\varphi_{n}&\phi_{n}\nonumber\\
\end{vmatrix},
\end{equation}
\begin{equation}\label{fsstt2}
\widehat{(T_{n})_{11}}=\begin{vmatrix}
\widetilde{\lambda}^{n}&0&\ldots&\widetilde{\lambda}^{3}&0&\widetilde{\lambda}&0\\
\widetilde{\lambda_{1}}^{n}\phi_{1}&\widetilde{\lambda_{1}}^{n-1}\varphi_{1}&\ldots&\widetilde{\lambda_{1}}^{3}\phi_{1}&\widetilde{\lambda_{1}}^{2}\varphi_{1}&\widetilde{\lambda_{1}}\phi_{1}&-\varphi_{1}\exp(-i(\dfrac{b}{a}x+\dfrac{b^2}{a^2}t))\\
\widetilde{\lambda_{2}}^{n}\phi_{2}&\widetilde{\lambda_{2}}^{n-1}\varphi_{2}&\ldots&\widetilde{\lambda_{2}}^{3}\phi_{2}&\widetilde{\lambda_{2}}^{2}\varphi_{2}&\widetilde{\lambda_{2}}\phi_{2}&-\varphi_{2}\exp(-i(\dfrac{b}{a}x+\dfrac{b^2}{a^2}t))\\
\vdots&\vdots&\vdots&\vdots&\vdots&\vdots&\vdots\\
\widetilde{\lambda_{n}}^{n}\phi_{n}&\widetilde{\lambda_{n}}^{n-1}\varphi_{n}&\ldots&\widetilde{\lambda_{n}}^{3}\phi_{n}&\widetilde{\lambda_{n}}^{2}\varphi_{n}&\widetilde{\lambda_{n}}\phi_{n}&-\varphi_{n}\exp(-i(\dfrac{b}{a}x+\dfrac{b^2}{a^2}t))\nonumber\\
\end{vmatrix},
\end{equation}
\begin{equation}\label{fsstt3}
\widehat{(T_{n})_{12}}=\begin{vmatrix}
0&\widetilde{\lambda}^{n-1}&...&0&\widetilde{\lambda}^{2}&0&-1\\
\widetilde{\lambda_{1}}^{n}\phi_{1}&\widetilde{\lambda_{1}}^{n-1}\varphi_{1}&\ldots&\widetilde{\lambda_{1}}^{3}\phi_{1}&\widetilde{\lambda_{1}}^{2}\varphi_{1}&\widetilde{\lambda_{1}}\phi_{1}&-\varphi_{1}\exp(-i(\dfrac{b}{a}x+\dfrac{b^2}{a^2}t))\\
\widetilde{\lambda_{2}}^{n}\phi_{2}&\widetilde{\lambda_{2}}^{n-1}\varphi_{2}&\ldots&\widetilde{\lambda_{2}}^{3}\phi_{2}&\widetilde{\lambda_{2}}^{2}\varphi_{2}&\widetilde{\lambda_{2}}\phi_{2}&-\varphi_{2}\exp(-i(\dfrac{b}{a}x+\dfrac{b^2}{a^2}t))\\
\vdots&\vdots&\vdots&\vdots&\vdots&\vdots&\vdots\\
\widetilde{\lambda_{n}}^{n}\phi_{n}&\widetilde{\lambda_{n}}^{n-1}\varphi_{n}&\ldots&\widetilde{\lambda_{n}}^{3}\phi_{n}&\widetilde{\lambda_{n}}^{2}\varphi_{n}&\widetilde{\lambda_{n}}\phi_{n}&-\varphi_{n}\exp(-i(\dfrac{b}{a}x+\dfrac{b^2}{a^2}t))\nonumber\\
\end{vmatrix},
\end{equation}
\begin{equation}\label{fsstt4}
\widehat{Q_{n}}=\begin{vmatrix}
\widetilde{\lambda_{1}}^{n-1}\varphi_{1}&\widetilde{\lambda_{1}}^{n-2}\phi_{1}&\ldots&\widetilde{\lambda_{1}}^{2}\varphi_{1}&\widetilde{\lambda_{1}}\phi_{1}&\varphi_{1}\\
\widetilde{\lambda_{2}}^{n-1}\varphi_{2}&\widetilde{\lambda_{2}}^{n-2}\phi_{2}&\ldots&\widetilde{\lambda_{2}}^{2}\varphi_{2}&\widetilde{\lambda_{2}}\phi_{2}&\varphi_{2}\\
\vdots&\vdots&\vdots&\vdots&\vdots&\vdots\\
\widetilde{\lambda_{n}}^{n-1}\varphi_{n}&\widetilde{\lambda_{n}}^{n-2}\phi_{n}&\ldots&\widetilde{\lambda_{n}}^{2}\varphi_{n}&\widetilde{\lambda_{n}}\phi_{n}&\varphi_{n}\nonumber\\
\end{vmatrix},
\end{equation}
\begin{equation}\label{fsstt5}
\widehat{(T_{n})_{21}}=\begin{vmatrix}
0&\widetilde{\lambda}^{n-1}&...&0&\widetilde{\lambda}^{2}&0&-1\\
\widetilde{\lambda_{1}}^{n}\varphi_{1}&\widetilde{\lambda_{1}}^{n-1}\phi_{1}&\ldots&\widetilde{\lambda_{1}}^{3}\varphi_{1}&\widetilde{\lambda_{1}}^{2}\phi_{1}&\widetilde{\lambda_{1}}\varphi_{1}&-\phi_{1}\exp(i(\dfrac{b}{a}x+\dfrac{b^2}{a^2}t))\\
\widetilde{\lambda_{2}}^{n}\varphi_{2}&\widetilde{\lambda_{2}}^{n-1}\phi_{2}&\ldots&\widetilde{\lambda_{2}}^{3}\varphi_{2}&\widetilde{\lambda_{2}}^{2}\phi_{2}&\widetilde{\lambda_{2}}\varphi_{2}&-\phi_{2}\exp(i(\dfrac{b}{a}x+\dfrac{b^2}{a^2}t))\\
\vdots&\vdots&\vdots&\vdots&\vdots&\vdots&\vdots\\
\widetilde{\lambda_{n}}^{n}\varphi_{n}&\widetilde{\lambda_{n}}^{n-1}\phi_{n}&\ldots&\widetilde{\lambda_{n}}^{3}\varphi_{n}&\widetilde{\lambda_{n}}^{2}\phi_{n}&\widetilde{\lambda_{n}}\varphi_{n}&-\phi_{n}\exp(i(\dfrac{b}{a}x+\dfrac{b^2}{a^2}t))\nonumber\\
\end{vmatrix},
\end{equation}
\begin{equation}\label{fsstt6}
\widehat{(T_{n})_{22}}=\begin{vmatrix}
\widetilde{\lambda}^{n}&0&\ldots&\widetilde{\lambda}^{3}&0&\widetilde{\lambda}&0\\
\widetilde{\lambda_{1}}^{n}\varphi_{1}&\widetilde{\lambda_{1}}^{n-1}\phi_{1}&\ldots&\widetilde{\lambda_{1}}^{3}\varphi_{1}&\widetilde{\lambda_{1}}^{2}\phi_{1}&\widetilde{\lambda_{1}}\varphi_{1}&-\phi_{1}\exp(i(\dfrac{b}{a}x+\dfrac{b^2}{a^2}t))\\
\widetilde{\lambda_{2}}^{n}\varphi_{2}&\widetilde{\lambda_{2}}^{n-1}\phi_{2}&\ldots&\widetilde{\lambda_{2}}^{3}\varphi_{2}&\widetilde{\lambda_{2}}^{2}\phi_{2}&\widetilde{\lambda_{2}}\varphi_{2}&-\phi_{2}\exp(i(\dfrac{b}{a}x+\dfrac{b^2}{a^2}t))\\\
\vdots&\vdots&\vdots&\vdots&\vdots&\vdots&\vdots\\
\widetilde{\lambda_{n}}^{n}\varphi_{n}&\widetilde{\lambda_{n}}^{n-1}\phi_{n}&\ldots&\widetilde{\lambda_{n}}^{3}\varphi_{n}&\widetilde{\lambda_{n}}^{2}\phi_{n}&\widetilde{\lambda_{n}}\varphi_{n}&-\phi_{n}\exp(i(\dfrac{b}{a}x+\dfrac{b^2}{a^2}t))\nonumber\\
\end{vmatrix}.
\end{equation}}

Next, we consider the transformed new solutions
($q^{[n]},r^{[n]}$) of WKI system corresponding to the n-fold DT.
Under covariant requirement of spectral problem of the WKI system,
the transformed form should be
\begin{equation}\label{nsys11}
\pa_{x}\psi^{[n]}=(-aJ\lambda^2+{Q_1}^{[n]}\lambda+{Q_0}^{[n]})\psi=U^{[n]}\psi,
\end{equation}
with
\begin{equation}\label{nfj1}
     \psi=\left( \begin{array}{c}
          \phi   \\
          \varphi \\
      \end{array} \right),
    \quad J= \left( \begin{array}{cc}
       i &0 \\
       0 &-i\\
     \end{array} \right),
  \quad {Q_1}^{[n]}=\left( \begin{array}{cc}
     \sqrt{2b} &aq^{[n]} \\
     ar^{[n]} & -\sqrt{2b}\\
  \end{array} \right),
   \quad {Q_0}^{[n]}=\left( \begin{array}{cc}
     0 &i\sqrt{\dfrac{b}{2}}q^{[n]} \\
     i\sqrt{\dfrac{b}{2}}r^{[n]} & 0\\
  \end{array} \right),
\end{equation}
and then  \begin{equation}\label{ntt2} {T_{n}}_{x}+T_{n}~U=U^{[n]}~T_{n}.
\end{equation}
Substituting $T_n$ given by eq.(\ref{tnss}) into eq.(\ref{ntt2}), and then comparing the
coefficients of $\lambda^{n+1}$, it yields
\begin{eqnarray}\label{ntt3}
&&q^{[n]}=\dfrac{a_{n}}{d_{n}}q+2i\dfrac{b_{n-1}}{d_{n}}, \ \
r^{[n]}=\dfrac{d_{n}}{a_{n}}r-2i\dfrac{c_{n-1}}{a_{n}}.
\end{eqnarray}
Furthermore, substitute $a_{n},d_{n},b_{n-1},c_{n-1}$ from eq.(\ref{fss1}) for $n=2k$ and
from eq.(\ref{fss2}) for $n=2k+1$, into (\ref{ntt3}), we get new solutions ($q^{[n]},r^{[n]}$) of couple
system in eq.(\ref{sy1}) and eq.(\ref{sy2}):\\
\noindent {\bf Theorem 3.}  {\sl Starting from a seed $q$, the n-fold DT $T_n$ in theorem 2 generates
new solutions
\begin{eqnarray}\label{ntt4}
&&q^{[n]}=\dfrac{\Omega_{n1}^{2}}{\Omega_{n3}^{2}}q-2i\dfrac{\Omega_{n1}\Omega_{n2}}{\Omega_{n3}^{2}},
\ \ r^{[n]}=\dfrac{\Omega_{n3}^{2}}
{\Omega_{n1}^{2}}r+2i\dfrac{\Omega_{n3}\Omega_{n4}}{\Omega_{n1}^{2}}.
\end{eqnarray}
Here, (1)for $n=2k$,
\begin{equation}\label{ntt5}
\Omega_{n1}=\begin{vmatrix}
\widetilde{\lambda_{1}}^{n-1}\varphi_{1}&\widetilde{\lambda_{1}}^{n-2}\phi_{1}&\ldots&\widetilde{\lambda_{1}}\varphi_{1}&\phi_{1}\\
\widetilde{\lambda_{2}}^{n-1}\varphi_{2}&\widetilde{\lambda_{2}}^{n-2}\phi_{2}&\ldots&\widetilde{\lambda_{2}}\varphi_{2}&\phi_{2}\\
\vdots&\vdots&\vdots&\vdots&\vdots\\
\widetilde{\lambda_{n-1}}^{n-1}\varphi_{n-1}&\widetilde{\lambda_{n-1}}^{n-2}\phi_{n-1}&\ldots&\widetilde{\lambda_{n-1}}\varphi_{n-1}&\phi_{n-1}\\
\widetilde{\lambda_{n}}^{n-1}\varphi_{n}&\widetilde{\lambda_{n}}^{n-2}\phi_{n}&\ldots&\widetilde{\lambda_{n}}\varphi_{n}&\phi_{n}\\
\end{vmatrix},
\end{equation}
\begin{equation*}
\Omega_{n2}=\begin{vmatrix}
\widetilde{\lambda_{1}}^{n}\phi_{1}&\widetilde{\lambda_{1}}^{n-2}\phi_{1}&\ldots&\widetilde{\lambda_{1}}\varphi_{1}&\phi_{1}\\
\widetilde{\lambda_{2}}^{n}\phi_{2}&\widetilde{\lambda_{2}}^{n-2}\phi_{2}&\ldots&\widetilde{\lambda_{2}}\varphi_{2}&\phi_{2}\\
\vdots&\vdots&\vdots&\vdots&\vdots\\
\widetilde{\lambda_{n-1}}^{n}\phi_{n-1}&\widetilde{\lambda_{n-1}}^{n-2}\phi_{n-1}&\ldots&\widetilde{\lambda_{n-1}}\varphi_{n-1}&\phi_{n-1}\\
\widetilde{\lambda_{n}}^{n}\phi_{n}&\widetilde{\lambda_{n}}^{n-2}\phi_{n}&\ldots&\widetilde{\lambda_{n}}\varphi_{n}&\phi_{n}\\
\end{vmatrix},
\end{equation*}
\begin{equation*}
\Omega_{n3}=\begin{vmatrix}
\widetilde{\lambda_{1}}^{n-1}\phi_{1}&\widetilde{\lambda_{1}}^{n-2}\varphi_{1}&\ldots&\widetilde{\lambda_{1}}\phi_{1}&\varphi_{1}\\
\widetilde{\lambda_{2}}^{n-1}\phi_{2}&\widetilde{\lambda_{2}}^{n-2}\varphi_{2}&\ldots&\widetilde{\lambda_{2}}\phi_{2}&\varphi_{2}\\
\vdots&\vdots&\vdots&\vdots&\vdots\\
\widetilde{\lambda_{n-1}}^{n-1}\phi_{n-1}&\widetilde{\lambda_{n-1}}^{n-2}\varphi_{n-1}&\ldots&\widetilde{\lambda_{n-1}}\phi_{n-1}&\varphi_{n-1}\\
\widetilde{\lambda_{n}}^{n-1}\phi_{n}&\widetilde{\lambda_{n}}^{n-2}\varphi_{n}&\ldots&\widetilde{\lambda_{n}}\phi_{n}&\varphi_{n}\\
\end{vmatrix},
\end{equation*}
\begin{equation*}
\Omega_{n4}=\begin{vmatrix}
\widetilde{\lambda_{1}}^{n}\varphi_{1}&\widetilde{\lambda_{1}}^{n-2}\varphi_{1}&\ldots&\widetilde{\lambda_{1}}\phi_{1}&\varphi_{1}\\
\widetilde{\lambda_{2}}^{n}\varphi_{2}&\widetilde{\lambda_{2}}^{n-2}\varphi_{2}&\ldots&\widetilde{\lambda_{2}}\phi_{2}&\varphi_{2}\\
\vdots&\vdots&\vdots&\vdots&\vdots\\
\widetilde{\lambda_{n-1}}^{n}\varphi_{n-1}&\widetilde{\lambda_{n-1}}^{n-2}\varphi_{n-1}&\ldots&\widetilde{\lambda_{n-1}}\phi_{n-1}&\varphi_{n-1}\\
\widetilde{\lambda_{n}}^{n}\varphi_{n}&\widetilde{\lambda_{n}}^{n-2}\varphi_{n}&\ldots&\widetilde{\lambda_{n}}\phi_{n}&\varphi_{n}\\
\end{vmatrix};
\end{equation*}
(2) for $n=2k+1$,
\begin{equation}\label{ntt6}
\Omega_{n1}=\begin{vmatrix}
\widetilde{\lambda_{1}}^{n-1}\varphi_{1}&\widetilde{\lambda_{1}}^{n-2}\phi_{1}&\widetilde{\lambda_{1}}^{n-3}\varphi_{1}&\ldots&\widetilde{\lambda_{1}}\phi_{1}&\varphi_{1}\exp(-i(\dfrac{b}{a}x+\dfrac{b^2}{a^2}t))\\
\widetilde{\lambda_{2}}^{n-1}\varphi_{2}&\widetilde{\lambda_{2}}^{n-2}\phi_{2}&\widetilde{\lambda_{2}}^{n-3}\varphi_{2}&\ldots&\widetilde{\lambda_{2}}\phi_{2}&\varphi_{2}\exp(-i(\dfrac{b}{a}x+\dfrac{b^2}{a^2}t))\\
\vdots&\vdots&\vdots&\vdots&\vdots&\vdots\\
\widetilde{\lambda_{n}}^{n-1}\varphi_{n}&\widetilde{\lambda_{n}}^{n-2}\phi_{n}&\widetilde{\lambda_{n}}^{n-3}\varphi_{n}&\ldots&\widetilde{\lambda_{n}}\phi_{n}&\varphi_{n}\exp(-i(\dfrac{b}{a}x+\dfrac{b^2}{a^2}t))\\
\end{vmatrix},
\end{equation}
\begin{equation*}
\Omega_{n2}=\begin{vmatrix}
\widetilde{\lambda_{1}}^{n}\phi_{1}&\widetilde{\lambda_{1}}^{n-2}\phi_{1}&\widetilde{\lambda_{1}}^{n-3}\varphi_{1}&\ldots&\widetilde{\lambda_{1}}\phi_{1}&\varphi_{1}\exp(-i(\dfrac{b}{a}x+\dfrac{b^2}{a^2}t))\\
\widetilde{\lambda_{2}}^{n}\phi_{2}&\widetilde{\lambda_{2}}^{n-2}\phi_{2}&\widetilde{\lambda_{2}}^{n-3}\varphi_{2}&\ldots&\widetilde{\lambda_{2}}\phi_{2}&\varphi_{2}\exp(-i(\dfrac{b}{a}x+\dfrac{b^2}{a^2}t))\\
\vdots&\vdots&\vdots&\vdots&\vdots&\vdots\\
\widetilde{\lambda_{n}}^{n}\phi_{n}&\widetilde{\lambda_{n}}^{n-2}\phi_{n}&\widetilde{\lambda_{n}}^{n-3}\varphi_{n}&\ldots&\widetilde{\lambda_{n}}\phi_{n}&\varphi_{n}\exp(-i(\dfrac{b}{a}x+\dfrac{b^2}{a^2}t))\\
\end{vmatrix},
\end{equation*}
\begin{equation*}
\Omega_{n3}=\begin{vmatrix}
\widetilde{\lambda_{1}}^{n-1}\phi_{1}&\widetilde{\lambda_{1}}^{n-2}\varphi_{1}&\widetilde{\lambda_{1}}^{n-3}\phi_{1}&\ldots&\widetilde{\lambda_{1}}\varphi_{1}&\phi_{1}\\
\widetilde{\lambda_{2}}^{n-1}\phi_{2}&\widetilde{\lambda_{2}}^{n-2}\varphi_{2}&\widetilde{\lambda_{2}}^{n-3}\phi_{2}&\ldots&\widetilde{\lambda_{2}}\varphi_{2}&\phi_{2}\\
\vdots&\vdots&\vdots&\vdots&\vdots&\vdots\\
\widetilde{\lambda_{n}}^{n-1}\phi_{n}&\widetilde{\lambda_{n}}^{n-2}\varphi_{n}&\widetilde{\lambda_{n}}^{n-3}\phi_{n}&\ldots&\widetilde{\lambda_{n}}\varphi_{n}&\phi_{n}\\
\end{vmatrix},
\end{equation*}
\begin{equation*}
\Omega_{n4}=\begin{vmatrix}
\widetilde{\lambda_{1}}^{n}\varphi_{1}&\widetilde{\lambda_{1}}^{n-2}\varphi_{1}&\widetilde{\lambda_{1}}^{n-3}\phi_{1}&\ldots&\widetilde{\lambda_{1}}\varphi_{1}&\phi_{1}\\
\widetilde{\lambda_{2}}^{n}\varphi_{2}&\widetilde{\lambda_{2}}^{n-2}\varphi_{2}&\widetilde{\lambda_{2}}^{n-3}\phi_{2}&\ldots&\widetilde{\lambda_{2}}\varphi_{2}&\phi_{2}\\
\vdots&\vdots&\vdots&\vdots&\vdots&\vdots\\
\widetilde{\lambda_{n}}^{n}\varphi_{n}&\widetilde{\lambda_{n}}^{n-2}\varphi_{n}&\widetilde{\lambda_{n}}^{n-3}\phi_{n}&\ldots&\widetilde{\lambda_{n}}\varphi_{n}&\phi_{n}\\
\end{vmatrix}.
\end{equation*}}
In order to consider the smoothness of the solution of MNLS equation in the next section,
it is necessary to get the determinant representation of the transformed eigenfunction
$\psi_j^{[n]}$ associated with $q^{[n]}$ and $r^{[n]}$.\\
{\bf Corollary  4}{\sl\  The determinant representation of transformed
eigenfunction $\psi_j^{[n]}=(T_n|_{\lambda=\lambda_j})
\psi_j(j\geq n+1)$ is expressed by following formulas.\\
1. If  $n=2k$, then
\begin{equation}\label{psineven}
\psi^{[n]}_j= \left(
\begin{array}{c}
\dfrac{\begin{vmatrix}
\widetilde{\lambda_{j}}^{n}\phi_{j}&\widetilde{\lambda_{j}}^{n-1}\varphi_{j}&\ldots&\widetilde{\lambda_{j}}^{2}\phi_{j}&\widetilde{\lambda_{j}}\varphi_{j}&\phi_{j}\\
\widetilde{\lambda_{1}}^{n}\phi_{1}&\widetilde{\lambda_{1}}^{n-1}\varphi_{1}&\ldots&\widetilde{\lambda_{1}}^{2}\phi_{1}&\widetilde{\lambda_{1}}\varphi_{1}&\phi_{1}\\
\widetilde{\lambda_{2}}^{n}\phi_{2}&\widetilde{\lambda_{2}}^{n-1}\varphi_{2}&\ldots&\widetilde{\lambda_{2}}^{2}\phi_{2}&\widetilde{\lambda_{2}}\varphi_{2}&\phi_{2}\\
\vdots&\vdots&\vdots&\vdots&\vdots&\vdots\\
\widetilde{\lambda_{n-1}}^{n}\phi_{n-1}&\widetilde{\lambda_{n-1}}^{n-1}\varphi_{n-1}&\ldots&\widetilde{\lambda_{n-1}}^{2}\phi_{n-1}&\widetilde{\lambda_{n-1}}\varphi_{n-1}&\phi_{n-1}\\
\widetilde{\lambda_{n}^{n}}\phi_{n}&\widetilde{\lambda_{n}}^{n-1}\varphi_{n}&\ldots&\widetilde{\lambda_{n}}^{2}\phi_{n}&\widetilde{\lambda_{n}}\varphi_{n}&\phi_{n}\nonumber\\
\end{vmatrix}}{\Omega_{n3}}\\ \\
\dfrac{\begin{vmatrix}
\widetilde{\lambda_{j}}^{n}\varphi_{j}&\widetilde{\lambda_{j}}^{n-1}\phi_{j}&\ldots&\widetilde{\lambda_{j}}^{2}\varphi_{j}&\widetilde{\lambda_{j}}\phi_{j}&\varphi_{j}\\
\widetilde{\lambda_{1}}^{n}\varphi_{1}&\widetilde{\lambda_{1}}^{n-1}\phi_{1}&\ldots&\widetilde{\lambda_{1}}^{2}\varphi_{1}&\widetilde{\lambda_{1}}\phi_{1}&\varphi_{1}\\
\widetilde{\lambda_{2}}^{n}\varphi_{2}&\widetilde{\lambda_{2}}^{n-1}\phi_{2}&\ldots&\widetilde{\lambda_{2}}^{2}\varphi_{2}&\widetilde{\lambda_{2}}\phi_{2}&\varphi_{2}\\
\vdots&\vdots&\vdots&\vdots&\vdots&\vdots\\
\widetilde{\lambda_{n-1}}^{n}\varphi_{n-1}&\widetilde{\lambda_{n-1}}^{n-1}\phi_{n-1}&\ldots&\widetilde{\lambda_{n-1}}^{2}\varphi_{n-1}&\widetilde{\lambda_{n-1}}\phi_{n-1}&\varphi_{n-1}\\
\widetilde{\lambda_{n}}^{n}\varphi_{n}&\widetilde{\lambda_{n}}^{n-1}\phi_{n}&\ldots&\widetilde{\lambda_{n}}^{2}\varphi_{n}&\widetilde{\lambda_{n}}\phi_{n}&\varphi_{n}\nonumber\\
\end{vmatrix}}{\Omega_{n1}}\\
\end{array} \right).
\end{equation}
2.If $n=2k+1$,then
\begin{equation} \label{psinodd}
\psi^{[n]}_j= \left(
\begin{array}{c}
\dfrac{\begin{vmatrix}
\widetilde{\lambda_{j}}^{n}\phi_{j}&\widetilde{\lambda_{j}}^{n-1}\varphi_{j}&\ldots&\widetilde{\lambda_{j}}^{3}\phi_{j}&\widetilde{\lambda_{j}}^{2}\varphi_{j}&\widetilde{\lambda_{j}}\phi_{j}&-\varphi_{j}\exp(-i(\dfrac{b}{a}x+\dfrac{b^2}{a^2}t))\\
\widetilde{\lambda_{1}}^{n}\phi_{1}&\widetilde{\lambda_{1}}^{n-1}\varphi_{1}&\ldots&\widetilde{\lambda_{1}}^{3}\phi_{1}&\widetilde{\lambda_{1}}^{2}\varphi_{1}&\widetilde{\lambda_{1}}\phi_{1}&-\varphi_{1}\exp(-i(\dfrac{b}{a}x+\dfrac{b^2}{a^2}t))\\
\widetilde{\lambda_{2}}^{n}\phi_{2}&\widetilde{\lambda_{2}}^{n-1}\varphi_{2}&\ldots&\widetilde{\lambda_{2}}^{3}\phi_{2}&\widetilde{\lambda_{2}}^{2}\varphi_{2}&\widetilde{\lambda_{2}}\phi_{2}&-\varphi_{2}\exp(-i(\dfrac{b}{a}x+\dfrac{b^2}{a^2}t))\\
\vdots&\vdots&\vdots&\vdots&\vdots&\vdots&\vdots\\
\widetilde{\lambda_{n}}^{n}\phi_{n}&\widetilde{\lambda_{n}}^{n-1}\varphi_{n}&\ldots&\widetilde{\lambda_{n}}^{3}\phi_{n}&\widetilde{\lambda_{n}}^{2}\varphi_{n}&\widetilde{\lambda_{n}}\phi_{n}&-\varphi_{n}\exp(-i(\dfrac{b}{a}x+\dfrac{b^2}{a^2}t))\nonumber\\
\end{vmatrix}}{\Omega_{n3}}\\ \\
\dfrac{\begin{vmatrix}
\widetilde{\lambda_{j}}^{n}\varphi_{j}&\widetilde{\lambda_{j}}^{n-1}\phi_{j}&\ldots&\widetilde{\lambda_{j}}^{3}\varphi_{j}&\widetilde{\lambda_{j}}^{2}\phi_{j}&\widetilde{\lambda_{j}}\varphi_{j}&-\phi_{j}\\
\widetilde{\lambda_{1}}^{n}\varphi_{1}&\widetilde{\lambda_{1}}^{n-1}\phi_{1}&\ldots&\widetilde{\lambda_{1}}^{3}\varphi_{1}&\widetilde{\lambda_{1}}^{2}\phi_{1}&\widetilde{\lambda_{1}}\varphi_{1}&-\phi_{1}\\
\widetilde{\lambda_{2}}^{n}\varphi_{2}&\widetilde{\lambda_{2}}^{n-1}\phi_{2}&\ldots&\widetilde{\lambda_{2}}^{3}\varphi_{2}&\widetilde{\lambda_{2}}^{2}\phi_{2}&\widetilde{\lambda_{2}}\varphi_{2}&-\phi_{2}\\
\vdots&\vdots&\vdots&\vdots&\vdots&\vdots&\vdots\\
\widetilde{\lambda_{n}}^{n}\varphi_{n}&\widetilde{\lambda_{n}}^{n-1}\phi_{n}&\ldots&\widetilde{\lambda_{n}}^{3}\varphi_{n}&\widetilde{\lambda_{n}}^{2}\phi_{n}&\widetilde{\lambda_{n}}\varphi_{n}&-\phi_{n}\nonumber\\
\end{vmatrix}}{\Omega_{n1}}\\
\end{array} \right).
\end{equation}}
\\
\\
{\bf 2.3 Reduction of the Darboux transformation for WKI system.}
\\
\\
\indent The solutions $q^{[n]}$ and $r^{[n]}$ in theorem 3 generated by the n-fold DT $T_n$ of WKI system are solutions of
the coupled system in eq.(\ref{sy1}) and eq.(\ref{sy2}). If it keeps the reduction condition, i.e.,$q^{[n]}=-(r^{[n]})^*$, DT of the WKI system in theorem 2 reduces to the DT of the MNLS equation, and then $q^{[n]}$ in theorem 3 implies automatically a new solution of the MNLS equation. In this subsection, we shall show how to choose
the eigenvalues and eigenfunctions in the determinant representations of the $T_n$ in order to realize  the reduction.

Under the reduction condition $q=-r^*$, the eigenfunction $\psi_k=\left(
\begin{array}{c}
\phi_k\\
\varphi_k
\end{array} \right)$ associated with eigenvalue $\lambda_k$ has following properties,
\begin{description}
\item[(i)] $\phi_{k}^{\ast}=\varphi_{k}$, $\lambda_{k}=-{\lambda_{k}}^{\ast}$;
\item[(ii)] ${\phi_{k}}^{\ast}=\varphi_{l}, \ {\varphi_{k}}^{\ast}=\phi_{l}$,
${\lambda_{k}}^{\ast}=-\lambda_{l}$, where $k\neq l$.
\end{description}
It is trivial to verify above properties in Lax pair,  eq.(\ref{sys11}) and  eq.(\ref{sys22}), of the WKI system by a straightforward calculation. These properties of eigenfunctions for the WKI system  provides one kind of possibility
for choosing suitable eigenvalues and eigenfunctions such that the reduction condition
$q^{[n]}=-(r^{[n]})^*$ holds in n-fold DT.
\\
\\
{\bf Lemma 5}   {\sl Let
\begin{equation}\label{onefoldredu}
 \lambda_1=i\beta_1 (\beta_1\in  R),\quad \quad
\psi_1=\left( \begin{array}{c}
\phi_1\\
\phi_1^*
\end{array} \right),
\end{equation}
in $T_1$, then $q^{[1]}=-(r^{[1]})^*$  and $T_1$ reduces to a  one-fold  DT of the MNLS.
}
\\
{\bf Proof} According to property (i), it is suitable to let
$\lambda_1$ and $\psi_1$ as eq.(\ref{onefoldredu}).
Under this choice, $\varphi^*_1=\phi_1$ and $\lambda_1^*=-\lambda_1$. Substituting this
relation   and $q=-r^*$ into eq.(\ref{sTT}), then
\begin{eqnarray*}
&&(q^{[1]})^*=(\frac{\phi_{1}}{\varphi_{1}})^2\exp(2i(\dfrac{b}{a}x+
\dfrac{b^2}{a^2}t))q^*-2i\frac{\phi_{1}}{\varphi_{1}}(\lambda_1+i\dfrac{\sqrt{2b}}{2a})
\exp(2i(\dfrac{b}{a}x+\dfrac{b^2}{a^2}t)),\nonumber\\
&&=-(\frac{\phi_{1}}{\varphi_{1}})^2\exp(2i(\dfrac{b}{a}x+
\dfrac{b^2}{a^2}t))r-2i\frac{\phi_{1}}{\varphi_{1}}(\lambda_1+i\dfrac{\sqrt{2b}}{2a})
\exp(2i(\dfrac{b}{a}x+\dfrac{b^2}{a^2}t))=-r^{[1]}.\Box \\
\end{eqnarray*}
{\bf Lemma 6} {\sl  Let
$
\psi_1=\left( \begin{array}{c}
\phi_1\\
\varphi_1
\end{array} \right) \text{be an eigenfunction of\ } \lambda_1$,\ and
\begin{equation}\label{twofoldredu1}
\lambda_2=-\lambda_1^*, \quad
\psi_2=\left( \begin{array}{c}
\varphi_1^*\\
\phi_1^*
\end{array} \right),
\end{equation}
in $T_2$ given by eq.(\ref{fss1}), then $q^{[2]}=-(r^{[2]})^*$ and $T_2$  reduces to a one-fold DT of the MNLS.
}
\\
{\bf Proof}  According to the above property (ii), it is suitable to choose
$\psi_2$ and $\lambda_2$ as eq.(\ref{twofoldredu1}). Let $n=2$ in eq.(\ref{ntt4}), then
\begin{eqnarray}
&q^{[2]}=\dfrac{\Omega_{21}^2}{\Omega_{23}^2}q-2i\dfrac{\Omega_{21}\Omega_{n2}}{\Omega_{23}^2}, \label{q2eq}\\
&r^{[2]}=\dfrac{\Omega_{23}^2}{\Omega_{21}^2}r+2i\dfrac{\Omega_{23}\Omega_{24}}{\Omega_{21}^2}, \label{r2eq}\\
&\text{with\
}\Omega_{21}=\widetilde{\lambda_1}\varphi_1\phi_2-\widetilde{\lambda_2}\phi_1\varphi_2,
\Omega_{22}=(\widetilde{\lambda_1}^2-\widetilde{\lambda_2}^2)\phi_1\phi_2,\label{omega21112}\\
&\Omega_{23}=\widetilde{\lambda_1}\phi_1\varphi_2-\widetilde{\lambda_2}\varphi_1\phi_2,
\Omega_{24}=(\widetilde{\lambda_1}^2-\widetilde{\lambda_2}^2)\varphi_1\varphi_2.
\label{omega22122}
\end{eqnarray}
Using the special choice of $\lambda_2$ and its eigenfunction in
eq.(\ref{twofoldredu1}), we have $\Omega_{21}^*=\Omega_{23},
\Omega_{22}^*=-\Omega_{24}$ from eq.(\ref{omega21112}) and
eq.(\ref{omega22122}). These two relations imply
$q^{[2]}=-(r^{[2]})^*$ from eq.(\ref{q2eq}) and eq.(\ref{r2eq}) with
the help of  original reduction condition $q=-r^*$. Therefore, under
the choice in eq.(\ref{twofoldredu1}), $T_2$ reduces to a one-fold DT
of the MNLS equation.$\Box$
\\
Furthermore, by repeatedly iterating $T_2$ as Lemma 6 for k times with
paired-eigenvalues and corresponding paired eigenfunctions, we have following results of the $T_{2k}$.
\\
{\bf Theorem 7} {\sl  Let $\lambda_{2l-1}$($l=1,2,3\dots,k$) be $k$ distinct eigenvalues for eq.(\ref{sys11}) and
eq.(\ref{sys22}),
$\psi_{2l-1}=\left( \begin{array}{c}
\phi_{2l-1}\\
\varphi_{2l-1}\end{array} \right)
$
be their associated eigenfunctions,  and
\begin{equation}\label{2nfoldredu}
 \lambda_{2l}= -\lambda_{2l-1}^*,
\psi_{2l}=\left( \begin{array}{c}
\varphi_{2l-1}^*\\
\phi_{2l-1}^*
\end{array} \right),
\end{equation}
in the (2k)-fold DT $T_{2k}$ of WKI system, then  $q^{[2k]}=-(r^{[2k]})^*$ in eq.(\ref{ntt4}),
and $T_{2k}$ in eq.(\ref{fss1}) reduces to a k-fold DT of the MNLS equation.
}\\
Thus $q^{[2k]}$ is called k-order solution of the MNLS.

       Similar to the reduction of the $T_{2k}$ for the WKI system, $T_{2k+1}$ in eq.(\ref{fss2}) can also be reduced
to the (k+1)-fold DT of the MNLS by choosing one pure imaginary
$\lambda_{2k+1}=i\beta_{2k+1}$ and  $k$
paired-eigenvalues $\lambda_{2l}= -\lambda_{2l-1}^*(l=1,2,\cdots,k)$
with corresponding eigenfunctions according to properties (i) and
(ii). Of course, there are many other ways to select eigenvalues and eigenfunctions
in order to do reduction of n-fold DT for the WKI system.
\section{Smoothness of the solutions $q^{[2k]}$}
The smoothness of the $q^{[n]}$ generated by DT is an important property of the solution
of the MNLS equation. In this section,we shall study this property for the solution
$q^{[2k]}$ through the $T_{2k}$ with the paired eigenfunctions and eigenvalues.
\\
\\
{\bf 3.1 Non-degenerate case }\\

   In this subsection, we shall show the smoothness of the new solution $q^{[2k]}(k=1,2,3\cdots)$ of the
MNLS equation under non-degenerate case: $\psi_j\not=0$ and $\lambda_i\not= \lambda_j(i \not= j)$. This solution is generated by $T_{2k}$ from a seed solution $q$ with the paired eigenvalues
and eigenfunctions in Theorem 7.
\\
{\bf Lemma 8}{\sl  ,
Let  $\lambda_1=\alpha_1+i \beta_1$,$\alpha_1\in R$ and $\beta_1\in R$,
$|\phi_1|+ |\varphi_1|>0$ in the
$q^{[2]}$ generated by the $T_2$ of Lemma 6.
If $\alpha_1\not=0$, then $q^{[2]}$ is a smooth function on whole (x,t)-plane}.
\\
{\bf Proof} From Lemma 6, by a straightforward calculation,  we have
\begin{equation}\label{q2phi1varphi1}
q^{[2]}=\dfrac{(\widetilde{\lambda_1}\varphi_1\varphi_1^*+ \widetilde{\lambda_1}^*\phi_1\phi_1^*)^2}
{(\widetilde{\lambda_1}\phi_1\phi_1^*+ \widetilde{\lambda_1}^*\varphi_1\varphi_1^*)^2}q-
2i\dfrac{(\widetilde{\lambda_1}^2-\widetilde{\lambda_1}^{*2})(\widetilde{\lambda_1}\varphi_1\varphi_1^*+
\widetilde{\lambda_1}^*\phi_1\phi_1^*)
\phi_1\varphi_1^*}
{(\widetilde{\lambda_1}\phi_1\phi_1^*+\widetilde{\lambda_1}^*\varphi_1\varphi_1^*)^2}.
\end{equation}
Note  that in the denominator of $q^{[2]}$, $\Omega_{21}
=\widetilde{\lambda_1}\phi_1\phi_1^*+\widetilde{\lambda_1}^*\varphi_1\varphi_1^*
=\alpha_1(|\phi_1|^2+|\varphi_1|^2)+i (\beta_1+\dfrac{\sqrt{2b}}{2a})
(|\phi_1|^2-|\varphi_1|^2)$.  $\Omega_{21}\not=0$ if
$\alpha_1\not=0$ and $|\phi_1|+ |\varphi_1|>0$. Therefore, $q^{[2]}$
is a smooth function under this condition. $\Box$
\\
{\bf Theorem 9} Let $\lambda_{2l-1}=\alpha_{2l-1}+i\beta_{2l-1}(l=1,2,3,\cdots,k)$ be k
distinct eigenvalues in the $q^{[2k]}$ generated by $T_{2k}$ of Theorem 7. If $\alpha_{2l-1}
\not=0(l=1,2,3,\cdots,k)$ and $|\phi_{2l-1}|+ |\varphi_{2l-1}|>0$, then $q^{[2k]}$ is a smooth function on whole (x,t)-plane.\\
{\bf Proof}
1) Using $T_2$, a new eigenfunction associated with $\lambda_3$ is
$\psi_3^{[2]}=\left
(\begin{array}{c}
\phi^{[2]}_3\\
\varphi^{[2]}_3
\end{array}\right)=
T_2(\widetilde{\lambda_3};\widetilde{\lambda_1})\psi_3
$. Furthermore  $\psi_3^{[2]}\not=0$ because $T_2$ is not a degenerate linear transformation
and $\psi_3\not=0$. Thus $|\phi^{[2]}_3|+ |\varphi^{[2]}_3|\not=0$.\\
2) By iteration of $T_2$ once with generating function $\psi_3^{[2]}$, a new solution of the MNLS equation is given by
\begin{equation}\label{q4phi2varphi2}
q^{[4]}=\dfrac{(\widetilde{\lambda_3}\varphi^{[2]}_3(\varphi^{[2]}_3)^*
+ \widetilde{\lambda_3}^*\phi^{[2]}_3(\phi^{[2]}_3)^*)^2}
{(\widetilde{\lambda_3}\phi^{[2]}_3(\phi^{[2]}_3)^*+
 \widetilde{\lambda_3}^*\varphi^{[2]}_3(\varphi^{[2]}_3)^*)^2}q^{[2]}
-
2i\dfrac{(\widetilde{\lambda_3}^2-\widetilde{\lambda_3}^{*2})
(\widetilde{\lambda_3}\varphi^{[2]}_3(\varphi^{[2]}_3)^*+
\widetilde{\lambda_3}^*\phi^{[2]}_3(\phi^{[2]}_3)^*)
\phi^{[2]}_3(\varphi^{[2]}_3)^*}
{(\widetilde{\lambda_3}\phi^{[2]}_3(\phi^{[2]}_3)^*
+\widetilde{\lambda_3}^*\varphi^{[2]}_3(\varphi^{[2]}_3)^*)^2}.
\end{equation}
3) By using Lemma 8, $q^{[4]}$ is a smooth function on (x,t) plane.\\
4) By this pattern,  k-time iteration of $T_2$ with generating functions
$\psi_{2l-1}(l=1,2,3,\cdots,k)$ implies $T_{2k}$ and corresponding new solution $q^{[2k]}$ of
the MNLS equation. Note that each step of iterations generates
smooth solution according to Lemma 8. Thus $q^{[2k]}$, the final solution generated by
k-time iteration, is a smooth function on the (x,t)-plane. $\Box$

   It is easy to check that the k-soliton solution is generated by $T_{2k}$ in Theorem 9 from
a trivial seed solution--zero solution. It is an interesting problem to study degenerate
cases and to apply it to get smooth solutions of the MNLS equations.
\\
\\
{\bf 3.2 Double degeneration case }\\

   It is easy to see from the determinant representations that there are two degenerate cases
of $T_{2k}$: 1) degenerate eigenvalues: $\lambda_{2l-1} \longrightarrow \lambda_1 (l=1,2,3,
\cdots,k) $; 2) degenerate eigenfunctions: $\psi_{2l-1}(\lambda_{2l-1};x,t)=0\ ( l=1,2,3,
\cdots,k)$ under certain values of parameters. We assume that $\lambda_0$ is only one zero point of the eigenfunction
$\psi_1$.  In this subsection, we shall apply $T_{2k}$ with double degeneration, i.e.,
 $\lambda_{2l-1} \longrightarrow \lambda_0 (l=1,2,3, \cdots,k)$ and  $\psi_{2l-1}
 (\lambda_0;x,t)=0\ ( l=1,2,3, \cdots,k)$, to get smooth solution of
 the MNLS equation. Specifically, under the double degeneration,
  $q^{[2k]}$ generated by $T_{2k}$ is expressed by
 an indeterminate form $\dfrac{0}{0}$. Thus,it is possible to get a smooth solution of the
 MNLS by higher order Taylor expansion of it at
a special value $\lambda_0$.

 To deal with the degeneration of $T_{2k}$, we begin with the $T_2$ under the reduction condition
in lemma 6.  Let $\psi_1(\lambda_0)$ be a smooth eigenfunction of Lax pair of the MNLS associated with non-zero
eigenvalue $\lambda_0$, which has also continuous dependence on the $\lambda_0$.  Under a small shift of
 eigenvalue,  it can be expanded as
 $\psi_1(\lambda_0+\epsilon;x,t)=\left(
\begin{array}{c}
\phi_1(\lambda_0+\epsilon;x,t)=\phi_1(\lambda_0;x,t)+\sum\limits_{l=1}^{k} b_l\epsilon^l+O(\epsilon^k)\\
\varphi_1(\lambda_0+\epsilon;x,t)=\varphi_1(\lambda_0;x,t)+\sum\limits_{l=1}^{k} a_l\epsilon^l+O(\epsilon^k)
\end{array} \right)$,
$b_l=\dfrac{\partial \phi_1(\lambda_0+\epsilon)}{\partial\epsilon}|_{\epsilon=0},a_l=\dfrac{\partial
 \varphi_1(\lambda_0+\epsilon)}{\partial\epsilon}|_{\epsilon=0}$, $|a_l|$ and $|b_l| (l=1,2,\cdots,k)$ are not both zero
. If $\lambda_0$ is  the only one zero point of $\psi_1$,i.e., $\psi_1(\lambda_0)=0$, then it is also a zero point of $W_2$ in $T_2$. Thus $\lambda_0$ is a singularity of  $T_2$  and the $q^{[2]}$.
 However this singularity  is removable.\\
{\bf Lemma 10} {\sl Let $\lambda_0$ be the only one zero point of eigenfunction
$\psi_1(\lambda_1)$, $\lambda_0=\alpha_0+i\beta_0$. If $\alpha_0\not=0$,  then $q^{[2]}$ in
eq.(\ref{q2phi1varphi1}) is a smooth solution, and the singular $T_2$ infers a smooth two-fold DT $T_2^{\prime}$}.
\\
{\bf Proof} By Taylor expansion, $\phi_1(\lambda_0+\epsilon)=\phi_1(\lambda_0)+ \dfrac{\partial \phi_1(\lambda_0+\epsilon)}{\partial\epsilon}|_{\epsilon=0}\epsilon+O(\epsilon^2)=b_1\epsilon+O(\epsilon^2)$, and
$\varphi_1(\lambda_0+\epsilon)=\varphi_1(\lambda_0)+ \dfrac{\partial \varphi_1(\lambda_0+\epsilon)}{\partial\epsilon}|_{\epsilon=0}\epsilon+O(\epsilon^2)=a_1\epsilon+O(\epsilon^2)$.\\
1) Submitting these expansion and $\lambda_1=\lambda_0+\epsilon$ into $q^{[2]}$ in eq.(\ref{q2phi1varphi1}), then
\begin{equation}\label{q2expansion}
q^{[2]}=\dfrac{(\widetilde{\lambda_0}|a_1|^2+ \widetilde{\lambda_0}^*|b_1|^2)^2\epsilon^4+O(\epsilon^5)}
{(\widetilde{\lambda_0}|b_1|^2+ \widetilde{\lambda_0}^*|a_1|^2)^2\epsilon^4+O(\epsilon^5)}q-
2i\dfrac{(\widetilde{\lambda_0}^2-\widetilde{\lambda_0}^{*2})(\widetilde{\lambda_0}|a_1|^2+
\widetilde{\lambda_0}^*|b_1|^2)
b_1a_1^*\epsilon^4+O(\epsilon^5) }
{(\widetilde{\lambda_0}|b_1|^2+\widetilde{\lambda_0}^*|a_1|^2)^2\epsilon^4+O(\epsilon^5) }.
\end{equation}
This formula provides a smooth solution
\begin{equation}\label{q2smooth}
q^{[2]}=\dfrac{(\widetilde{\lambda_0}|a_1|^2+ \widetilde{\lambda_0}^*|b_1|^2)^2}
{(\widetilde{\lambda_0}|b_1|^2+ \widetilde{\lambda_0}^*|a_1|^2)^2}q-
2i\dfrac{(\widetilde{\lambda_0}^2-\widetilde{\lambda_0}^{*2})(\widetilde{\lambda_0}|a_1|^2+
\widetilde{\lambda_0}^*|b_1|^2)
b_1a_1^* }
{(\widetilde{\lambda_0}|b_1|^2+\widetilde{\lambda_0}^*|a_1|^2)^2 }
\end{equation}
by setting $\epsilon\rightarrow 0$ if  $|a_1|+|b_1|\not=0$ and $\alpha_0\not=0$.\\
2)  According to
\begin{equation}\label{psij2}
\psi^{[2]}_j=T_2(\lambda_j;\lambda_1,\lambda_2)\psi_j= \left(
\begin{array}{c}
\dfrac{\begin{vmatrix}
\widetilde{\lambda_{j}}^{2}\phi_{j}&\widetilde{\lambda_{j}}\varphi_{j}&\phi_{j}\\
\widetilde{\lambda_{1}}^{2}\phi_{1}&\widetilde{\lambda_{1}}\varphi_{1}&\phi_{1}\\
\widetilde{\lambda_{2}}^{2}\phi_{2}&\widetilde{\lambda_{2}}\varphi_{2}&\phi_{2}\\
\end{vmatrix}}{\begin{vmatrix}
\widetilde{\lambda_{1}}\phi_{1}&\varphi_{1}\\
\widetilde{\lambda_{2}}\phi_{2}&\varphi_{2}
\end{vmatrix}}\\ \\
\dfrac{\begin{vmatrix}
\widetilde{\lambda_{j}}^{2}\varphi_{j}&\widetilde{\lambda_{j}}\phi_{j}&\varphi_{j}\\
\widetilde{\lambda_{1}}^{2}\varphi_{1}&\widetilde{\lambda_{1}}\phi_{1}&\varphi_{1}\\
\widetilde{\lambda_{2}}^{2}\varphi_{2}&\widetilde{\lambda_{2}}\phi_{2}&\varphi_{2}
\end{vmatrix}}{\begin{vmatrix}
\widetilde{\lambda_{1}}\varphi_{1}&\phi_{1}\\
\widetilde{\lambda_{2}}\varphi_{2}&\phi_{2}
\end{vmatrix}}\\
\end{array} \right),
\end{equation}
 $T_2$ is singular if $\phi_1(\lambda_0)=\varphi_1(\lambda_0)=0$.
Taking  above Taylor expansions of $\phi_1(\lambda_0+\epsilon)$ and $\varphi_1(\lambda_0+\epsilon)$ and $\lambda_1=\lambda_0+\epsilon$ into $\psi_j^{[2]}$ under the reduction conditions given by  eq.(\ref{twofoldredu1}), then
\begin{eqnarray}\label{psij2nondeg}
&&\psi^{[2]}_j=\mbox{\hspace{-0.2cm}} \left(\mbox{\hspace{-0.2cm}}
\begin{array}{c}
\dfrac{\begin{vmatrix}
\widetilde{\lambda_{j}}^{2}\phi_{j}&\widetilde{\lambda_{j}}\varphi_{j}&\phi_{j}\\
\widetilde{\lambda_{0}}^{2}b_1\epsilon&\widetilde{\lambda_{0}}a_1\epsilon &b_1 \epsilon\\
(\widetilde{\lambda_{0}}^{2}a_1)^*\epsilon&-(\widetilde{\lambda_{0}}b_1)^*\epsilon &a_1^*\epsilon \nonumber\\
\end{vmatrix}+O(\epsilon^3)}{\begin{vmatrix}
\widetilde{\lambda_{0}}b_1\epsilon&a_1\epsilon\\
-(\widetilde{\lambda_{0}}a_1)^*\epsilon&b_1^*\epsilon \nonumber\\
\end{vmatrix}+O(\epsilon^3) }\\ \\
\dfrac{\begin{vmatrix}
\widetilde{\lambda_{j}}^{2}\varphi_{j}&\widetilde{\lambda_{j}}\phi_{j}&\varphi_{j}\\
\widetilde{\lambda_{0}}^{2}a_1\epsilon &\widetilde{\lambda_{0}}b_1\epsilon&a_1\epsilon\\
(\widetilde{\lambda_{0}}^{2}b_1)^*\epsilon &-(\widetilde{\lambda_{0}}a_1)^*\epsilon&b_1^*\epsilon\nonumber\\
\end{vmatrix}+O(\epsilon^3)}{\begin{vmatrix}
\widetilde{\lambda_{0}}a_1\epsilon &b_1\epsilon\\
-(\widetilde{\lambda_{0}}b_1)^*\epsilon&a_1^*\epsilon\nonumber\\
\end{vmatrix}+O(\epsilon^3) }\\
\end{array} \mbox{\hspace{-0.2cm}}\right)\stackrel{\epsilon\rightarrow 0}{=}
 \left(\mbox{\hspace{-0.2cm}}
\begin{array}{c}
\dfrac{\begin{vmatrix}
\widetilde{\lambda_{j}}^{2}\phi_{j}&\widetilde{\lambda_{j}}\varphi_{j}&\phi_{j}\\
\widetilde{\lambda_{0}}^{2}b_1&\widetilde{\lambda_{0}}a_1 &b_1\\
(\widetilde{\lambda_{0}}^{2}a_1)^*&-(\widetilde{\lambda_{0}}b_1)^* &a_1^* \nonumber\\
\end{vmatrix}}{\begin{vmatrix}
\widetilde{\lambda_{0}}b_1&a_1\\
-(\widetilde{\lambda_{0}}a_1)^*&b_1^*\nonumber\\
\end{vmatrix} }\\ \\
\dfrac{\begin{vmatrix}
\widetilde{\lambda_{j}}^{2}\varphi_{j}&\widetilde{\lambda_{j}}\phi_{j}&\varphi_{j}\\
\widetilde{\lambda_{0}}^{2}a_1 &\widetilde{\lambda_{0}}b_1&a_1\\
(\widetilde{\lambda_{0}}^{2}b_1)^* &-(\widetilde{\lambda_{0}}a_1)^*&b_1^*\nonumber\\
\end{vmatrix}}{\begin{vmatrix}
\widetilde{\lambda_{0}}a_1 &b_1\\
-(\widetilde{\lambda_{0}}b_1)^*&a_1^*\nonumber\\
\end{vmatrix} }\\
\end{array}\mbox{\hspace{-0.2cm}} \right)\\
&&\triangleq T^{\prime}_2(\lambda_j;\lambda_0)\psi_j.
\end{eqnarray}
Note that the denominator in above formula is a non-zero function of (x,t). This shows $\psi_j^{[2]}$ and $T_2^{\prime}$ are smooth. $\square$\\
{\bf Remark} If $a_1=b_1=0$ in $\psi_1(\lambda_0+\epsilon)$, we can use next non-zero coefficient of higher order  in its expansion to generate smooth $q^{[2]}$ and $T_2^{\prime}$ as the same manner in the above Lemma.

Because $T_2$ just includes one eigenvalue $\lambda_1$, it does not have double degeneration.
The first non-trivial example of double degeneration of DT is $T_4$ by setting $\lambda_3\rightarrow \lambda_0$ and
$\psi_3=\psi_1(\lambda_0)=\left(
\begin{array}{c}
\phi_1(\lambda_0)\\
\varphi_1(\lambda_0)
\end{array} \right)=0$.
Thus it is easy to find $\psi_3^{[2]}=0$ from eq.(\ref{psij2nondeg}),and we can not do DT again along the
eigenvalue $\lambda_0$. But it can be re-obtained by following limit method.\\
{\bf Lemma 11} {\sl According to lemma 10,
\begin{equation}
\psi^{[2]}_3|_{\lambda_3=\lambda_0+\epsilon}= \left(
\begin{array}{c}
b^{[2]}_2\epsilon^2+O(\epsilon^3)\\
a^{[2]}_2\epsilon^2+O(\epsilon^3)
\end{array} \right),
\end{equation}
 $|b_2^{[2]}|$ and  $|a^{[2]}_2|$ are not both zero. Further, $q^{[4]}$ in  eq.(\ref{q4phi2varphi2}) provides a smooth solution of
the MNLS equation.
}\\
{\bf Proof.} Let  $\lambda_3=\lambda_0+\epsilon$ and $\psi_3=\psi_1(\lambda_0+\epsilon)$ in eq.(\ref{psij2nondeg}),
then $\psi^{[2]}_3|_{\lambda_3=\lambda_0+\epsilon}$ is a non-zero smooth function because $T_2^{\prime}(\lambda_0+\epsilon;\lambda_0)$ is  a non- degenerate linear transformation. Thus
$\psi^{[2]}_3|_{\lambda_3=\lambda_0+\epsilon}$ can be expanded as following form
\begin{eqnarray}
&&\psi^{[2]}_3=\left(\mbox{\hspace{-0.2cm}}
\begin{array}{l}
\dfrac{\begin{vmatrix}
\dfrac{\partial^2 }{\partial\epsilon^2}((\widetilde{\lambda_0}+\epsilon)^{2}\phi_{1}(\lambda_0+\epsilon))|_{\epsilon=0}
&\dfrac{\partial^2 }{\partial\epsilon^2}((\widetilde{\lambda_0}+\epsilon)\varphi_{1}(\lambda_0+\epsilon))|_{\epsilon=0}&\dfrac{\partial^2 }{\partial\epsilon^2}\phi_{1}(\lambda_0+\epsilon)|_{\epsilon=0}\\
\widetilde{\lambda_{0}}^{2}b_1&\widetilde{\lambda_{0}}a_1 &b_1\\
(\widetilde{\lambda_{0}}^{2}a_1)^*&-(\widetilde{\lambda_{0}}b_1)^* &a_1^* \nonumber\\
\end{vmatrix} \epsilon^2+O(\epsilon^3)}
{\begin{vmatrix}
\widetilde{\lambda_{0}}b_1&a_1\\
-(\widetilde{\lambda_{0}}a_1)^*&b_1^*\nonumber\\
\end{vmatrix} }\\ \\
\dfrac{\begin{vmatrix}
\dfrac{\partial^2 }{\partial\epsilon^2}((\widetilde{\lambda_0}+\epsilon)^{2}\varphi_{1}(\lambda_0+\epsilon))|_{\epsilon=0}
&\dfrac{\partial^2 }{\partial\epsilon^2}((\widetilde{\lambda_0}+\epsilon)\phi_{1}(\lambda_0+\epsilon))|_{\epsilon=0}&\dfrac{\partial^2 }{\partial\epsilon^2}\varphi_{1}(\lambda_0+\epsilon)|_{\epsilon=0}\\
\widetilde{\lambda_{0}}^{2}a_1 &\widetilde{\lambda_{0}}b_1&a_1\\
(\widetilde{\lambda_{0}}^{2}b_1)^* &-(\widetilde{\lambda_{0}}a_1)^*&b_1^*\nonumber\\
\end{vmatrix}\epsilon^2+O(\epsilon^3) }{\begin{vmatrix}
\widetilde{\lambda_{0}}a_1 &b_1\\
-(\widetilde{\lambda_{0}}b_1)^*&a_1^*\nonumber\\
\end{vmatrix} }
\end{array}\mbox{\hspace{-0.2cm}} \right)\\
&&=\left(
\begin{array}{c}
b^{[2]}_2\epsilon^2+O(\epsilon^3)\\
a^{[2]}_2\epsilon^2+O(\epsilon^3)
\end{array} \right).\label{psi32expansion}
\end{eqnarray}
 Here $b^{[2]}_2$ and $a^{[2]}_2$ are smooth and  are not both zero function.
 In the expansion of $\psi^{[2]}_3|_{\lambda_3=\lambda_0+\epsilon}$, there exist as least one term
 $|b^{[2]}_l|+|a^{[2]}_l|\not=0(l\geqslant 2)$. If all coefficients of expansion are zero, $\psi^{[2]}_3|_{\lambda_3
 =\lambda_0+\epsilon}=0$. Thus  it does not lost any generality by setting that
     $|b^{[2]}_2|$ and  $|a^{[2]}_2|$  are not both zero.
 Furthermore, take this expansion back into eq.(\ref{q4phi2varphi2}) with
$\lambda_3=\lambda_0+\epsilon$, then
\begin{eqnarray}\label{q4smooth}
&q^{[4]}\mbox{\hspace{-0.4cm}}&=\dfrac{(\widetilde{\lambda_0}|a^{[2]}_2|^2+ \widetilde{\lambda_0}^*|b^{[2]}_2|^2)^2\epsilon^8+O(\epsilon^9)}
{(\widetilde{\lambda_0}|b^{[2]}_2|^2+ \widetilde{\lambda_0}^*|a^{[2]}_2|^2)^2\epsilon^8+O(\epsilon^9)}q^{[2]}-
2i\dfrac{(\widetilde{\lambda_0}^2-\widetilde{\lambda_0}^{*2})(\widetilde{\lambda_0}|a^{[2]}_2|^2+
\widetilde{\lambda_0}^*|b^{[2]}_2|^2)
b^{[2]}_2(a^{[2]}_2)^*\epsilon^8+O(\epsilon^9) }
{(\widetilde{\lambda_0}|b^{[2]}_2|^2+\widetilde{\lambda_0}^*|a^{[2]}_2|^2)^2\epsilon^8+O(\epsilon^9) }\nonumber\\
&&
\stackrel{\epsilon\rightarrow 0}{=}\dfrac{(\widetilde{\lambda_0}|a^{[2]}_2|^2+ \widetilde{\lambda_0}^*|b^{[2]}_2|^2)^2}
{(\widetilde{\lambda_0}|b^{[2]}_2|^2+ \widetilde{\lambda_0}^*|a^{[2]}_2|^2)^2}q^{[2]}-
2i\dfrac{(\widetilde{\lambda_0}^2-\widetilde{\lambda_0}^{*2})(\widetilde{\lambda_0}|a^{[2]}_2|^2+
\widetilde{\lambda_0}^*|b^{[2]}_2|^2)
b^{[2]}_2(a^{[2]}_2)^* }
{(\widetilde{\lambda_0}|b^{[2]}_2|^2+\widetilde{\lambda_0}^*|a^{[2]}_2|^2)^2}\triangleq q^{[4]}_{smooth},
\end{eqnarray}
which is a smooth solution because $\alpha_0\not=0$, $|b^{[2]}_2|$ and $|a^{[2]}_2|$ are not both zero.
Here $q^{[2]}$ is given by  eq.(\ref{q2smooth}).
 \; \; \;$\square$\\

It is trivial to find that $q^{[4]}$ in eq.(\ref{q4smooth}) is
generated from $q^{[2]}$ by iteration of $T_2^{\prime}$ with new
generating function ($a_2^{[2]},b_2^{[2]}$). In other words,
$q^{[4]}$ is obtained from $q^{[2]}$ through a non-degenerate
$T_2^{\prime}(\lambda_0;a^{[2]}_2,b^{[2]}_2)$. Thus there exist a
non-degenerate four-fold DT
$T_4^{\prime}=T_2^{\prime}(\lambda_0;a^{[2]}_2,b^{[2]}_2)T_2^{\prime}(\lambda_0;
a_1,b_1)$, which gives  this smooth $q^{[4]}$ from seed $q$.
Therefore we overcome the problem of double degeneration in $T_4$
when   $\lambda_3=\lambda_1$ and
$\psi_3(\lambda_0)=\psi_1(\lambda_0)=0$. Moreover, let
 $\lambda_3\mapsto \lambda_1=\lambda_0+\epsilon$ and $\psi_3(\lambda_0)=\psi_1(\lambda_0)=0$ in
$\psi_j^{[4]}$ in eq.(\ref{psineven}), perform Taylor expansion at the
rows associated with $\lambda_1$ and $\lambda_3$,  we have
\begin{eqnarray}\label{psi4smooth}
\psi^{[4]}_j|_{\lambda_j=\lambda_0+\epsilon}=\left(
\begin{array}{c}
\dfrac{\begin{vmatrix}
h_{41}^{3}&h_{32}^{3}&h_{21}^{3}&h_{12}^{3}&h_{01}^{3}\\
h_{41}^{1}&h_{32}^{1}&h_{21}^{1}&h_{12}^{1}&h_{01}^{1}\\
{h^{1}_{42}}^{\ast}&-{h^{1}_{31}}^{\ast}&{h^{1}_{22}}^{\ast}&-{h^{1}_{11}}^{\ast}&{h^{1}_{02}}^{\ast}\\
h_{41}^{2}&h_{32}^{2}&h_{21}^{2}&h_{12}^{2}&h_{01}^{2}\\
{h^{2}_{42}}^{\ast}&-{h^{2}_{31}}^{\ast}&{h^{2}_{22}}^{\ast}&-{h^{2}_{11}}^{\ast}&{h^{2}_{02}}^{\ast}\\
\end{vmatrix}\epsilon^3+O(\epsilon^4)}{\begin{vmatrix}
h_{31}^{1}&h_{22}^{1}&h_{11}^{1}&h_{02}^{1}\\
-{h^{1}_{32}}^{\ast}&{h^{1}_{21}}^{\ast}&-{h^{1}_{12}}^{\ast}&{h^{1}_{01}}^{\ast}\\
h_{31}^{2}&h_{22}^{2}&h_{11}^{2}&h_{02}^{2}\\
-{h^{2}_{32}}^{\ast}&{h^{2}_{21}}^{\ast}&-{h^{2}_{12}}^{\ast}&{h^{2}_{01}}^{\ast}\\
\end{vmatrix}}\\ \\
\dfrac{\begin{vmatrix}
h_{42}^{3}&h_{31}^{3}&h_{22}^{3}&h_{11}^{3}&h_{02}^{3}\\
h_{42}^{1}&h_{31}^{1}&h_{22}^{1}&h_{11}^{1}&h_{02}^{1}\\
{h^{1}_{41}}^{\ast}&-{h^{1}_{32}}^{\ast}&{h^{1}_{21}}^{\ast}&-{h^{1}_{12}}^{\ast}&{h^{1}_{01}}^{\ast}\\
h_{42}^{2}&h_{31}^{2}&h_{22}^{2}&h_{11}^{2}&h_{02}^{2}\\
{h^{2}_{41}}^{\ast}&-{h^{2}_{32}}^{\ast}&{h^{2}_{21}}^{\ast}&-{h^{2}_{12}}^{\ast}&{h^{2}_{01}}^{\ast}\\
\end{vmatrix}\epsilon^3+O(\epsilon^4)}{\begin{vmatrix}
h_{32}^{1}&h_{21}^{1}&h_{12}^{1}&h_{01}^{1}\\
-{h^{1}_{31}}^{\ast}&{h^{1}_{22}}^{\ast}&-{h^{1}_{11}}^{\ast}&{h^{1}_{02}}^{\ast}\\
h_{32}^{2}&h_{21}^{2}&h_{12}^{2}&h_{01}^{2}\\
-{h^{2}_{31}}^{\ast}&{h^{2}_{22}}^{\ast}&-{h^{2}_{11}}^{\ast}&{h^{2}_{02}}^{\ast}\\
\end{vmatrix}}\\
\end{array} \right) \triangleq (T_4^{\prime}(\lambda_j;\lambda_0)\psi_j)|_{\lambda_j=\lambda_0+\epsilon},
\end{eqnarray}
which provides a determinant representation of $T_4^{\prime}$. In the above process,
reduction conditions in eq.(\ref{2nfoldredu}) is used to calculate  the elements of the rows associated with $\lambda_2$ and
$\lambda_4$.
Here
\begin{eqnarray}\label{hm}
h_{m1}^{l}=\dfrac{\partial^{l}}{\partial{\epsilon}^{l}}((\widetilde{\lambda_0}
+\epsilon)^{m}\phi_1(\lambda_1=\lambda_0+\epsilon))|_{\epsilon=0},
h_{m2}^{l}=\dfrac{\partial^{l}}{\partial{\epsilon}^{l}}((\widetilde{\lambda_0}
+\epsilon)^{m}\varphi_1(\lambda_1=\lambda_0+\epsilon))|_{\epsilon=0}
\end{eqnarray}
with $\ m=0,1,\cdots,4, \ l=1,2$. Similarly, let $\lambda_3=\lambda_1=\lambda_0+\epsilon$
in eq.(\ref{ntt4}), then Taylor expansion with respect to $\epsilon$ gives a representation
of $q^{[4]}_{smooth}$ in eq.(\ref{q4smooth}) as follows.
\begin{equation}
q^{[4]}_{smooth}=\dfrac{\Omega_{41}^{2}}{\Omega_{43}^{2}}q-2i\dfrac{\Omega_{41}\Omega_{42}}{\Omega_{43}^{2}},
\end{equation}
and
\begin{equation*}
\Omega_{41}=\begin{vmatrix}
{h^{1}}_{32}&{h^{1}}_{21}&{h^{1}}_{12}&{h^{1}}_{01}\\
-{h^{1}_{31}}^{\ast}&{h^{1}_{22}}^{\ast}&-{h^{1}_{11}}^{\ast}&{h^{1}_{02}}^{\ast}\\
{h^{2}}_{32}&{h^{2}}_{21}&{h^{2}}_{12}&{h^{2}}_{01}\\
-{h^{2}_{31}}^{\ast}&{h^{2}_{22}}^{\ast}&-{h^{2}_{11}}^{\ast}&{h^{2}_{02}}^{\ast}\\
\end{vmatrix},
\end{equation*}
\begin{equation*}
\Omega_{42}=\begin{vmatrix}
{h^{1}}_{41}&{h^{1}}_{21}&{h^{1}}_{12}&{h^{1}}_{01}\\
{h^{1}_{42}}^{\ast}&{h^{1}_{22}}^{\ast}&-{h^{1}_{11}}^{\ast}&{h^{1}_{02}}^{\ast}\\
{h^{2}}_{41}&{h^{2}}_{21}&{h^{2}}_{12}&{h^{2}}_{01}\\
{h^{2}_{42}}^{\ast}&{h^{2}_{22}}^{\ast}&-{h^{2}_{11}}^{\ast}&{h^{2}_{02}}^{\ast}\\
\end{vmatrix},
\end{equation*}
\begin{equation*}
\Omega_{43}=\begin{vmatrix}
{h^{1}}_{31}&{h^{1}}_{22}&{h^{1}}_{11}&{h^{1}}_{02}\\
-{h^{1}_{32}}^{\ast}&{h^{1}_{21}}^{\ast}&-{h^{1}_{12}}^{\ast}&{h^{1}_{01}}^{\ast}\\
{h^{2}}_{31}&{h^{2}}_{22}&{h^{2}}_{11}&{h^{2}}_{02}\\
-{h^{2}_{32}}^{\ast}&{h^{2}_{21}}^{\ast}&-{h^{2}_{12}}^{\ast}&{h^{2}_{01}}^{\ast}\\
\end{vmatrix}.
\end{equation*}
Note that the smoothness of this solution is analyzed in eq.(\ref{q4smooth}) according to the iteration of
non-degenerate
$T_2^{\prime}$.

 On the one side, by (k-1) times iteration of $T_2^{\prime}$ with a fixed eigenvalue $\lambda_0$ but
  different eigenfunctions, a non-degenerate (2k-2) fold DT
$T_{2k-2}^{\prime}$ is obtained  because each step of iteration is non-degenerate.
On the other hand, setting double degeneration, i.e.,$\lambda_i=\lambda_0$ in $\psi_{2k-1}^{[2k-2]}$ and
$\psi_i(\lambda_0)=0(i=1,3,5,
\cdots,2k-3)$ in eq.(\ref{psineven}), and using reduction conditions eq.(\ref{2nfoldredu}),  then this $T_{2k-2}^{\prime}$ also can
be presented after performing Taylor expansion  with respect to $\epsilon$. The smoothness of $q^{[2k-2]}$ is provided
by the smoothness of  each step of iteration as we shown in eq.(\ref{q4smooth}).
By the determinant representation of $\psi_j^{[2k-2]}(j\geq 2k-1)$ with double degeneration and reduction condition, we have following Lemma. \\
{\bf Lemma 12} {\sl The non-degenerate $T_{2k-2}^{\prime}$ generates
a smooth eigenfunction of $\lambda_{2k-1}=\lambda_0+\epsilon$ from $\psi_{2k-1}(\lambda_0+\epsilon)$ as
\begin{equation}\label{psi2kmins2expansion}
\psi^{[2k-2]}_{2k-1}|_{\lambda_{2k-1}=\lambda_0+\epsilon}= \left(
\begin{array}{c}
b^{[2k-2]}_k\epsilon^k+O(\epsilon^{k+1})\\
a^{[2k-2]}_k\epsilon^k+O(\epsilon^{k+1})
\end{array} \right).
\end{equation}
Here  $|b_k^{[2k-2]}|$ and $|a^{[2k-2]}_k|$ are smooth and  not both zero.
}\\
{\bf Lemma 13} {\sl   Set $\psi^{[2k-2]}_{2k-1}|_{\lambda_{2k-1}=\lambda_0+\epsilon}$ be generating function of a two-fold DT
$\hat{T}_2=T_2(\lambda_0+\epsilon;\psi^{[2k-2]}_{2k-1})$ with the help of the reduction conditions,
then  $q^{[2k]}$, generated from $q^{[2k-2]}$, is a smooth rational k-order  solution of the MNLS equation.}
\\
{\bf Proof}  Replace eigenvalue $\lambda_1$, eigenfunction $\psi_1$ and seed solution $q$
by  $\lambda_0+\epsilon$, $ \psi^{[2k-2]}_{2k-1}|_{\lambda_{2k-1}=\lambda_0+\epsilon}$  and $q^{[2k-2]}$  in the formula of two-fold DT in
eq.(\ref{q2phi1varphi1}), respectively, then
\begin{eqnarray}\label{q2ksmooth}
q^{[2k]}\mbox{\hspace{-0.5cm}}&=&\mbox{\hspace{-0.5cm}}
\dfrac{(\widetilde{\lambda_0}|a^{[2k-2]}_k|^2+ \widetilde{\lambda_0}^*|b^{[2k-2]}_k|^2)^2\epsilon^{4k}+O(\epsilon^{4k+1})}
{(\widetilde{\lambda_0}|b^{[2k-2]}_k|^2+ \widetilde{\lambda_0}^*|a^{[2k-2]}_k|^2)^2
\epsilon^{4k}+O(\epsilon^{4k+1})}q^{[2k-2]}\nonumber\\
&\mbox{\hspace{0.5cm}}-&2i\dfrac{(\widetilde{\lambda_0}^2-\widetilde{\lambda_0}^{*2})
(\widetilde{\lambda_0}|a^{[2k-2]}_k|^2+\widetilde{\lambda_0}^*|b^{[2k-2]}_k|^2)
b^{[2k-2]}_k(a^{[2k-2]}_k)^*\epsilon^{4k}+O(\epsilon^{4k+1}) }
{(\widetilde{\lambda_0}|b^{[2k-2]}_k|^2+\widetilde{\lambda_0}^*|a^{[2k-2]}_k|^2)^2\epsilon^{4k}+O(\epsilon^{4k+1}) }
\nonumber\\
&\stackrel{\epsilon\rightarrow 0}{=}&\mbox{\hspace{-0.2cm}}
\dfrac{(\widetilde{\lambda_0}|a^{[2k-2]}_k|^2+ \widetilde{\lambda_0}^*|b^{[2k-2]}_k|^2)^2}
{(\widetilde{\lambda_0}|b^{[2k-2]}_k|^2+ \widetilde{\lambda_0}^*|a^{[2k-2]}_k|^2)^2}q^{[2k-2]}\nonumber \\
&\mbox{\hspace{0.5cm}}-&
2i\dfrac{(\widetilde{\lambda_0}^2-\widetilde{\lambda_0}^{*2})(\widetilde{\lambda_0}|a^{[2k-2]}_k|^2+
\widetilde{\lambda_0}^*|b^{[2k-2]}_k|^2)
b^{[2k-2]}_k(a^{[2k-2]}_k)^* }
{(\widetilde{\lambda_0}|b^{[2k-2]}_k|^2+\widetilde{\lambda_0}^*|a^{[2k-2]}_k|^2)^2}\triangleq q^{[2k]}_{smooth},
\end{eqnarray}
which is a smooth solution because $\alpha_0\not=0$, $|b^{[2k-2]}_k|$ and $|a^{[2k-2]}_k|$ are not both zero.
 \; \; \;$\square$\\

Under double degeneration mentioned above, $T_{2k}$ is degenerate. Thus solution in eq.(\ref{ntt4}) is singular
at $\lambda_0$. This lemma shows that its singularity is removable. We shall give a explicit representation
of $q^{[2k]}_{smooth}$ by Taylor expansion as the same manner of $q^{[4]}_{smooth}$. In other words,
$T_{2k}$ gives a non-degenerate 2k-fold DT $T^{\prime}_{2k}$ by Taylor expansion, and $q^{[2k]}_{smooth}$ is generated
from $q$ by  this non-degenerate DT.
\\
\noindent {\bf Theorem 14.}  {\sl Let $\psi_1$ be an smooth eigenfunction of Lax pair of the MNLS, has also continuous
dependence on the $\lambda_0$.  Here $\lambda_0$ is only one zero point of $\psi_1$.
Setting  $\lambda_{2l-1}\rightarrow \lambda_1=\lambda_0(l=1,2,3,\cdots,k)$ in  $q^{[2k]}$  of  eq.(\ref{ntt4}),
then higher order Taylor expansion in it with respect to $\epsilon$ leads to a  smooth k-order solution $q^{[2k]}_{smooth}$
of the MNLS possessing following formula.
\begin{eqnarray}\label{rntt4}
&&{q}^{[2k]}_{smooth}=\dfrac{\Omega_{k1}^{2}}{\Omega_{k3}^{2}}q-2i\dfrac{\Omega_{k1}\Omega_{k2}}{\Omega_{k3}^{2}}.
\end{eqnarray}
Here
\begin{equation*}
\Omega_{k1}=\begin{vmatrix}
{h^{1}}_{2k-12}&{h^{1}}_{2k-21}&{h^{1}}_{2k-32}&{h^{1}}_{2k-41}&\ldots&{h^{1}}_{12}&{h^{1}}_{01}\\
-{h^{1}_{2k-11}}^{\ast}&{h^{1}_{2k-22}}^{\ast}&-{h^{1}_{2k-31}}^{\ast}&{h^{1}_{2k-42}}^{\ast}&\ldots&-{h^{1}_{11}}^{\ast}&{h^{1}_{02}}^{\ast}\\
\vdots&\vdots&\vdots&\vdots&\vdots\\
{h^{k}}_{2k-12}&{h^{k}}_{2k-21}&{h^{k}}_{2k-32}&{h^{k}}_{2k-41}&\ldots&{h^{k}}_{12}&{h^{k}}_{01}\\
-{h^{k}_{2k-11}}^{\ast}&{h^{k}_{2k-22}}^{\ast}&-{h^{k}_{2k-31}}^{\ast}&{h^{k}_{2k-42}}^{\ast}&\ldots&-{h^{k}_{11}}^{\ast}&{h^{k}_{02}}^{\ast}\\
\end{vmatrix},
\end{equation*}
\begin{equation*}
\Omega_{k2}=\begin{vmatrix}
{h^{1}}_{2k1}&{h^{1}}_{2k-21}&{h^{1}}_{2k-32}&{h^{1}}_{2k-41}&\ldots&{h^{1}}_{12}&{h^{1}}_{01}\\
{h^{1}_{2k2}}^{\ast}&{h^{1}_{2k-22}}^{\ast}&-{h^{1}_{2k-31}}^{\ast}&{h^{1}_{2k-42}}^{\ast}&\ldots&-{h^{1}_{11}}^{\ast}&{h^{1}_{02}}^{\ast}\\
\vdots&\vdots&\vdots&\vdots&\vdots\\
{h^{k}}_{2k1}&{h^{k}}_{2k-21}&{h^{k}}_{2k-32}&{h^{k}}_{2k-41}&\ldots&{h^{k}}_{12}&{h^{k}}_{01}\\
{h^{k}_{2k2}}^{\ast}&{h^{k}_{2k-22}}^{\ast}&-{h^{k}_{2k-31}}^{\ast}&{h^{k}_{2k-42}}^{\ast}&\ldots&-{h^{k}_{11}}^{\ast}&{h^{k}_{02}}^{\ast}\\
\end{vmatrix},
\end{equation*}
\begin{equation*}
\Omega_{k3}=\begin{vmatrix}
{h^{1}}_{2k-11}&{h^{1}}_{2k-22}&{h^{1}}_{2k-31}&{h^{1}}_{2k-42}&\ldots&{h^{1}}_{11}&{h^{1}}_{02}\\
-{h^{1}_{2k-12}}^{\ast}&{h^{1}_{2k-21}}^{\ast}&-{h^{1}_{2k-32}}^{\ast}&{h^{1}_{2k-41}}^{\ast}&\ldots&-{h^{1}_{12}}^{\ast}&{h^{1}_{01}}^{\ast}\\
\vdots&\vdots&\vdots&\vdots&\vdots\\
{h^{k}}_{2k-11}&{h^{k}}_{2k-22}&{h^{k}}_{2k-31}&{h^{k}}_{2k-42}&\ldots&{h^{k}}_{11}&{h^{k}}_{02}\\
-{h^{k}_{2k-12}}^{\ast}&{h^{k}_{2k-21}}^{\ast}&-{h^{k}_{2k-32}}^{\ast}&{h^{k}_{2k-41}}^{\ast}&\ldots&-{h^{k}_{12}}^{\ast}&{h^{k}_{01}}^{\ast}\\
\end{vmatrix},
\end{equation*}
$h_{m1}^{l}$ and  $h_{m2}^{l}$ are defined by eq.(\ref{hm})}, $m=0,1,\cdots,2k, \ l=1,2,\cdots k$.\\
Note again as theorem 7 that k-order just denotes the k-fold of DT for the MNLS,which is not related to the smoothness of the
solution.

\section{Rational solutions generated by 2k-fold degenerate Darboux transformation}

  According to theorem 14, it is a crucial step to find a suitable zero point $\lambda_0$ of eigenfunction such that
$h_{m1}^{l}$ and  $h_{m2}^{l}$ are both polynomials in  $x$ and $t$,
which will be given from an explicit formula of $\psi_1$ in this
section. We shall find explicit forms of the eigenfunction $\psi_j$
associated with a periodic seed solution, and then present smooth
rational $k$-order solutions according to theorem 14. Next, the
explicit representations of the first order and the second order
rational solutions
 of the MNLS equation are constructed. The explicit forms of the first four order
  rogue waves are also provided. Furthermore,  localization of the first order rational
 solution is analyzed  and  several new patterns of the higher order rogue waves are presented.
  We shall show an unusual  result: for a given value of $a$, the increasing value of $b$ can
  damage gradually the localization of the rational solution.

Let $a_1$ and  $c_1$ be two complex constants, then $ q=c_1
\exp{(i(a_1 x+(-b{c_1}^{2}-{a_1}^{2}-aa_1{c_1}^{2})t))}$ is a
periodic solution of the MNLS equation, which will be used as a seed
solution of the DT. Substituting $q=c_1 \exp{(i(a_1
x+(-b{c_1}^{2}-{a_1}^{2}-aa_1{c_1}^{2})t))}$ into the spectral
problem eq.(\ref{sys11})
 and eq.(\ref{sys22}), and using the method of separation of variables and the superposition
principle, the eigenfunction $\psi_{2k-1}$ associated with
$\lambda_{2k-1}$ is given by
{\small
\begin{eqnarray}\label{eigenfunfornonzeroseed}
\left(\mbox{\hspace{-0.2cm}} \begin{array}{c}
 \mbox{\hspace{-0.1cm}}\phi_{2k-1}(x,t,\lambda_{2k-1})\mbox{\hspace{-0.1cm}}\\
 \mbox{\hspace{-0.1cm}}\varphi_{2k-1}(x,t,\lambda_{2k-1})\mbox{\hspace{-0.1cm}}\\
\end{array}\mbox{\hspace{-0.2cm}}\right)\mbox{\hspace{-0.25cm}}=\mbox{\hspace{-0.25cm}}\left(\mbox{\hspace{-0.2cm}}\begin{array}{c}
 \mbox{\hspace{-0.1cm}}C_1\varpi_1(x,t,\lambda_{2k-1})[1]\mbox{\hspace{-0.15cm}}+\mbox{\hspace{-0.15cm}}C_2\varpi_2(x,t,\lambda_{2k-1})[1]\mbox{\hspace{-0.15cm}}+\mbox{\hspace{-0.15cm}}C_3\varpi_1^{\ast}(x,t,\mbox{\hspace{-0.2cm}}-\mbox{\hspace{-0.15cm}}{\lambda_{2k-1}^{\ast})}[2]\mbox{\hspace{-0.15cm}}+\mbox{\hspace{-0.15cm}}C_4\varpi_2^{\ast}(x,t,\mbox{\hspace{-0.2cm}}-\mbox{\hspace{-0.15cm}}{\lambda_{2k-1}^{\ast})}[2]\mbox{\hspace{-0.1cm}}\\
 \mbox{\hspace{-0.1cm}}C_1\varpi_1(x,t,\lambda_{2k-1})[2]\mbox{\hspace{-0.15cm}}+\mbox{\hspace{-0.15cm}}C_2\varpi_2(x,t,\lambda_{2k-1})[2]\mbox{\hspace{-0.15cm}}+\mbox{\hspace{-0.15cm}}C_3\varpi_1^{\ast}(x,t,\mbox{\hspace{-0.2cm}}-\mbox{\hspace{-0.15cm}}{\lambda_{2k-1}^{\ast})}[1]\mbox{\hspace{-0.15cm}}+\mbox{\hspace{-0.15cm}}C_4\varpi_2^{\ast}(x,t,\mbox{\hspace{-0.2cm}}-\mbox{\hspace{-0.15cm}}{\lambda_{2k-1}^{\ast})}[1]\mbox{\hspace{-0.1cm}}\\
\end{array}\mbox{\hspace{-0.1cm}}\right).
\end{eqnarray}
}
Here
{\small
\begin{eqnarray*}
\left(\mbox{\hspace{-0.2cm}}\begin{array}{c}
 \varpi_1(x,t,\lambda_{2k-1})[1]\\
 \varpi_1(x,t,\lambda_{2k-1})[2]\\
\end{array}\mbox{\hspace{-0.2cm}}\right)\mbox{\hspace{-0.25cm}}=\mbox{\hspace{-0.25cm}}\left(\mbox{\hspace{-0.2cm}}\begin{array}{c}
\exp(-\sqrt{S(\lambda_{2k-1})}P(\lambda_{2k-1})+\dfrac{1}{2}i\theta) \\
-\dfrac{a_1+2a{\lambda_{2k-1}}^{2}+2i\sqrt{2b}\lambda_{2k-1}-\sqrt{S(\lambda_{2k-1}})}{(-\sqrt{2b}+2ia\lambda_{2k-1})c_1}\exp(-\sqrt{S(\lambda_{2k-1})}P(\lambda_{2k-1})-\dfrac{1}{2}i\theta )\\
\end{array}\mbox{\hspace{-0.2cm}}\right),
\end{eqnarray*}
\begin{eqnarray*}
\left(\mbox{\hspace{-0.2cm}}\begin{array}{c}
 \varpi_2(x,t,\lambda_{2k-1})[1]\\
 \varpi_2(x,t,\lambda_{2k-1})[2]\\
\end{array}\mbox{\hspace{-0.2cm}}\right)\mbox{\hspace{-0.25cm}}=\mbox{\hspace{-0.25cm}}\left(\mbox{\hspace{-0.2cm}}\begin{array}{c}
\exp(\sqrt{S(\lambda_{2k-1})}P(\lambda_{2k-1})+\dfrac{1}{2}i\theta) \\
-\dfrac{a_1+2a{\lambda_{2k-1}}^{2}+2i\sqrt{2b}\lambda_{2k-1}+\sqrt{S(\lambda_{2k-1}})}{(-\sqrt{2b}+2ia\lambda_{2k-1})c_1}\exp(\sqrt{S(\lambda_{2k-1})}P(\lambda_{2k-1})-\dfrac{1}{2}i\theta )\\
\end{array}\mbox{\hspace{-0.2cm}}\right),
\end{eqnarray*}
}
\begin{eqnarray*}
\varpi_1(x,t,\lambda_{2k-1})= \left( \begin{array}{c}
 \varpi_1(x,t,\lambda_{2k-1})[1]\\
 \varpi_1(x,t,\lambda_{2k-1})[2]\\
\end{array} \right),~~~~~
\varpi_2(x,t,\lambda_{2k-1})= \left( \begin{array}{c}
 \varpi_2(x,t,\lambda_{2k-1})[1]\\
 \varpi_2(x,t,\lambda_{2k-1})[2]
\end{array} \right),
\end{eqnarray*}
\begin{eqnarray*}
&&S(\lambda_{2k-1})=4a^2{\lambda_{2k-1}}^{4}+8ia\sqrt{2b}{\lambda_{2k-1}}^{3}+(4a^2c_1^2+4aa_1-8b){\lambda_{2k-1}}^2\\\nonumber
&&~~~~~~~~~~~~~~~~~+(4i\sqrt{2b}a_1+4ia\sqrt{2b}{c_1}^{2})\lambda_{2k-1}+{a_1}^2-2b{c_1}^2,\\
\nonumber
&&P(\lambda_{2k-1})=((4\lambda_{2k-1}b-6i\sqrt{2b}a{\lambda_{2k-1}}^{2}\mbox{\hspace{-0.2cm}}+{c_1}^{2}\sqrt{2b}ai-4a^2{\lambda_{2k-1}}^{3}
\mbox{\hspace{-0.2cm}}+\sqrt{2b}a_1i+2a_1a\lambda_{2k-1}\\\nonumber
&&~~~~~~~~~~~~~~~~+2{c_1}^{2}a^2\lambda_{2k-1})t+(-2a\lambda_{2k-1}-\sqrt{2b}i)x)\dfrac{1}{2(2ia\lambda_{2k-1}-\sqrt{2b})},
\ (k=1,2,\dots,n), \\ \nonumber &&\theta=a_1
x+(-b{c_1}^{2}-{a_1}^{2}-aa_1{c_1}^{2})t, \nonumber
\end{eqnarray*}
and $a,b,a_1,c_1,x,t \in \Bbb R$, $C_1, C_2, C_3, C_4 \in \Bbb C$. Note
that $\varpi_1(x,t,\lambda_{2k-1})$ and
$\varpi_2(x,t,\lambda_{2k-1})$ are two linear independent solutions of the
spectral problem eq.(\ref{sys11}) and eq.(\ref{sys22}).  We   can
only get the  trivial solutions through DT of the MNLS equation by
setting eigenfunction $\psi_{2k-1}$ be one of them. This is the reason of setting
$\psi_{2k-1}$ as the linear  superposition in eq.(\ref{eigenfunfornonzeroseed}).\\

   In order to make higher order rational solution of the MNLS,
a crucial step is to find the zero point of $S$ and the eigenfunctions $\psi_{l}$
such that exponential functions vanish  and the indeterminate form $\dfrac{0}{0}$ appear in
the $q^{[2k]}$.  By tedious calculation, we have following lemma concerning of this fact.\\ \\
{\bf Lemma 15} {\sl Let
\begin{eqnarray}\label{coefficientofeigenfunction}
&&C_1=-(K_0+2)+\exp(\dfrac{1}{2}iS(\lambda_{2k-1})\sum_{j=0}^{k-1}S_j(\lambda_{2k-1}-\lambda_0)^{j}),\nonumber\\
&&C_2=-(K_0+2)+\exp(-\dfrac{1}{2}iS(\lambda_{2k-1})\sum_{j=0}^{k-1}S_j(\lambda_{2k-1}-\lambda_0)^{j}),\nonumber\\
&&C_3=K_0+\exp(\dfrac{1}{2}iS(\lambda_{2k-1})\sum_{j=0}^{k-1}L_j(\lambda_{2k-1}-\lambda_0)^{j}),\nonumber\\
&&C_4=K_0+\exp(-\dfrac{1}{2}iS(\lambda_{2k-1})\sum_{j=0}^{k-1}L_j(\lambda_{2k-1}-\lambda_0)^{j}),\label{chooseCC}
\end{eqnarray}
then $\lambda_{2k-1}=\lambda_0=-i\dfrac{\sqrt{2b}-\sqrt{2aa_1+a^2{c_1}^{2}+2b}+ac_1}{2a}$
is only one zero point of $S$ and eigenfunction $\psi_{2k-1}$ in
eq.(\ref{eigenfunfornonzeroseed}).  Here $K_0,S_j,L_j\in \Bbb C$.}\\
\\
{\bf Theorem 16} {\sl For the eigenfunction $\psi_{1}$ defined by
eq.(\ref{eigenfunfornonzeroseed}) and
eq.(\ref{coefficientofeigenfunction}), $h_{m1}^{l}$ and
$h_{m2}^{l}$ are both polynomials in variables $x$ and $t$.
Furthermore, ${q}^{[2k]}_{smooth}$ generates smooth rational k-order
solutions of the MNLS equation.}
 \\
{\bf Proof}.  Take the eigenfunction $\psi_{2l-1}(l=1,2,3,\cdots,k)$ defined by eq.(\ref{eigenfunfornonzeroseed})
and eq.(\ref{coefficientofeigenfunction}) into  eq.(\ref{ntt4}), $q^{[2k]}$ provides a solution expressed by an
 indeterminate form $\dfrac{0}{0}$ of the MNLS equation when  $\lambda_{2k-1}\rightarrow-i\dfrac{\sqrt{2b}
-\sqrt{2aa_1+a^2{c_1}^{2}+2b}+ac_1}{2a}$. Obviously, this is a case
of  double degeneration,and then we can apply Theorem 14 here.
Because $\lambda_0$ is a zero point of $S$ and $\psi_{1}$ in
eq.(\ref{eigenfunfornonzeroseed}), exp(S) will disappear when
$\lambda_1=\lambda_0$, $h_{m1}^{l}$ and  $h_{m2}^{l}$ are both
polynomials in $x$ and $t$. Therefore,  take the eigenfunction
$\psi_{1}$ defined by eq.(\ref{eigenfunfornonzeroseed}) and
eq.(\ref{coefficientofeigenfunction}) into theorem 14,
${q}^{[2k]}_{smooth}$ are smooth rational solutions of the MNLS.
$\square$
\\
Note again as theorem 7 that k-order just denotes the k-fold of DT for the MNLS,which is not related to the smoothness
 of  the solution or the order of polynomials.
\\
{\bf 4.1 Asymptotic behavior of rational 1-order solution}
\\
For simplicity, set $k=1, a_1=-1,c_1=1,K_0=1, S_0=L_0=0$ in theorem 16, the rational 1-order
solution $q^{[2]}_{smooth}$ is given by the form as follows.
\begin{equation}\label{firstrw}
{q^{[1]}_{rational}}=-\exp((-x-tb-t+ta)i)\dfrac{H_1H_2}{{H_1^{\ast}}^{2}},
\end{equation}
\begin{eqnarray*}
\text{with}
&&H_1=2(a-b)((-2b+3a^2-6a+4)t^2-4(-1+a)
xt+x^2)+1-2ia(x-(3a-2)t),\\
&&H_2=2(a-b)((-2b+3a^2-6a+4
)t^2-4(-1+a)xt+x^2)-3-2i(ax-(-6a+3a^2+4b)t).
\end{eqnarray*}
By letting $ x \rightarrow {\infty}, \ t \rightarrow {\infty}$, $|q^{[1]}_{rw}|^{2}\rightarrow 1$.
Moreover, $q^{[1]}_{rational}$ is not a traveling wave except $a=b$. Further more, set imaginary part of $H_1$ be zero,
i.e.  $x=(3a-2)t$, but its real part $4t^2(a-b)^2+1>0$,so this solution is smooth on whole plane
of $x$ and $t$.

In order to show the novel properties, we shall  discuss the asymptotic behaviors of the rational
 1-order solution.\\
\noindent {\bf Case a).} Let $a< b$, there is only one saddle point of the profile for the
$q^{[1]}_{rational}$  at point $(0,0)$ . It is a nonlocal solution with two peaks. \\
\noindent {\bf Case b).} Let $a= b$, the maximum amplitude of $|q^{[1]}_{rational}|^{2}$ is
 equal to 9 and the trajectory is defined by $x=(-2+3b)t$. It is a line soliton with only one peak.
 Of course it is nonlocal. \\
\noindent {\bf Case c).} Let $a> b\geq a-\dfrac{3}{8}a^2$, there is  only one maximum at
$(0, 0)$ in the profile of $|q^{[1]}_{rational}|^{2}$, which is 9.
This solution has one peak and one vale, and then
nonlocal.    \\
 \noindent {\bf Case d).} Let $b< a-\dfrac{3}{8}a^2$,  there are  one  maximum  at (0,0) and
 and two minima at ( $x=\mp\dfrac{-6a+3a^2+4b}{a-b}\sqrt{\dfrac{3}{32a-12a^2-32b}},
  t = \pm\dfrac{a}{a-b}\sqrt{\dfrac{3}{32a-12a^2-32b}}$ ) in the profile of the
  $|q^{[1]}_{rational}|^2$. The maximum is equal to 9 and the minimum is $0$. This solution
  has one dominant peak and two hollows. It is localized in both $x$ and $t$ directions,
   and then is called the first order rogue wave solution.   The dynamical evolution of the
  $|q^{[1]}_{rational}|^{2}$ is in accord with   the classical "Peregrine soliton" in the NLS equation
  \cite{Peregrine} and DNLS equation \cite{xuhe,xxuhe}. \\
 This discussion implies an unusual result: for a given value of $a$, the increasing value of $b$ can
 damage gradually the  localization of the rational solution.  This is in contrast to the usual conjecture that that the localization of this solution
 will be enhanced because of the appearance of the two nonlinear effects represented by $a$ and $b$, according to a common understanding of the role for the nonlinear effects in wave propagation.

For a given value $a=1$, the profiles of $|q^{[1]}_{rational}|^{2}$
in case d),case c) and case  a) are given in Figure 1 with
$b=\frac{1}{3},\frac{3}{4}, 3$, respectively. This figure shows
visually the lost of the localization of the solution due to the
increasing value of $b$. We omit the picture  of case b) because it
is a standard soliton. The density plots of the solutions of case
d), i.e.,rogue wave solutions, of the MNLS equation are given in
figure 2 with $b=0,\frac{1}{3},\frac{7}{15}$. It is clear to see the
diffusion of the peak and hollows of the first order rogue wave
when the value of $b$ is increasing in Figure 2. The explicit form
of the rogue wave in Figure 2(b) is
\begin{equation}\label{q1rwb}
q^{[1]}_{rw}=\frac {\exp^{(-i\dfrac{3\,x + t}{3})}\,(
9 - 18\,i\,x + 4\,t^{2} + 18\,i\,t + 12\,x^{2})\,(27 - 4\,t^{2}
 + 18\,i\,x - 12\,x^{2} + 30\,i\,t)}{( - 9 - 18\,i\,x - 4\,t^{2}
 + 18\,i\,t - 12\,x^{2})^{2}},
\end{equation}
which is obtained from $q^{[1]}_{rational}$ by setting $a=1$ and $b=1/3$.  The orders of polynomial in
numerator and  denominator of $q^{[1]}_{rw}$ are both 4.
 Figure 3(a) is plotted for the contour line at height 5 of
the rogue wave $|q^{[1]}_{rational}|^{2}$  with different values of  $b$ in Figure 2. Note that height 5
is half value of the peak over the asymptotic plane. Let $a=1$, then $d=\dfrac{1}{2}\sqrt{\dfrac{3((4b-3)^2+1)}{(b-1)^2(5-8b)}}$ is the
distance from the minimum point of the $|q^{[1]}_{rational}|^{2}$ to the coordinate origin, which is plotted in Figure 3(b). Figure 3 shows the remarkable decrease in localization of
the first order rogue wave. In particular, the distance $d$ goes to infinity when $b\rightarrow \frac{5}{8}$ in
Figure 3(b),  then $|q^{[1]}_{rational}|^{2}$ loses completely the localization in $x$ and $t$. This fact is consistent with
the limit value of $b$ in case d).

To see the novelty of the two-peak solution for case a), we would like to present its explicit form as
\begin{equation}\label{twopeak}
|q^{[1]}_{two peak}|^2=\dfrac{8+96t^2+32x^2-32t(2x-2t)}{(20t^2+1-4x^2)^2+(2x-2t)^2}+1,
\end{equation}
which is obtained by setting $a=1$ from eq.(\ref{firstrw}) and  is plotted in Figure 1(c). The approximate trajectory of two peaks in the profile of $|q^{[1]}_{twopeak}|^2$ are two curves defined by $20t^2+1-4x^2=0$ if $x>\frac{1}{2}$.
It is interesting to note that the height of two peaks is gradually increasing (decreasing).
The maximum height of peak along the trajectory is  $21+4\sqrt{5}$ and the minimum height is
$21-4\sqrt{5}$.  This two-peak solution with a variable height and a non-vanishing boundary of soliton equation has never been
discovered, to the best of our knowledge.  Besides,  two peak soliton in present paper can not be a
usual double-soliton because the rational soliton just has one eigenvalue $\lambda_0$.  However a double-soliton has two
eigenvalues in general.
\\
{\bf 4.2 Analytical forms and localization of the higher order rogue wave solutions}
\\
Let $k=2, a_1=-1,c_1=1,K_0=1, S_0=L_0=S_1=L_1=0$ in theorem 16, the rational 2-order solution
$q^{[4]}_{smooth}$ becomes
\begin{eqnarray}
{q^{[2]}_{rational}}=\exp{(i(-x-bt-t+at))}\dfrac{(iI_1+R_1)
(iI_1+R_1+iI_2+R_2)}{(-R_1+iI_1)^{2}}.
\end{eqnarray}
Here $I_1$, $R_1$, $I_2$ and $R_2$  are given in the appendix III because the formula is complicated.
Let $k=2, a_1=-1, c_1=1, a=1, b=\frac{1}{3}, K_0=1, S_0=L_0, S_1=L_1, L_0=1, L_1=30$ in theorem 16,
$q^{[4]}_{smooth}$ gives following 2-order rogue wave solution (Figure 4(b))
\begin{eqnarray}\label{2rw}
q^{[2]}_{rw}=-\exp^{(-\,i\,\dfrac{3\,x + t}{3})}\dfrac{v_{21}v_{22}}{v_{23}^2}
\end{eqnarray}
with
\begin{eqnarray*}
&&v_{21}=-{v_{23}}^{\ast},\\
&&v_{22}= - 300348\,x\,t - 204444\,x^{2}\,t  + 41364\,t^{2}\,x - 1080\,\sqrt{3}\,x^{4} - 24\,t^{4}\,
\sqrt{3} + 6756\,\sqrt{3}\,t^{3} - 864\,x^{4}\,t^{2}  \\
&&\mbox{\hspace{0.9cm}} - 288\,x^{2}
\,t^{4}- 48348\,\sqrt{3}\,x^{3} - 48\,t^{5}\,\sqrt{3} - 432\,x^{
5}\,\sqrt{3} + 55386\,x\,\sqrt{3} + 16110\,t\,\sqrt{3} + 19008\,x
^{3}\,t \\
&&\mbox{\hspace{0.9cm}} + 41040\,x^{2}\,t^{2} + 6336\,t^{3}\,x - 1728\,x^{3}\,t^{
2} - 288\,x\,t^{4} + 53262\,t^{2}\,\sqrt{3} - 129546\,x^{2}\,
\sqrt{3} \\
&&\mbox{\hspace{0.9cm}} - 181440\,i\,t - 325782\,i\,t^{2} - 864\,x^{6} - 432\,x^{
2}\,t^{2}\,\sqrt{3} + 67644\,x\,\sqrt{3}\,t^{2} - 288\,x\,\sqrt{3
}\,t^{3} \\
&&\mbox{\hspace{0.9cm}} - 864\,\sqrt{3}\,x^{3}\,t - 288\,x^{2}\,\sqrt{3}\,t^{3}
 - 48\,x\,\sqrt{3}\,t^{4} - 77868\,x\,\sqrt{3}\,t - 432\,x^{4}\,
\sqrt{3}\,t \\
&&\mbox{\hspace{0.9cm}} - 39204\,x^{2}\,\sqrt{3}\,t + 868482\,i\,x^{2} + 720\,i\,
t^{5} + 562653\,i\,\sqrt{3} + 757836\,i\,x + 67104\,i\,t^{3} \\
&&\mbox{\hspace{0.9cm}} + 9720\,i\,x^{4} + 3888\,i\,x^{3} + 3888\,i\,x^{5} + 216
\,i\,t^{4} - 288\,t^{2}\,\sqrt{3}\,x^{3} - 41472\,i\,x^{2}\,t \\
&&\mbox{\hspace{0.9cm}} - 68688\,i\,x\,t^{2} - 299268\,i\,t\,\sqrt{3} - 79110\,i
\,\sqrt{3}\,t^{2} - 32\,t^{6} - 2441511\,\sqrt{3} - 1032\,t^{4}
 \\
&&\mbox{\hspace{0.9cm}} + 7992\,x^{4} + 1272726\,x + 1051650\,t + 702\,x^{2} -
2592\,x^{5} + 20844\,x^{3} + 29052\,t^{3} \\
&&\mbox{\hspace{0.9cm}} + 3888\,i\,\sqrt{3}\,x^{2}\,t^{2} - 8482662 - 382077\,i
 + 1152\,i\,x\,\sqrt{3}\,t^{3} + 2592\,i\,t^{2}\,x^{3} + 432\,i\,
x\,t^{4} \\
&&\mbox{\hspace{0.9cm}} + 3888\,i\,x^{2}\,t^{2} + 4320\,i\,x\,t^{3} + 6480\,i\,x
^{4}\,t + 936\,i\,\sqrt{3}\,t^{4} + 588384\,i\,x\,\sqrt{3} + 576
\,i\,\sqrt{3}\,t^{3} \\
&&\mbox{\hspace{0.9cm}} + 305532\,i\,x\,t + 12960\,i\,x^{3}\,t + 4320\,i\,x^{2}\,
t^{3} + 6480\,i\,\sqrt{3}\,x^{3} + 357858\,i\,\sqrt{3}\,x^{2} \\
&&\mbox{\hspace{0.9cm}} + 3240\,i\,\sqrt{3}\,x^{4} - 364932\,i\,\sqrt{3}\,x\,t +
134838\,t^{2} + 3456\,i\,\sqrt{3}\,x^{3}\,t + 5184\,i\,\sqrt{3}\,
x^{2}\,t  + 3888\,i\,\sqrt{3}\,x\,t^{2},\\
&&v_{23}= - 160380\,x\,
t - 251100\,x^{2}\,t + 25812\,t^{2}\,x - 1080\,\sqrt{3}\,x^{4}
 \\
&&\mbox{\hspace{0.9cm}} - 24\,t^{4}\,\sqrt{3} + 1572\,\sqrt{3}\,t^{3} - 864\,x^{4
}\,t^{2} - 288\,x^{2}\,t^{4} - 58716\,\sqrt{3}\,x^{3} - 48\,t^{5}
\,\sqrt{3} \\
&&\mbox{\hspace{0.9cm}} - 432\,x^{5}\,\sqrt{3} - 3582\,x\,\sqrt{3} + 609030\,t\,
\sqrt{3} - 12096\,x^{3}\,t + 25488\,x^{2}\,t^{2} - 4032\,t^{3}\,x
 \\
&&\mbox{\hspace{0.9cm}} - 1728\,x^{3}\,t^{2} - 288\,x\,t^{4} + 48078\,t^{2}\,
\sqrt{3} - 145098\,x^{2}\,\sqrt{3} - 186921\,i\,\sqrt{3} - 161352
\,i\,t \\
&&\mbox{\hspace{0.9cm}} - 24624\,i\,x^{3} - 9720\,i\,x^{4} - 3888\,i\,x^{5} - 216
\,i\,t^{4} - 150498\,i\,t^{2} - 864\,x^{6} \\
&&\mbox{\hspace{0.9cm}} - 432\,x^{2}\,t^{2}\,\sqrt{3} + 57276\,x\,\sqrt{3}\,t^{2}
 - 288\,x\,\sqrt{3}\,t^{3} - 864\,\sqrt{3}\,x^{3}\,t - 288\,x^{2}
\,\sqrt{3}\,t^{3} \\
&&\mbox{\hspace{0.9cm}} - 48\,x\,\sqrt{3}\,t^{4} - 93420\,x\,\sqrt{3}\,t - 432\,x
^{4}\,\sqrt{3}\,t - 54756\,x^{2}\,\sqrt{3}\,t + 251100\,i\,x +
50976\,i\,t^{3} \\
&&\mbox{\hspace{0.9cm}} + 432\,i\,t^{5} + 498150\,i\,x^{2} - 648\,i\,\sqrt{3}\,x
^{4} - 2592\,i\,t^{2}\,x^{3} - 277668\,i\,t\,\sqrt{3} - 288\,t^{2
}\,\sqrt{3}\,x^{3} \\
&&\mbox{\hspace{0.9cm}} - 3888\,i\,x^{2}\,t^{2} - 1296\,i\,\sqrt{3}\,x^{3} -
335340\,i\,x\,t - 432\,i\,x\,t^{4} - 32\,t^{6} - 2470023\,\sqrt{3
} \\
&&\mbox{\hspace{0.9cm}} - 5352\,t^{4} + 216\,x^{4} - 929826\,x + 1823418\,t -
108162\,x^{2} - 2592\,x^{5} + 5292\,x^{3} \\
&&\mbox{\hspace{0.9cm}} + 23868\,t^{3} + 1296\,i\,\sqrt{3}\,x\,t^{2} + 1296\,i\,
\sqrt{3}\,x^{2}\,t^{2} + 15552\,i\,x^{2}\,t + 47952\,i\,x\,t^{2}
 + 2592\,i\,x\,t^{3} \\
&&\mbox{\hspace{0.9cm}} + 3888\,i\,x^{4}\,t + 504\,i\,\sqrt{3}\,t^{4} + 18792\,i
\,x\,\sqrt{3} + 7776\,i\,x^{3}\,t + 2592\,i\,x^{2}\,t^{3} + 35262
\,i\,\sqrt{3}\,t^{2} \\
&&\mbox{\hspace{0.9cm}} + 17334\,i\,\sqrt{3}\,x^{2} - 322380\,i\,\sqrt{3}\,x\,t
 - 103626\,t^{2} - 9075582 - 43335\,i.
\end{eqnarray*}
Furthermore, the explicit formulas, $q^{[3]}_{rw}$ and
$q^{[4]}_{rw}$,  of the third and fourth rogue waves
 are obtained from $q^{[2k]}_{smooth}$ with
$k=3$ and $4$, respectively. However only $q^{[3]}_{rw}$  is given
 in appendix IV, and we have to omit $q^{[4]}_{rw}$ because it is
12 pages long. They are plotted in Figure 5(b) and
6(b),respectively. Of course, they are local solutions. Obviously,
the degrees of polynomials in above $q^{[k]}_{rw}(k=2,3,4)$ are
12,24,40. This fact supports following conjecture: In general, the
degree of the polynomial of the denominator for the rational
$k$-order solution in theorem 16 is  $2k(k+1)$. It is a double of
the corresponding degree \cite{Dubard2} of the rogue wave  for the
NLS equation due to the contribution of the square of $\Omega_{n3}$
in theorem 3.

  Figures 4, 5 and 6 are plotted for the second order, third order and fourth order rogue
waves from $q^{[2k]}_{smooth}(k=2,3,4)$.
These figures show that they are localized in both $x$ and $t$ direction and peaks are
diffused dramatically when the value of $b$ is increased. This observation is a
strong  support to  show that the localization of the rogue wave for the MNLS equation is
decreased remarkably by
increasing the value of $b$. This is a unique phenomenon in a  rogue wave solution of the
MNLS because of the
appearance of the two nonlinear terms. However, it can not happen in the rogue wave of the
NLS equation.

Our method can also be applied to get other patterns of the rogue wave by selecting different values of the parameters.
 For example, the fundamental patterns (a simple central highest peak surrounded by several
gradually decreasing peaks in two sides) of the second order, the third order and the fourth order rogue wave are plotted
in Figure 7 with the help of $|q^{[2k]}_{smooth}|^2$. Similarly, a triangle pattern,
a ring-decomposition pattern ( a second order rogue wave surrounded by
seven first rogue wave) and a pentagon pattern of the fourth order rogue wave
 are plotted in Figure 8(a),8(b) and 8(c),
respectively. Note that  Figures 7-8  are density plots of $|q^{[2k]}_{smooth}|^2$
with $a_1=-1,c_1=1,a=1,b=\frac{1}{3}$. All figures in the paper are plotted by using the analytical
 and exact forms  of the solutions of the MNLS. The validity of theses exact solutions is
verified by symbolic computation with  a computer.
\section{Conclusions and Discussions}
In this paper,  the determinant representation of the
n-fold DT for the WKI system is given  in theorem 2.
By choosing paired eigenvalues and paired eigenfunctions
in the form $ \lambda_l
\leftrightarrow \psi_l=\left( \begin{array}{c}
\phi_l\\
\varphi_l\end{array} \right) , \text{and} ~\lambda_{2l}=
-\lambda_{2l-1}^*,\leftrightarrow \psi_{2l}=\left( \begin{array}{c}
\varphi_{2l-1}^*\\
\phi_{2l-1}^*
\end{array} \right)$,
the $2k$-fold DT $T_{2k}$ of the MNLS equation are derived in
theorem 7. The smoothness of the $q^{[2k]}$ is given in theorem 9
for the non-degenerate case and in theorem 14 for the double
degenerate case through the iteration and determinant
representation. Furthermore, the smoothness of the rational
solutions $q^{[2k]}_{smooth}$ is given in theorem 16. By a detailed
analysis of the localization of the rational solutions and the rogue
waves, we get an unusual result: for a given value of $a$, the
increasing value of $b$ can damage gradually the localization of the
rational solution, and a novel two-peak rational solution with
a variable height and a non-vanish boundary in section 4. Note that this
two peak rational soliton just has one eigenvalue $\lambda_0$, which  can not be
a usual double-soliton.  We have verified the validity of theses exact and analytical solutions
by symbolic computation with  a computer.

Finally we would like to stress that there is no doubt of the
novelty of the rational solutions presented in this paper  although
there exists a simple gauge transformation, for example, see
refs.\cite{mihalache,Lenells}, between the MNLS equation and the
DNLS equation. To illustrate this statement clearly, we shall use following form of the
DNLS equation of $\tilde{q}=\tilde{q}(X,T)$:
\begin{equation}\label{dnlsI}
i\tilde{q}_{_T}-\tilde{q}_{_{XX}}+i(\tilde{q}^2\tilde{q}^*)_{_X}=0.
\end{equation}
The explicit form of this gauge transformation \cite{mihalache,Lenells} from the MNLS to the DNLS is
\begin{equation}\label{transformation}
\tilde{q}=-q(x,t)a^{\frac{2}{5}}\exp(i\frac{abx+b^2t}{a^2}),
\end{equation}
with $t=-\dfrac{T}{a^{\frac{2}{5}}}, x=\dfrac{-a^{\frac{6}{5}}X+2bT }{a^{\frac{7}{5}}}$.
The inverse transformation maps the DNLS to the MNLS, which is given by
\begin{equation}\label{inversetransformation}
q=-\tilde{q}(X,T)a^{-\frac{2}{5}}\exp(-i\frac{-a^{\frac{6}{5}}bX+b^2T}{a^{\frac{12}{5}}}),
\end{equation}
and $T=-a^{\frac{2}{5}}t, X=\dfrac{-ax-2bt}{a^{\frac{4}{5}}}$.
Accordingly, there exists following  transformation
\begin{eqnarray}
\left( \begin{matrix}
|q|_x\\ \\
|q|_t
\end{matrix} \right)
=a^{-\frac{2}{5}}\left( \begin{matrix}
-a^{\frac{1}{5}} & 0 \\ \\
-2b a^{-\frac{4}{5}}&  -a^{\frac{2}{5}}
\end{matrix} \right)
\left( \begin{matrix}
|\tilde{q}|_{_X}\\ \\
|\tilde{q}|_{_T}
\end{matrix} \right)
\end{eqnarray}
between $(|q|_x,|q|_t)$ and  $(|\tilde{q}|_{_X},|\tilde{q}|_{_T})$, which shows that
invertible transformations in  eq.(\ref{transformation}) and eq.(\ref{inversetransformation})
preserve the numbers of extreme value points for a given $t$ (or $T$) and stationary points
in profiles of $|q|$ and $\tilde{q}$.  In other words, this simple gauge transformation can not change
the numbers of peaks or valleys in $|q|$ and $|\tilde{q}|$.  Therefore,
the first-order  RW  and  the first-order   rational  soliton of the DNLS \cite{xuhe,blguo,ljguo} can not be mapped to the  two-peak  rational solution of the MNLS by this transformation. By the gauge transformation eq.(\ref{inversetransformation}),
two solutions of the MNLS  with $a=1$ and $b=3$ are generated from corresponding solutions of the DNLS
in eq. $(56)$ with  $\alpha_1=\beta_1=\dfrac{1}{2}$ and in  eq. $(46)$ with  $\beta_1=\dfrac{1}{2}$ of Ref. \cite{xuhe},
which are plotted in  Figure 9.   Similarly,  two peak rational solution of the DNLS can be  generated from
 a corresponding solution in eq. (\ref{twopeak}) of the MNLS, which is plotted in Figure 10. The two peak rational solution of the
DNLS has only one eigenvalue and a non-vanishing boundary, which has never been reported in literatures.

{\bf Acknowledgments} {\noindent \small This work is supported by
the NSF of China under Grant No.10971109 and 11271210. Jingsong He
is also supported by the Natural Science Foundation of Ningbo under
Grant No.2011A610179, and K.C. Wong Magna Fund in Ningbo University.
J. He thanks sincerely Prof. A.S. Fokas for arranging the visit to
Cambridge University in 2012-2013 and for many useful discussions.}

\section*{Appendix I:The matrix form of the one-fold Darboux matrix}
\setcounter{equation}{0}
\renewcommand{\theequation}{A.\arabic{equation}}
Consider the universality of DT, suppose that a trial Darboux matrix $T$ in eq.(\ref{bh3}) is of
\begin{equation}\label{tt1}
T=T(\lambda)=\left( \begin{array}{cc}
a_{1}&b_{1} \\
c_{1} &d_{1}\\
\end{array} \right)\lambda+\left( \begin{array}{cc}
a_{0}&b_{0}\\
c_{0} &d_{0}\\
\end{array} \right).
\end{equation}
Here $a_{0},  b_{0},  c_{0},   d_{0},   a_{1},   b_{1},   c_{1},   d_{1}$ are undetermined functions
of $x$ and $t$. From \begin{equation}\label{tt2} T_{x}+T~U=U^{[1]}~T,
\end{equation}
comparing the coefficients of $\lambda^{j}, j=3, 2, 1, 0$, it yields
\begin{eqnarray}\label{xx1}
&&\lambda^{3}: b_{1}=0,\ c_{1}=0,\nonumber\\
&&\lambda^{2}: q~a_{1}+2~i~b_{0}-q^{[1]}d_{1}=0,\ -r^{[1]}~a_{1}+r~d_{1}-2~i~c_{0}=0,\nonumber\\
&&\lambda^{1}: {a_{1}}_{x}+ar~b_{0}-aq^{[1]}c_{0}=0,\ a_0aq+\dfrac{1}{2}ia_1\sqrt{2b}q-\dfrac{1}{2}i\sqrt{2b}q^{[1]}d_1-2b_0\sqrt{2b}-aq^{[1]}d_0=0,\nonumber\\
&&~~~~~~~{d_{1}}_{x}+a q c_{0}-a r^{[1]}b_{0}=0,\
-ar^{[1]}a_0+2c_0\sqrt{2b}+d_0ar+\dfrac{1}{2}id_1\sqrt{2b}r-\dfrac{1}{2}i
\sqrt{2b}r^{[1]}a_1=0,\nonumber\\
&&\lambda^{0}:
{a_0}_x+ib_0\sqrt{\dfrac{b}{2}}r-i\sqrt{\dfrac{b}{2}}q^{[1]}c_0=0,\
{b_0}_x-i\sqrt{\dfrac{b}{2}}q^{[1]}d_0+i\sqrt{\dfrac{b}{2}}qa_0=0,\nonumber\\
&&~~~~~~~i\sqrt{\dfrac{b}{2}}d_0r+{c_0}_x-i\sqrt{\dfrac{b}{2}}r^{[1]}a_0=0,\
{d_0}_x+i\sqrt{\dfrac{b}{2}}c_0q-i\sqrt{\dfrac{b}{2}}r^{[1]}b_0=0.
\end{eqnarray}
Similarly, from  \begin{equation}\label{tt3} T_{t}+T~V=V^{[1]}~T,
\end{equation} comparing the coefficients of $\lambda^{j} ,j =4,3, 2, 1, 0$,it implies
\begin{eqnarray}\label{ttt1}
&&\lambda^{4}: q~a_{1}+2~i~b_{0}-q^{[1]}d_{1}=0,\ -r^{[1]}~a_{1}+r~d_{1}-2~i~c_{0}=0,\nonumber\\
&&\lambda^{3}:
-ia_1a^2rq+ia^2r^{[1]}q^{[1]}a_1-2a^2q^{[1]}c_0+2b_0a^2r=0,\  2c_0a^2q+id_1a^2rq-2a^2r^{[1]}b_0-ia^2r^{[1]}q^{[1]}d_1=0, \nonumber\\
&&~~~~~~~3ia_1\sqrt{2b}a q+2a_0a^2q-8b_0a\sqrt{2b}-3i\sqrt{2b}aq^{[1]}d_1-2a^2q^{[1]}d_0=0, \nonumber\\
&&~~~~~~~-2a^2r^{[1]}a_0-3i\sqrt{2b}ar^{[1]}a_1+8c_0a\sqrt{2b}+3id_1\sqrt{2b}ar+2d_0a^2r=0, \nonumber\\
&&\lambda^{2}:-\sqrt{2b}ar^{[1]}q^{[1]}a_1-3i\sqrt{2b}aq^{[1]}c_0+3ib_0\sqrt{2b}ar-ia_0a^2rq+ia^2r^{[1]}q^{[1]}a_0+a_1\sqrt{2b}a
r q=0, \nonumber\\
&&~~~~~~~3ic_0\sqrt{2b}aq+id_0 a^2 r q+\sqrt{2b}ar^{[1]}q^{[1]}d_1-3i\sqrt{2b}ar^{[1]}b_0-d_1\sqrt{2b}a r q-ia^2r^{[1]}q^{[1]}d_0=0,\nonumber\\
&&~~~~~~~3ia_0\sqrt{2b}a q+a_1a^2 r q^2+i b_0 a^2 r q-8ib_0 b-3i\sqrt{2b}a q^{[1]} d_0-2a_1 b q+ia_1 a q_x+i a^2 r^{[1]} q^{[1]}b_0\nonumber\\
&&~~~~~~~-i a{q^{[1]}}_x d_1+2 b q^{[1]}d_1-a^2 r^{[1]} {q^{[1]}}^2 d_1=0, \nonumber\\
&&~~~~~~~2 b r^{[1]} a_1-i c_0 a^2 r q+8i c_0 b-2 d_1 b r-3i\sqrt{2b}a r^{[1]} a_0-i d_1 a r_x+d_1 a^2 q r^2-a^2  q^{[1]}{r^{[1]}}^2 a_1\nonumber\\
&&~~~~~~~+ia{r^{[1]}}_x a_1-ia^2r^{[1]}q^{[1]}c_0+3id_0\sqrt{2b}ar=0,\nonumber\\
&&\lambda^{1}: -\sqrt{2b}ar^{[1]}q^{[1]}a_0+a_0\sqrt{2b}a r
q+{a_1}_t+\dfrac{1}{2}i a_1 b q r-ia {q^{[1]}}_x c_0-2 b_0 b r+2 b
q^{[1]}c_0-ib_0 a r_x\nonumber\\
&&~~~~~~-a^2r^{[1]}
{q^{[1]}}^2 c_0-\dfrac{1}{2}ibq^{[1]}r^{[1]}a_1+b_0 a^2 q r^2=0,\nonumber\\
&&~~~~~~-2a_0 b q+2
bq^{[1]}d_0-i\sqrt{\dfrac{b}{2}}ar^{[1]}{q^{[1]}}^2d_1-a_1\sqrt{\dfrac{b}{2}}q_x+a_0
a^2 r q^2-a^2 r^{[1]}{q^{[1]}}^2 d_0+ia_0 a
q_x\nonumber\\
&&~~~~~~+i\sqrt{\dfrac{b}{2}}a_1 a r q^2-ia {q^{[1]}}_x
d_0-b_0\sqrt{2b}a r q-\sqrt{2b}a r^{[1]}q^{[1]}b_0+\sqrt{\dfrac{b}{2}}{q^{[1]}}_xd_1=0,\nonumber\\
&&~~~~~~-i d_0 a r_x+d_0 a^2 q r^2+ia {r^{[1]}}_x
a_0+\sqrt{2b}ar^{[1]}
q^{[1]}c_0-i\sqrt{\dfrac{b}{2}}aq^{[1]}{r^{[1]}}^2a_1+i\sqrt{\dfrac{b}{2}}d_1a
qr^2\nonumber\\
&&~~~~~~+\sqrt{\dfrac{b}{2}}d_1r_x-2d_0 b r-a^2
q^{[1]}{r^{[1]}}^2a_0-\sqrt{\dfrac{b}{2}}{r^{[1]}}_xa_1+c_0\sqrt{2b}a r q+2b r^{[1]}a_0=0,\nonumber\\
&&~~~~~~ia{r^{[1]}}_x
b_0-a^2q^{[1]}{r^{[1]}}^2b_0+\sqrt{2b}ar^{[1]}q^{[1]}d_0-d_0\sqrt{2b}a
r q-2c_0 b
q+{d_1}_t+\dfrac{1}{2}ibq^{[1]}r^{[1]}d_1\nonumber\\
&&~~~~~~-\dfrac{1}{2}ibqrd_1+2
br^{[1]}b_0+c_0a^2rq^2+ic_0aq_x=0,\nonumber\\
&&\lambda^{0}:
i\sqrt{\dfrac{b}{2}}b_0aqr^2-i\sqrt{\dfrac{b}{2}}ar^{[1]}{q^{[1]}}^2c_0+b_0\sqrt{\dfrac{b}{2}}r_x+\sqrt{\dfrac{b}{2}}{q^{[1]}}_xc_0
+{a_0}_t+\dfrac{1}{2}ia_0 b q  r-\dfrac{1}{2}ibq^{[1]}r^{[1]}a_0=0,\nonumber\\
&&~~~~~~-\sqrt{\dfrac{b}{2}}a_0q_x+ia0\sqrt{\dfrac{b}{2}}a r
q^2-\dfrac{1}{2}ib_0bqr-i\sqrt{\dfrac{b}{2}}ar^{[1]}{q^{[1]}}^2d_0-\dfrac{1}{2}ibq^{[1]}r^{[1]}b_0
+\sqrt{\dfrac{b}{2}}{q^{[1]}}_xd_0+{b_0}_t=0,\nonumber\\
&&~~~~~~{c_0}_t-\sqrt{\dfrac{b}{2}}{r^{[1]}}_xa_0+id_0\sqrt{\dfrac{b}{2}}
aqr^2+\dfrac{1}{2}ibq^{[1]}r^{[1]}c_0-i\sqrt{\dfrac{b}{2}}aq^{[1]}{r^{[1]}}^2a0+d0i\sqrt{\dfrac{b}{2}}r_x
+\dfrac{1}{2}ic_0bqr=0,\nonumber\\
&&~~~~~~{d_0}_t-c0\sqrt{\dfrac{b}{2}}q_x-\sqrt{\dfrac{b}{2}}
{r^{[1]}}_xb_0-\dfrac{1}{2}id_0bqr+i\sqrt{\dfrac{b}{2}}c_0arq^2-i\sqrt{\dfrac{b}{2}}aq^{[1]}{r^{[1]}}^2b_0
+\dfrac{1}{2}ibq^{[1]}r^{[1]}d_0=0.
\end{eqnarray}

 We shall construct a basic(or one-fold ) Darboux transformation matrix $T$ by solving eqs.(\ref{xx1}) and
 eqs.(\ref{ttt1}). Let $a_{1}d_{1}a\neq0$,
 substituting $q^{[1]}=q\dfrac{a_{1}}{d_{1}}+\dfrac{2~i~b_{0}}{d_{1}}$
 from eq.(\ref{xx1}) into
 $a_0aq+\dfrac{1}{2}ia_1\sqrt{2b}q-\dfrac{1}{2}i\sqrt{2b}q^{[1]}d_1-2b_0\sqrt{2b}-aq^{[1]}d_0=0$
 of eq.(\ref{xx1}), then
 \begin{equation}\label{DTelement1}
  -a_0aqd_1+b_0\sqrt{2b}d_1+2iad_0b_0+ad_0a_1q=0;
  \end{equation}
 substituting $r^{[1]}=r\dfrac{d_{1}}{a_{1}}-\dfrac{2~i~c_{0}}{a_{1}}$ from eq.(\ref{ttt1})  into
 $-ar^{[1]}a_0+2c_0\sqrt{2b}+d_0ar+\dfrac{1}{2}id_1\sqrt{2b}r-\dfrac{1}{2}i
\sqrt{2b}r^{[1]}a_1=0$ of  eq.(\ref{xx1}), then
\begin{equation}\label{DTelement2}
-a_0ad_1r+2ia_0ac_0+\sqrt{2b}c_0a_1+d_0ara_1=0.
 \end{equation}
 Assume $2iad_0+\sqrt{2b}d_1\neq0$ in eq.(\ref{DTelement1}) and $2iaa_0+\sqrt{2b}a_1\neq0$ in eq.(\ref{DTelement2}),
we have  $b_0=-\dfrac{aq(-a_0d_1+d_0a_1)}{2iad_0+\sqrt{2b}d_1}$ and
$c_0=-\dfrac{ar(-a_0d_1+d_0a_1)}{2iaa_0+\sqrt{2b}a_1}$. Taking these values of $b_0$ and $c_0$ into  other eqations
in eq.(\ref{xx1}) and eq.(\ref{ttt1}),  we find that  $a_1, d_1, a_0,d_0$ are only
function of $t$, which gives trivial solutions of the MNLS eqution by DT.
This means we can choose
$2iad_0+\sqrt{2b}d_1=0$  and $2iaa_0+\sqrt{2b}a_1\neq0$ in eq.(\ref{DTelement1}) and eq.(\ref{DTelement2}) without loss any
generality, then  $a_0=i\sqrt{\dfrac{b}{2}}\dfrac{a_1}{a}$ and
$d_0=i\sqrt{\dfrac{b}{2}}\dfrac{d_1}{a}$.  Furthermore, taking $q^{[1]}$,$r^{[1]}$,
 $a_0$ and $d_0$ into eq.(\ref{xx1}) and eq.(\ref{ttt1}), then  ${b_0}_x=-\dfrac{ibb_0}{a}$,
 ${c_0}_x=\dfrac{ibc_0}{a}$, ${b_0}_t=-\dfrac{ib^2b_0}{a^2}$ and
${c_0}_t=\dfrac{ib^2c_0}{a^2}$.
\section*{Appendix II:Proof of Theorem 1}
 Note that $(a_{1}d_{1})_{x}=0$ is derived from the eq.(\ref{xx1}) of appendix I, and
then we set $a_{1}=\dfrac{1}{d_{1}}$ in the following calculation.
By transformation eq.(\ref{TT}) and eq.(\ref{xx1}), new solutions
 \begin{eqnarray} \label{TT1}
q^{[1]}=\dfrac{a_{1}}{d_{1}}q+2~i~\dfrac{b_{0}}{d_{1}},r^{[1]}=\dfrac{d_{1}}{a_{1}}
r-2~i~\dfrac{c_{0}}{a_{1}}
\end{eqnarray}
are generated by $T_1$ from a seed solution $q$. We need to parameterize $T_1$ by the
eigenfunctions associated with $\lambda_1$. This purpose can be realized through a system of
equations defined by its kernel, i.e.,  $T_{1}(\lambda)|_{\lambda=\lambda_1}\psi_{1}=0$.
Solving this  system of algebraic equations on $(a_1,d_1,b_0,c_0)$, eq.(\ref{DT1aibi}) is
obtained. Next, substituting $(a_1,d_1,b_0,c_0)$ into eq.(\ref{TT1}), new solutions
$q^{[1]}$ and $r^{[1]}$ are given as eq.(\ref{sTT}). Further, by using explicit
matrix representation eq.(\ref{DT1matrix}) of $T_1$, the new eigenfunction $\psi^{[1]}_j
=T_1(\lambda;\lambda_1)|_{\lambda=\lambda_j} \psi_j$ for $j\geqq 2$ becomes
\begin{eqnarray*}
\psi^{[1]}_j=\left.\left(\mbox{\hspace{-0.2cm}}
\begin{array}{cc}
-\frac{\varphi_{1}}{\phi_{1}}\exp(-i(\dfrac{b}{a}x+\dfrac{b^2}{a^2}t))(\lambda+i\dfrac{\sqrt{2b}}{2a})& (\lambda_1+i\dfrac{\sqrt{2b}}{2a})\exp(-i(\dfrac{b}{a}x+\dfrac{b^2}{a^2}t))\\
(\lambda_1+i\dfrac{\sqrt{2b}}{2a})\exp(i(\dfrac{b}{a}x+\dfrac{b^2}{a^2}t)
&-\frac{\phi_{1}}{\varphi_{1}}\exp(i(\dfrac{b}{a}x+\dfrac{b^2}{a^2}t))(\lambda+i\dfrac{\sqrt{2b}}{2a})
\end{array}  \right)\right|_{\lambda=\lambda_j} \left( \begin{array} {c}
\phi_j\\
\varphi_j
\end{array} \right)
\end{eqnarray*}
\begin{eqnarray} =\left(
\begin{array}{c}
\dfrac{1}{\phi_1}\left|\begin{array}{cc}
-(\lambda_j+i\dfrac{\sqrt{2b}}{2a})\phi_j & \varphi_j\\
-(\lambda_1+i\dfrac{\sqrt{2b}}{2a})\phi_1 & \varphi_1
 \end{array}\right| \exp(-i(\dfrac{b}{a}x+\dfrac{b^2}{a^2}t))\\ \\
\dfrac{1}{\varphi_1} \left|\begin{array}{cc}
-(\lambda_j+i\dfrac{\sqrt{2b}}{2a})\varphi_j & \phi_j\\
-(\lambda_1+i\dfrac{\sqrt{2b}}{2a})\varphi_1 & \phi_1
 \end{array}\right|\exp(i(\dfrac{b}{a}x+\dfrac{b^2}{a^2}t))
\end{array}
\right).
\end{eqnarray}
Last, a tedious calculation verifies that $T_1$ in eq.(\ref{DT1matrix})
and new solutions indeed satisfy eq.(\ref{xx1}) and eq.(\ref{ttt1}) in appendix I.
In the process of verification, it is crucial to use the fact that
$\psi_1$ satisfies eq.(\ref{sys11}) and eq.(\ref{sys22}) of
the Lax pair associated with a seed solution $q$ and eigenvalue $\lambda_1$.
So WKI spectral  problem  is  covariant under transformation $T_1$
in eq.(\ref{DT1matrix}) and eq.(\ref{sTT}). Therefore  $T_1$ is the DT of
eq.(\ref{sy1}) and eq.(\ref{sy2}). $\square$
\section*{appendix III:The rational 2-order solution}
\begin{eqnarray}
&&{}\mbox{\hspace{-1cm}}R_1= ( - 54\,b + 54\,a + 36\,a^{2})\,x^{2} - 12\,(b -
a + 2\,a^{2})\,( - b + a)\,x^{4} + 8\,( - b + a)^{3}\,x^{6}\nonumber\\
&&{}\mbox{\hspace{0.9cm}} + ( - 216\,a^{3} - 216\,b + 216\,a + 216\,a\,b - 72\,a^{2
})\,t\,x \nonumber\\
&&{}\mbox{\hspace{0.9cm}} + 48\,(2\,a\,b - 2\,b + 5\,a^{3} + 2\,a - 6\,a^{2})\,( -
b + a)\,t\,x^{3} - 96\,(a - 1)\,( - b + a)^{3}\,t\,x^{5} \nonumber\\
&&{}\mbox{\hspace{0.9cm}} + (216\,a - 216\,b - 450\,a^{2}\,b + 108\,a^{2} + 18\,a^{
3} - 360\,a\,b + 396\,b^{2} + 324\,a^{4})\,t^{2} \nonumber\\
&&{}\mbox{\hspace{0.9cm}} - 72\,( - b + a)\,(12\,a^{4} - 23\,a^{3} + 18\,a^{2} + 3
\,a^{2}\,b - 12\,a\,b - 4\,a + 4\,b + 2\,b^{2})\,t^{2}\,x^{2} \nonumber\\
&&{}\mbox{\hspace{0.9cm}} + 24\,( - 2\,b + 19\,a^{2} - 38\,a + 20)\,( - b + a)^{3}
\,t^{2}\,x^{4} + 48\,( - b + a) \nonumber\\
&&{}\mbox{\hspace{0.9cm}}(27\,a^{5} - 72\,a^{4} + 90\,a^{3} - 52\,a^{2} - 42\,a^{2}\,b + 8
\,a + 48\,a\,b + 12\,b^{2}\,a - 8\,b - 12\,b^{2})\,t^{3}\,x \nonumber\\
&&{}\mbox{\hspace{0.9cm}} - 64\,(a - 1)\,(17\,a^{2} - 34\,a + 20 - 6\,b)\,( - b + a
)^{3}\,t^{3}\,x^{3} - 12\,( - b + a)(54\,a^{6} - 189\,a^{5} \nonumber\\
&&{}\mbox{\hspace{0.9cm}} - 27\,a^{4}\,b + 356\,a^{4} - 136\,a^{3}\,b - 364\,a^{3}
 + 372\,a^{2}\,b + 144\,a^{2} + 68\,a^{2}\,b^{2} - 204\,b^{2}\,a
 \nonumber\\
&&{}\mbox{\hspace{0.9cm}} - 160\,a\,b - 16\,a + 48\,b^{2} + 16\,b + 36\,b^{3})t^{4}
\nonumber\\
&&{}\mbox{\hspace{0.9cm}} + 24\,( - 2\,b + 3\,a^{2} - 6\,a + 4)\,( - 2\,b + 19\,a^{
2} - 38\,a + 20)\,( - b + a)^{3}\,t^{4}\,x^{2} \nonumber\\
&&{}\mbox{\hspace{0.9cm}} - 96\,(a - 1)\,( - 2\,b + 3\,a^{2} - 6\,a + 4)^{2}\,( - b
 + a)^{3}\,t^{5}\,x \nonumber\\
&&{}\mbox{\hspace{0.9cm}} + 8\,( - b + a)^{3}\,( - 2\,b + 3\,a^{2} - 6\,a + 4)^{3}
\,t^{6} + 9.
\end{eqnarray}
\begin{eqnarray}
&&{}\mbox{\hspace{-2cm}}I_1=  - 54\,a\,x - 24\,(a^{2} + a - b)\,a\,x^{3}
 - 24\,a\,( - b + a)^{2}\,x^{5} + 18\,a\,(17\,a - 6)\,t \nonumber\\
&&{}\mbox{\hspace{0.9cm}} + 72\,( - 2\,a - a^{2} + 3\,a^{3} + 2\,b - a\,b)\,a\,t
\,x^{2} + 24\,a\,(11\,a - 10)\,( - b + a)^{2}\,t\,x^{4} \nonumber\\
&&{}\mbox{\hspace{0.9cm}} - 72\,(9\,a^{4} - 15\,a^{3} - 2\,a^{2} + 4\,a - 4\,b
 + 8\,a\,b + 3\,a^{2}\,b - 2\,b^{2})\,a\,t^{2}\,x \nonumber\\
&&{}\mbox{\hspace{0.9cm}} - 48\,(23\,a^{2} - 42\,a - 2\,b + 20)\,a\,( - b + a)^{
2}\,t^{2}\,x^{3} + 24(27\,a^{5} - 81\,a^{4} + 60\,a^{3} \nonumber\\
&&{}\mbox{\hspace{0.9cm}} + 16\,a^{2} - 8\,a + 27\,a^{3}\,b - 30\,a^{2}\,b - 36\,a
\,b + 8\,b + 12\,b^{2} + 6\,b^{2}\,a)a\,t^{3} \nonumber\\
&&{}\mbox{\hspace{0.9cm}} + 48\,(45\,a^{3} - 124\,a^{2} + 120\,a - 14\,a\,b - 40
 + 12\,b)\,a\,( - b + a)^{2}\,t^{3}\,x^{2} \nonumber\\
&&{}\mbox{\hspace{0.9cm}} - 24\,( - 2\,b + 3\,a^{2} - 6\,a + 4)\,( - 2\,b + 27\,
a^{2} - 46\,a + 20)\,a\,( - b + a)^{2}\,t^{4}\,x \nonumber\\
&&{}\mbox{\hspace{0.9cm}} + 24\,a\,(3\,a - 2)\,( - 2\,b + 3\,a^{2} - 6\,a + 4)^{
2}\,( - b + a)^{2}\,t^{5}.
\end{eqnarray}
\begin{eqnarray}
&&R_2=(144\,b - 144\,a^{2} - 144\,a)\,x^{2} - 48\,( -
b + a)^{2}\,x^{4} + 288\,(3\,a - 2)\,(a^{2} + a - b)\,t\,x \nonumber\\
&&{}\mbox{\hspace{1cm}} + 192\,(a - 2)\,( - b + a)^{2}\,t\,x^{3}+ ( - 864\,b^{2} + 288\,a^{2} + 1296\,a^{2}\,b - 1296\,a
^{4} - 576\,a + 432\,a^{3} + 576\,b)\,t^{2}\nonumber\\
&&{}\mbox{\hspace{1cm}} + 288\,(a^{2} + 2\,a + 2\,b - 4)\,( - b + a)^{2}\,t^{2}\,
x^{2} - 192\,(9\,a^{3} - 16\,a^{2} + 10\,a\,b + 8 - 12\,b)\,(
 - b + a)^{2}\,x\,t^{3} \nonumber\\
&&{}\mbox{\hspace{1cm}} + 48\,(10\,b + 9\,a^{2} - 10\,a - 4)\,( - 2\,b + 3\,a^{2}
 - 6\,a + 4)\,( - b + a)^{2}\,t^{4} + 36.
\end{eqnarray}
\begin{eqnarray}
&&I_2= 96\,a\,( - b + a)\,x^{3} + 144\,a\,x - 24
(15\,b - 27\,a + 18\,a^{2})\,t - 288\,(b - 3\,a + 2\,a^{2})\,( - b + a)\,t\,x^{2}\nonumber\\
&&{}\mbox{\hspace{0.9cm}}  -
96( - b + a)^{3}\,t\,x^{4}+ 288(a - 2)\,(2\,b + 3\,a^{2} - 4\,a)\,( - b + a)\,
t^{2}\,x + 768(a - 1)\,( - b + a)^{3}\,t^{2}\,x^{3}\nonumber\\
&&{}\mbox{\hspace{0.9cm}}  + 96(9\,a^{3} - 38\,a^{2} + 9\,a^{2}\,b + 20\,a + 16
\,a\,b - 2\,b^{2} - 12\,b)\,( - b + a)\,t^{3}\nonumber\\
&&{}\mbox{\hspace{0.9cm}}- 192(11\,a^{2} - 22\,a + 12 - 2\,b)\,( - b + a)^{3}
\,t^{3}\,x^{2}+ 768(a - 1)\,( - 2\,b + 3\,a^{2} - 6\,a + 4)\,( - b
 + a)^{3}\,t^{4}\,x  \nonumber\\
&&{}\mbox{\hspace{0.9cm}}- 96( - 2\,b + 3\,a^{2} - 6\,a + 4)^{2}\,( - b + a)
^{3}\,t^{5}.
\end{eqnarray}
\section*{appendix IV:The third order and fourth order rogue waves}
Set $K_0=1,S_0=L_0=S_1=L_1=0,S_2=L_2,L_2=9000, a_1=-1,
c_1=1,a=1,b=\dfrac{1}{3}$ in $q^{[6]}_{smooth}$ of theorem 16,  the third order rogue wave is given by
\begin{eqnarray}
q^{[3]}_{rw}=\exp^{(-\,i\,\dfrac{3\,x + t}{3})}\dfrac{v_{31}v_{32}}{v_{33}^2}
\end{eqnarray}
with
\begin{eqnarray*}
&&v_{31}=-{v_{33}}^{\ast},\\
&&v_{32}=651440377800\,x\,t -
860934420000000\,x^{2}\,t \\
&&\mbox{\hspace{0.9cm}} + 562477154400000\,t^{2}\,x + 140867294400\,x^{4}\,t^{2}
 + 324911217600\,x^{2}\,t^{4} \\
&&\mbox{\hspace{0.9cm}} - 130096756800\,x^{3}\,t - 905575464000\,x^{2}\,t^{2} +
713406398400\,t^{3}\,x \\
&&\mbox{\hspace{0.9cm}} - 204073344000000\,x^{3}\,t^{2} - 47617113600000\,x\,t^{4
} + 30611001600000\,x^{4}\,t \\
&&\mbox{\hspace{0.9cm}} - 102036672000000\,x^{2}\,t^{3} + 25339106880\,x^{6} +
110592\,i\,x\,t^{10} + 1658880\,i\,x^{3}\,t^{8} \\
&&\mbox{\hspace{0.9cm}} + 9953280\,i\,x^{5}\,t^{6} + 29859840\,i\,x^{7}\,t^{4} +
44789760\,i\,x^{9}\,t^{2} + 1067558680800000\,i\,t\,x \\
&&\mbox{\hspace{0.9cm}} + 683645702400000\,i\,x\,t^{3} + 30611001600000\,i\,x^{3}
\,t + 314613072000\,i\,x^{3}\,t^{2} \\
&&\mbox{\hspace{0.9cm}} + 399856208400\,i\,t\,x^{2} + 606480469200\,i\,t^{2}\,x
 + 44789760\,i\,t\,x^{10} + 2764800\,i\,t^{9}\,x^{2} \\
&&\mbox{\hspace{0.9cm}} + 16588800\,i\,t^{7}\,x^{4} + 3627970560000\,i\,x\,t^{5}
 + 373248000\,i\,t^{7}\,x^{2} + 32878483200\,i\,x^{5}\,t^{2} \\
&&\mbox{\hspace{0.9cm}} + 10825125120\,i\,x\,t^{6} + 7936185600\,i\,t^{3}\,x^{4}
 + 150888303360\,i\,t^{5}\,x^{2} + 49766400\,i\,t^{5}\,x^{6} \\
&&\mbox{\hspace{0.9cm}} + 74649600\,i\,t^{3}\,x^{8} + 191545508160\,t^{6} -
847046095200\,t^{4} - 88006629600\,x^{4} \\
&&\mbox{\hspace{0.9cm}} - 4761711360000\,t^{5} + 645700815000000\,x +
1627166053800000\,t \\
&&\mbox{\hspace{0.9cm}} - 2703334078800000\,i\,t^{2} - 114791256000000\,i\,x^{4}
 - 2456745454800\,i\,t^{3} \\
&&\mbox{\hspace{0.9cm}} - 695984581440\,i\,t^{5} - 302330880\,i\,x^{9} -
4308215040\,i\,x^{7} - 59691453120\,i\,x^{5} \\
&&\mbox{\hspace{0.9cm}} + 1394713767933176175 - 387420489000000\,i -
619872892886583900\,x^{2} \\
&&\mbox{\hspace{0.9cm}} + 110199605760000\,x^{5} + 218103386400000\,x^{3} +
2529234007200000\,t^{3} \\
&&\mbox{\hspace{0.9cm}} + 391910400\,t^{7}\,x - 5553930240\,t^{6}\,x^{2} +
18662400\,t^{8}\,x^{2} - 3661562880\,t^{5}\,x^{3} \\
&&\mbox{\hspace{0.9cm}} + 216483840\,t^{6}\,x^{4} - 16426644480\,x^{5}\,t^{3} +
4870886400\,x^{4}\,t^{4} + 335923200\,t\,x^{9} \\
&&\mbox{\hspace{0.9cm}} - 5971968\,t^{2}\,x^{10} + 447897600\,t^{3}\,x^{7} -
4976640\,t^{4}\,x^{8} + 223948800\,t^{5}\,x^{5} \\
&&\mbox{\hspace{0.9cm}} - 2211840\,t^{6}\,x^{6} + 49766400\,t^{7}\,x^{3} - 552960
\,t^{8}\,x^{4} + 4147200\,t^{9}\,x - 73728\,t^{10}\,x^{2} \\
&&\mbox{\hspace{0.9cm}} + 243792063360\,t^{5}\,x - 107818750080\,x^{5}\,t -
291371385600\,x^{3}\,t^{3} \\
&&\mbox{\hspace{0.9cm}} - 12899450880\,x^{6}\,t^{2} + 724101120\,x^{6}\,t^{4} -
5744286720\,x^{7}\,t + 839808000\,x^{8}\,t^{2} \\
&&\mbox{\hspace{0.9cm}} - 251942400000\,t^{6}\,x + 1360488960000\,t^{5}\,x^{2} +
2267481600000\,x^{4}\,t^{3} \\
&&\mbox{\hspace{0.9cm}} + 755827200000\,x^{3}\,t^{4} + 4081466880000\,x^{5}\,t^{2
} - 6802444800000\,x^{6}\,t \\
&&\mbox{\hspace{0.9cm}} + 10617972480\,i\,t^{7} + 671528847600000\,i\,x^{2} +
114877440\,i\,t^{9} + 184320\,i\,t^{11} \\
&&\mbox{\hspace{0.9cm}} + 28697814000\,i\,x^{3} + 293355432000000\,i\,t^{4} +
929809203732704700\,i\,x \\
&&\mbox{\hspace{0.9cm}} + 1511654400000\,i\,t^{6} + 32651735040000\,i\,x^{6} +
1549682126751993300\,i\,t \\
&&\mbox{\hspace{0.9cm}} - 206625142301186700\,t^{2} - 50388480000\,t^{7} +
938718720\,t^{8} - 801792\,t^{10} \\
&&\mbox{\hspace{0.9cm}} - 2985984\,x^{12} - 4096\,t^{12} + 141087744\,x^{10} -
1360488960000\,x^{7} + 1058158080\,x^{8} \\
&&\mbox{\hspace{0.9cm}} + 26873856\,i\,x^{11} - 321415516800000\,i\,t^{2}\,x^{2}
 - 375551640000\,i\,x\,t^{4} \\
&&\mbox{\hspace{0.9cm}} - 48467419200\,i\,t\,x^{4} - 269263440000\,i\,t^{3}\,x^{2
} - 40814668800000\,i\,t^{2}\,x^{4} \\
&&\mbox{\hspace{0.9cm}} - 27209779200000\,i\,t^{4}\,x^{2} - 33592320\,i\,t^{8}\,x
 - 3986288640\,i\,t^{3}\,x^{6} \\
&&\mbox{\hspace{0.9cm}} - 1713208320\,i\,x^{8}\,t - 1052559360\,i\,x^{4}\,t^{5}
 - 671846400\,i\,x^{3}\,t^{6} \\
&&\mbox{\hspace{0.9cm}} - 4434186240\,i\,x^{7}\,t^{2} - 3157678080\,i\,x^{5}\,t^{
4} - 39932870400\,i\,t^{4}\,x^{3} \\
&&\mbox{\hspace{0.9cm}} - 18139852800000\,i\,x^{3}\,t^{3} - 2645395200\,i\,t\,x^{
6},\\
&&v_{33}=77484097800\,x\,t  + 792059666400000\,x^{2}\,t + 11479125600000\,t^{2}\,x -
74543457600\,x^{4}\,t^{2} \\
&&\mbox{\hspace{0.9cm}} - 416555265600\,x^{2}\,t^{4} - 191318760000\,x^{3}\,t +
696400286400\,x^{2}\,t^{2} \\
&&\mbox{\hspace{0.9cm}} - 1089241473600\,t^{3}\,x - 40814668800000\,x^{3}\,t^{2}
 - 20407334400000\,x\,t^{4} \\
&&\mbox{\hspace{0.9cm}} + 275499014400000\,x^{4}\,t + 61222003200000\,x^{2}\,t^{3
} - 9580109760\,x^{6} \\
&&\mbox{\hspace{0.9cm}} + 459165024000000\,i\,x\,t^{3} + 459165024000000\,i\,x^{3
}\,t + 120530818800\,i\,t\,x^{2} \\
&&\mbox{\hspace{0.9cm}} + 140300424000\,i\,t\,x^{4} + 389439964800\,i\,t^{3}\,x^{
2} + 327155079600\,i\,t^{2}\,x \\
&&\mbox{\hspace{0.9cm}} + 26873856\,i\,t\,x^{10} + 1658880\,i\,t^{9}\,x^{2} +
9953280\,i\,t^{7}\,x^{4} + 5441955840000\,i\,x\,t^{5} \\
&&\mbox{\hspace{0.9cm}} + 403107840\,i\,t^{7}\,x^{2} + 67184640\,i\,x^{4}\,t^{5}
 + 253808640\,i\,x^{3}\,t^{6} + 2821754880\,i\,x^{7}\,t^{2} \\
&&\mbox{\hspace{0.9cm}} + 1813985280\,i\,x^{5}\,t^{4} + 44971718400\,i\,t^{4}\,x
^{3} + 28343520000\,i\,t^{3}\,x^{4} \\
&&\mbox{\hspace{0.9cm}} + 102263420160\,i\,t^{5}\,x^{2} + 7482689280\,i\,t\,x^{6}
 + 29859840\,i\,t^{5}\,x^{6} + 44789760\,i\,t^{3}\,x^{8} \\
&&\mbox{\hspace{0.9cm}} + 68614413120\,t^{6} - 394116645600\,t^{4} + 29335543200
\,x^{4} - 21087578880000\,t^{5} \\
&&\mbox{\hspace{0.9cm}} + 439076554200000\,x - 1058949336600000\,t \\
&&\mbox{\hspace{0.9cm}} - 464904587876168025 - 77484097800000\,i -
619872812532704700\,x^{2} \\
&&\mbox{\hspace{0.9cm}} - 36733201920000\,x^{5} + 34437376800000\,x^{3} +
631351908000000\,t^{3} \\
&&\mbox{\hspace{0.9cm}} - 503884800\,t^{7}\,x - 9943326720\,t^{6}\,x^{2} -
11197440\,t^{8}\,x^{2} + 100776960\,t^{5}\,x^{3} \\
&&\mbox{\hspace{0.9cm}} + 77137920\,t^{6}\,x^{4} + 4534963200\,x^{5}\,t^{3} +
5542732800\,x^{4}\,t^{4} - 201553920\,t\,x^{9} \\
&&\mbox{\hspace{0.9cm}} - 5971968\,t^{2}\,x^{10} - 268738560\,t^{3}\,x^{7} -
4976640\,t^{4}\,x^{8} - 134369280\,t^{5}\,x^{5} \\
&&\mbox{\hspace{0.9cm}} - 2211840\,t^{6}\,x^{6} - 29859840\,t^{7}\,x^{3} - 552960
\,t^{8}\,x^{4} - 2488320\,t^{9}\,x - 73728\,t^{10}\,x^{2} \\
&&\mbox{\hspace{0.9cm}} - 156796352640\,t^{5}\,x + 9183300480\,x^{5}\,t +
98635449600\,x^{3}\,t^{3} - 7255941120\,x^{6}\,t^{2} \\
&&\mbox{\hspace{0.9cm}} + 425502720\,x^{6}\,t^{4} - 906992640\,x^{7}\,t +
571069440\,x^{8}\,t^{2} - 251942400000\,t^{6}\,x \\
&&\mbox{\hspace{0.9cm}} + 1360488960000\,t^{5}\,x^{2} + 2267481600000\,x^{4}\,t^{
3} + 755827200000\,x^{3}\,t^{4} \\
&&\mbox{\hspace{0.9cm}} + 4081466880000\,x^{5}\,t^{2} - 6802444800000\,x^{6}\,t
 - 44003314800\,i\,x^{3} \\
&&\mbox{\hspace{0.9cm}} - 929809186514016300\,i\,x - 26873856\,i\,x^{11} -
100776960\,i\,x^{9} \\
&&\mbox{\hspace{0.9cm}} - 4761711360\,i\,x^{7} - 36223018560\,i\,x^{5} -
68874753600000\,i\,x^{4} \\
&&\mbox{\hspace{0.9cm}} - 464904586800000\,i\,x^{2} + 176013259200000\,i\,t^{4}
 - 206624970114302700\,t^{2} \\
&&\mbox{\hspace{0.9cm}} - 50388480000\,t^{7} - 382579200\,t^{8} - 3234816\,t^{10}
 - 2985984\,x^{12} - 4096\,t^{12} \\
&&\mbox{\hspace{0.9cm}} + 87340032\,x^{10} - 1360488960000\,x^{7} + 453496320\,x
^{8} + 110592\,i\,t^{11} \\
&&\mbox{\hspace{0.9cm}} + 943201486800\,i\,t^{3} + 285645994560\,i\,t^{5} +
85847040\,i\,t^{9} \\
&&\mbox{\hspace{0.9cm}} + 929809272607458300\,i\,t + 8162933760000\,i\,x^{6} +
14016395520\,i\,t^{7} \\
&&\mbox{\hspace{0.9cm}} + 604661760000\,i\,t^{6} + 705966224400000\,i\,t^{2} -
780580540800000\,i\,t^{2}\,x^{2} \\
&&\mbox{\hspace{0.9cm}} - 103312130400000\,i\,t\,x - 784265198400\,i\,x\,t^{4} -
103737283200\,i\,x^{3}\,t^{2} \\
&&\mbox{\hspace{0.9cm}} - 13604889600000\,i\,t^{4}\,x^{2} - 11197440\,i\,t^{8}\,x
 - 1209323520\,i\,t^{3}\,x^{6} \\
&&\mbox{\hspace{0.9cm}} - 48977602560000\,i\,x^{5}\,t - 110592\,i\,x\,t^{10} -
1658880\,i\,x^{3}\,t^{8} - 9953280\,i\,x^{5}\,t^{6} \\
&&\mbox{\hspace{0.9cm}} - 29859840\,i\,x^{7}\,t^{4} - 44789760\,i\,x^{9}\,t^{2}
 - 302330880\,i\,x^{8}\,t - 12924645120\,i\,x^{5}\,t^{2}  - 15998342400\,i\,x\,t^{6}.\\
\end{eqnarray*}
\clearpage

\begin{figure}[htb]
\begin{center}
(a)\includegraphics*[height=4cm]{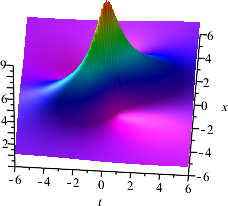}(b)\includegraphics*[height=4cm]{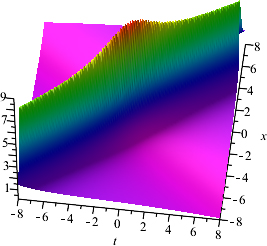}
(c)\includegraphics*[height=4.8cm]{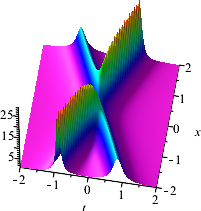}
\mbox{}\vspace{-.2cm} \caption{{(Color online)
Profiles of $|q^{[1]}_{rational}|^2$ with a given value $a=1$. From left to right,
$b=\dfrac{1}{3},b=\dfrac{3}{4}, b=3$, which shows visually the lost of the localization of this solution. Note that
there are two hollows in Fig(a),and there is a vale in Fig(b).}}
\end{center}
\end{figure}


\begin{figure}[htb]
\begin{center}
(a)\includegraphics*[height=4.5cm]{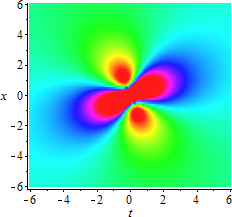}
(b)\includegraphics*[height=4.5cm]{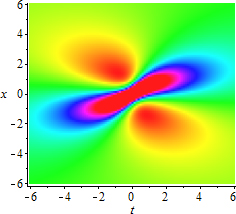}
(c)\includegraphics*[height=4.3cm]{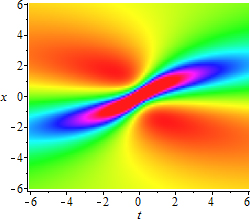}
\mbox{}\vspace{-.1cm} \caption{{(Color online) Density plot of
the rogue wave $|q^{[1]}_{rational}|^2$ with $a=1$. From left to right,
$b=0,\dfrac{1}{3},\dfrac{7}{15}$,which shows the diffusion of the peak and hollows.}}
\end{center}
\end{figure}

\begin{figure}[htb]
\begin{center}
(a)\includegraphics*[height=6cm]{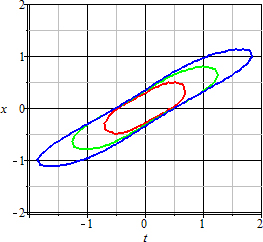}(b)\includegraphics*[height=6cm]{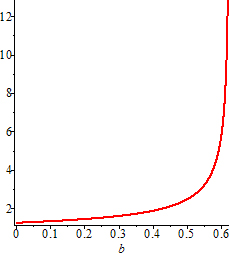}
\mbox{}\vspace{-.1cm} \caption{(Color online) The decrease in localization of
the first order rogue wave $|q^{[1]}_{rational}|^2$ with $a=1$. (a)  Contour line at height 5 with $b=0$ (inner,red),
$\dfrac{1}{3}$(middle,green),$\dfrac{7}{15}$(outer,blue). (b) The distance from minimum point of
$|q^{[1]}_{rational}|^2$ with $a=1$ to coordinate origin.}
\end{center}
\end{figure}

\begin{figure}[htb]
\begin{center}
(a)\includegraphics*[height=4.5cm]{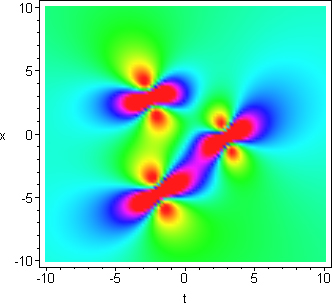}
(b)\includegraphics*[height=4.5cm]{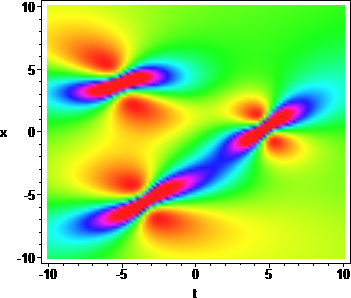}
(c)\includegraphics*[height=4.5cm]{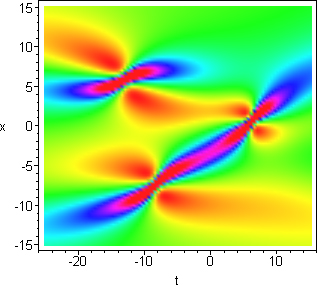}
\mbox{}\vspace{-.3cm} \caption{{(Color online) Diffusion of the second order rogue wave  through the density plot of
$|q^{[4]}_{smooth}|^2$ with $K_0=1, S_0=L_0, S_1=L_1, L_0=1, L_1=30$. From left to right, $b=0,\dfrac{1}{3},
\dfrac{7}{15}$. }}
\end{center}
\end{figure}


\begin{figure}[htb]
\begin{center}
(a)\includegraphics*[height=4cm]{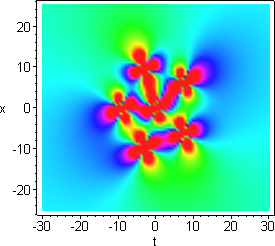}
(b)\includegraphics*[height=4cm]{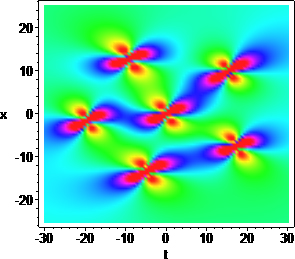}
(c)\includegraphics*[height=4cm]{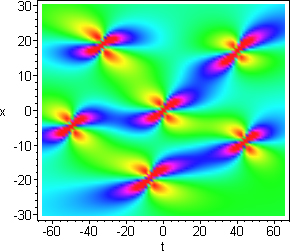}
\mbox{}\vspace{-.3cm} \caption{{(Color online)  Diffusion of the third order rogue wave through the density plot of
$|q^{[6]}_{smooth}|^2$  with $K_0=1,
S_0=L_0=S_1=L_1=0,S_2=L_2,L_2=9000$. From left to right,$b=0,\dfrac{1}{3},\dfrac{7}{15}$. }}
\end{center}
\end{figure}


\begin{figure}[htb]
\begin{center}
(a)\includegraphics*[height=4cm]{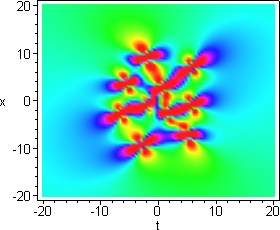}
(b)\includegraphics*[height=4cm]{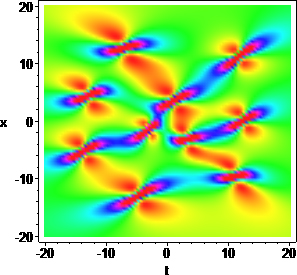}
(c)\includegraphics*[height=4cm]{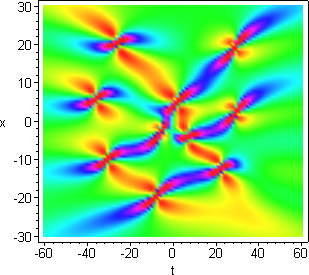}
\mbox{}\vspace{-.3cm} \caption{{(Color online)  Diffusion of the fourth order  rogue wave through the density plot of
$|q^{[8]}_{smooth}|^2$  with $K_0=1, S_0=L_0=0,
S_1=L_1=0, S_2=L_2=500, S_3=L_3=9000$. From left to right, $b=0,\dfrac{1}{3},\dfrac{7}{15}$. }}
\end{center}
\end{figure}

\begin{figure}[htb]
\begin{center}
(a)\includegraphics*[height=4.5cm]{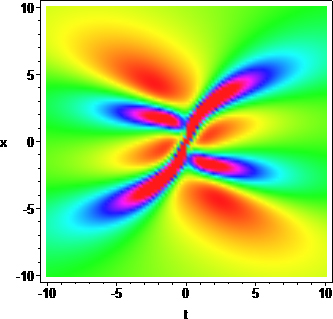}
(b)\includegraphics*[height=4.5cm]{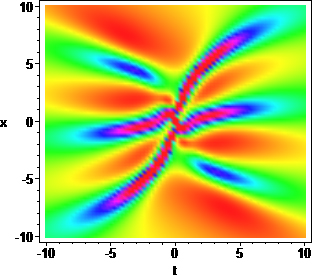}
(c)\includegraphics*[height=4.5cm]{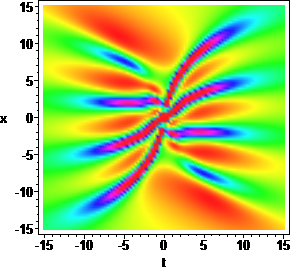}
\mbox{}\vspace{-.3cm} \caption{{(Color online) The density plot of
the fundamental pattern for rogue waves $|q^{[2k]}_{smooth}|^2(k=2,3,4)$.
a)The second order:$ K_0=1, S_0=L_0=S_1=L_1=0$;
b)The third order:$ K_0=1, S_0=L_0=S_1=L_1=S_2=L_2=0$;
c) The fourth order: $ K_0=1,
S_0=L_0=S_1=L_1=S_2=L_2=S_3=L_3=0$. }}
\end{center}
\end{figure}

\begin{figure}[htb]
\begin{center}
(a)\includegraphics*[height=4cm]{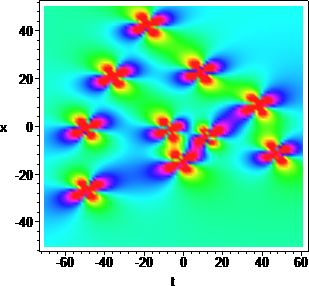}(b)\includegraphics*[height=4cm]{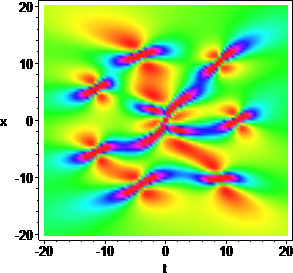} \;\;
(c)\includegraphics*[height=4cm]{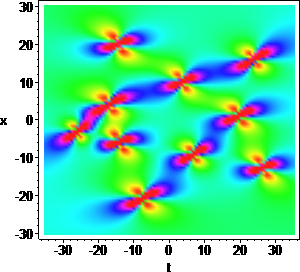}
\mbox{}\vspace{-.3cm} \caption{{(Color online)
The density plot of three patterns for the fourth order rogue wave $|q^{[8]}_{smooth}|^2$ .
a)triangle pattern with $K_0=6-2i,S_0=L_0=6+5i, S_1=L_1=600, S_2=L_2=0, S_3=L_3=0$;
b) ring-decomposition pattern with $K_0=1, S_0=L_0=0, S_1=L_1=0, S_2=L_2=0,
S_3=L_3=9000$; c) pentagon pattern $K_0=1, S_0=L_0=0, S_1=L_1=0, S_2=L_2=10000,
S_3=L_3=0$.}}
\end{center}
\end{figure}

\begin{figure}[htb]
\begin{center}
(a)\includegraphics*[height=5cm]{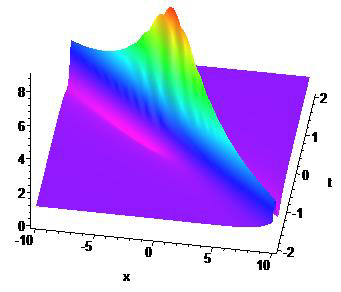}(b)\includegraphics*[height=6cm]{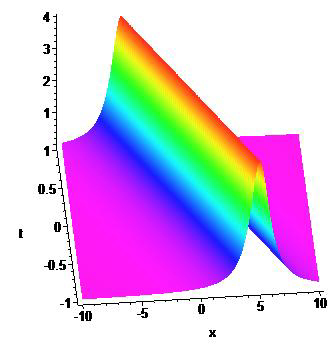} \;\;
\mbox{}\vspace{-.3cm} \caption{{(Color online)
Two solutions of the MNLS  generated from corresponding solutions of the DNLS.  The left panel is plotted for a RW, the right panel is plotted for a rational soliton.  }}
\end{center}
\end{figure}

\begin{figure}[htb]
\begin{center}
(a)\includegraphics*[height=5cm]{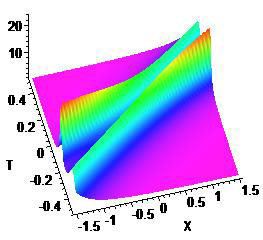}(b)\includegraphics*[height=5cm]{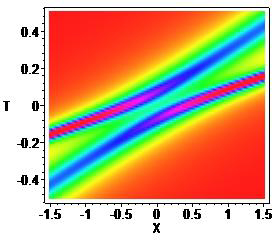} \;\;
\mbox{}\vspace{-.3cm} \caption{{(Color online)
Two peak solution of the DNLS equation generated from a corresponding solution of the MNLS in eq. (\ref{twopeak}).  The right panel is the density plot of the left panel. }}
\end{center}
\end{figure}


\begin{thebibliography}{99}
\bibitem{Kharif1} C.Kharif and E.Pelinovsky, 2003, Physical mechanisms of the rogue wave phenomenon, Eur.J.Mech.B (Fluids). 22, 603-634.
\bibitem{Kharif2}C.Kharif, E.Pelinovsky and A.Slunyaev, 2009, Rogue Waves in the Ocean (Berlin: Springer).
\bibitem{Optical1}
D. R. Solli, C. Ropers, P. Koonath and B. Jalali,  2007,  Optical
rogue waves,  Nature 450, 1054-1057.
\bibitem{Optical2}B. Kibler, J. Fatome, C. Finot, G. Millot,F. Dias,
G. Genty,N. Akhmediev and J. M. Dudley, 2010, The Peregrine soliton in
nonlinear fibre optics, Nature.Physics. 6, 790-795.
\bibitem{derman}
M.S. Ruderman, 2010,  Freak waves in laboratory and space plasmas,
Eur. Phys. J. Spec. Top, 185, 57-66.
\bibitem{Moslem}W.M.Moslem, P.K.Shukla and B.Eliasson, 2011,
Surface plasma rogue waves, Euro.Phys.Lett. 96, 25002 (5pp).
\bibitem{shukla} P.K. Shukla, W.M. Moslem, 2012,Alfv\'enic rogue waves,
Phys. Lett. A 376, 1125-1128.
\bibitem{konotop1}Yu. V. Bludov, V. V. Konotop,N. Akhmediev,2009,
Matter rogue waves,Phys.Rev.A.80,033610(5pp).
\bibitem{akhmediev1}A.Chabchoub, N.P.Hoffmann and N.Akhmediev, 2011,
 Rogue Wave Observation in a Water Wave Tank, Phys. Rev. Lett. 106, 204502 (4pp).
\bibitem{akhmediev2}A. Chabchoub, N. P. Hoffmann,
N. Akhmediev,2012, Observation of rogue wave holes in a water wave tank, J. Geophys. Res. 117, C00J02(5pp).
\bibitem{akhmediev3}
A. Chabchoub, N. Hoffmann,
M. Onorato, N. Akhmediev, 2012,
Super RogueWaves: Observation of a Higher-Order
Breather inWaterWaves, Phy.Rev. X 2, 011015(6pp).
\bibitem{Peregrine}D. H. Peregrine, 1983, Water waves, Nonlinear Schr\"odinger equations and their solutions, J. Aust. Math. Soc. Ser. B, Appl. Math. 25, 16-43.
\bibitem{Dysthe1}Kristian, B.Dysthe and K.Trulsen, 1999,
Note on Breather Type Solutions of the NLS as Models
 for Freak-Waves, Phys Scri. T82, 48-52.
\bibitem{shrira} V. I. Shrira, V. V. Geogjaev,2010
What makes the Peregrine soliton so special as a prototype
of freak waves?, J. Eng. Math. 67, 11-22.
\bibitem{zakharov1} V.E. Zakharov, 1968
Stability of periodic waves of finite amplitude on the surface of
a deep fluid,  J. Appl. Mech. and Tech. Phys. 9,190-194.
\bibitem{zakharov2} V.E.Zakharov and L.A.Ostrovsky,
2009, Modulation instability: the beginning, Phys.D. 238, 540 (8pp).
\bibitem{akhmediev4} N.N.Akhmediev and  V.I.Korneev, 1986, Modulation instability and periodic
solutions of the Nonlinear Schr\"odinger equation, Theor. Math.
Phys. 69, 1089-1093.
\bibitem{akhmediev5} M. Erkintalo, K. Hammani, B. Kibler, C.
Finot, N. Akhmediev, J. M.Dudley, and G. Genty, 2011,
Higher-Order Modulation Instability in Nonlinear Fiber Optics, Phys.
Rev. Lett. 107, 253901 (5pp).

\bibitem{Dubard2} P. Dubard, V.B. Matveev, 2011, Multi-rogue waves solutions to the focusing NLS equation and the KP-I equation, Nat. Hazards. Earth. Syst. Sci. 11, 667-672.
\bibitem{Gaillard} P. Gaillard, 2011, Families of quasi-rational solutions of the NLS equation and multi-rogue waves, J. Phys. A: Math.Theor. 44, 435204 (15pp).
\bibitem{triplets}A. Ankiewicz, J.D. Kedziora, N. Akhmediev, 2011, Rogue wave triplets, Phys. Lett.A. 375, 2782-2785.
\bibitem{ohtayang}Y. Ohta, J.K. Yang, 2012, General high-order rogue waves and their dynamics in the nonlinear Schr\"oedinger equation, Proc. R. Soc. A. 468, 1716-1740.
\bibitem{Circular}D.J. Kedziora, A. Ankiewicz, and N. Akhmediev, 2011, Circular rogue wave clusters, Phys. Rev. E. 84, 056611 (7pp).
\bibitem{guo}B.L.Guo,L.M.Ling and Q.P.Liu, 2012, Nonlinear Schr\"odinger Equation: Generalized Darboux Transformation and Rogue Wave Solutions, Phys.Rev.E 85, 026607 (9pp).
\bibitem{Gaillard2}P. Gaillard, 2013, Degenerate determinant representation of solutions
of the nonlinear Schr\"odinger equation, higher order Peregrine breathers and multi-rogue waves,
J.Math.Phys. 54,013504(32pp)
\bibitem{hegenerating2012}J.S.He, H.R. Zhang, L.H. Wang, K. Porsezian and
A.S.Fokas,2013, Generating mechanism for higher-order rogue waves,
Phys.Rev.E 87, 052914(10pp).
 \bibitem{matveev}V. B. Matveev, M.A. Salle,1991,
 Darboux Transfromations and Solitons(Springer-Verlag, Berlin).
\bibitem{hedt} J.S.He, L.Zhang, Y.Cheng and Y.S.Li, 2006, Determinant representation of
Darboux transformation for the AKNS system, Science in China Series
A: Mathematics. 49, 1867-1878.
\bibitem{kenji1}K. Imai, 1999,  Generlization of Kaup-Newell Inverse Scattering Formulation and
Darboux Transformation, J.Phys.Soc.Japan. 68, 355-359.
\bibitem{steduel} H. Steudel, 2003, The hierarchy of multi-soliton solutions of the
Derivative Nonlinear Schr\"odinger Equation, J.Phys.A: Math.Gen. 36,
1931-1946.
\bibitem{xuhe}S.W. Xu, J.S.He and L.H.Wang, 2011, The Darboux transformation of the
derivative nonlinear Schrodinger equation, J. Phys. A: Math. Theor.
44, 305203 (22pp).
\bibitem{xxuhe}S.W. Xu and J.S.He, 2012, The rogue wave and breather solution of the Gerdjikov-Ivanov
equation, J. Math. Phys. 53, 063507 (17pp).
\bibitem{gzxwhe2014}
L. J.Guo, Y. S. Zhang, S.W.Xu, Z.W. Wu and J.S. He, 2014,  The higher order rogue wave solutions of
the Gerdjikov-Ivanov equation, Phys. Scr. 89, 035501(11pp).
\bibitem{KMio}
K. Mio, T. Ogino, K. Minami and S. Takeda, 1976, Modified Nonlinear
Schr\"odinger Equation for Alfv\'en Waves Propagating along the
Magnetic Field in Cold Plasmas, J. Phys. Soc. Jpn. 41, 265-271.
\bibitem{Zabolotskii}
A. A. Zabolotskii, 1987, Self-induced transparency and quadratic
stark effect, Phys. Lett. A. 124, 500-502.
\bibitem{MikiWadati1} M. Wadati, K. Konno and Y.H. Ichikawa, 1979, A
Generalization of Inverse Scattering Method, J. Phys. Soc. Jpn. 46,
1965-1966.
\bibitem{AKundu}
A. Kundu, 1984, Landau-Lifshitz and higher-order nonlinear systems
gauge generated from nonlinear Schr\"odinger-type equations, J.
Math. Phys. 25, 3433-3438.
\bibitem{Seenuvasakumaran} K. Porsezian,
P. Seenuvasakumaran and K. Saravanan, 2000,
Next hierarchy of mixed derivative nonlinear Schr\"odinge equation, Chaos, Solitons and
Fractals. 11, 2223-2231.
\bibitem{QingDing}Q. Ding and Z.N. Zhu,  2002, On the gauge equivalent structure
of the modified nonlinear Schr\"odinger equation, Phys. Lett. A. 295,
192-197.
\bibitem{Kundu}A.Kundu, 2006, Integrable Hierarchy of Higher Nonlinear
Schr\"odinger Type Equations, Symmetry, Integrability and Geometry:
Methods and Applications. 2, 078(12pp).
\bibitem{TKawata}T.Kawata, J.I.Sakai and N.Kobayashi, 1980, Inverse Method for the
Mixed Nonlinear Schr\"odinger Equation and Soliton Solutions, J.
Phys. Soc. Jpn. 48, 1371-1379.
\bibitem{Chowdhury}
A. R. Chowdhury, S. Paul  and S. Sen, 1985, Periodic solutions of
the mixed nonlinear Schrodinger equation, Phys. Rev. D. 32,
3233-3237.
\bibitem{mihalache} D.Mihalache, N.Truta, N.C.Panoiu and
D.M..Baboiu, 1993, Analytic method for solving the modified
nonlinear schr\"odinger equation describing solition propagation
along optical fibers, Phys. Rev. A. 47, 3190-3194.
\bibitem{Doktorov2}
E. V. Doktorov, 2002, The modified nonlinear Schr\"odinger equation:
facts and artefacts, Eur. Phys. J. B. 29, 227-231.
\bibitem{Rangwala}
A.A.Rangwala and J.A. Rao, 1990,  Backlund transformations, soliton
solutions and wave functions of Kaup-Newell and
Wadati-Konno-Ichikawa systems, J. Math. Phys. 31, 1126-1132.
\bibitem{OCWrigh}
O.C. Wrigh, 2004, Homoclinic connections of unstable plane waves of
the modified nonlinear Schr\"odinger equation, Chaos, Solitons and
Fractals. 20, 735-749.
\bibitem{TiechengXia}T.C.Xia, X. H. Chen and D.Y.Chen, 2005, Darboux transformation and soliton-like solutions of
nonlinear Schr\"odinger equations, Chaos, Solitons and Fractals. 26,
889-896.
\bibitem{HaiQiangZhang} H.Q.Zhang, B.G.Zhai and X.L. Wang, 2012, Soliton and breather solutions of the modified nonlinear
Schr\"odinger equation, Phys. Scr. 85, 015007(8pp).
\bibitem{Shanliang} S. L. Liu and W. Z. Wang, 1993, Exact N-soliton solution of the modified nonlinear
Schr\"odinger equation, Phys. Rev. E. 48, 3054-3059.
\bibitem{MinLi}M. Li, B. Tian, W.J. Liu, H.Q. Zhang and P. Wang, 2010, Dark and antidark solitons in the modified nonlinear
Schr\"odinger equation accounting for the self-steepening effect,
Phys. Rev. E. 81, 046606(8pp).
\bibitem{Kitaevy}
A. V. Kitaevy and A. H. Vartanianz, 1997,  Leading-order temporal
asymptotics of the modified nonlinear Schr\" odinger equation:
solitonless sector, Inverse. Problems, 13, 1311-1339.
\bibitem{Vartaniana}
A.H. Vartaniana, 2000, Higher order asymptotics of the modified
non-linear schr\"odinger equation, Comm. PDE. 25, 1043-1098.
\bibitem{perturbation}
X. J. Chen and J.K. Yang, 2002, Direct perturbation theory for
solitons of the derivative nonlinear Schr\"odinger equation and the
modified nonlinear Schr\"odinger equation, Phys. Rev. E. 65,
066608(12pp).
\bibitem{Shchesnovich}
V.S. Shchesnovich and E.V. Doktorov, 1999, Perturbation theory for
the modified nonlinear Schr\"odinger solitons, Physica. D. 129,
115-129.
\bibitem{VMLashkin}
V. M. Lashkin, 2004, Soliton of modified nonlinear Schr\"odinger
equation with random perturbations, Phys. Rev. E. 69, 016611(11pp).
\bibitem{miller1} J. C. DiFranco and P. D. Miller, 2008,
The semiclassical modified nonlinear Schr\"odinger equation I:
Modulation theory and spectral analysis, Physica D 237,947-997.
\bibitem{miller2} J. C. DiFranco, P. D. Miller and B. K. Muite,
2011, On the modified nonlinear Schrodinger equation in the semicalassical
limit: supersonic,subsonic, and transsonic behavior,
Acta Mathematica Scientia 31B, 2343-2377.
\bibitem{miller3} J. C. DiFranco and P. D. Miller, 2012,
The semiclassical modified nonlinear Schr\"odinger equation II:
Asymptotic analysis of the Cauchy probblem. The elliptic region for
transsionic initial data, aXiv:arXiv:1208.1541v1.
\bibitem{Lenells}J. Lenells,
The derivative nonlinear Schr\"odinger equation on the half-line
Physica D 237 (2008), 3008-3019.
\bibitem{blguo}B.L. Guo, L.L.Ling and Q.P.Liu, 2013,
High-order solutions and generalized Darboux transformations of
derivative nonlinear Schrodinger equations, Stud. Appl. Math.
130, 317-344.
\bibitem{ljguo}Y.S.Zhang, L.J.Guo, S.W.Xu, Z.W.Wu and J.S.He,  2014
The hierarchy of higher order solutions of the derivative  nonlinear
Schr\"odinger equation, Commun. Nonlinear Sci. Numer. Simulat. 19, 1706-1722.
\end{thebibliography}
\end{document}